\documentclass[aps,prx,10pt,twocolumn,superscriptaddress,groupedaddress,longbibliography,showpacs]{revtex4-1}
\usepackage{bm,graphicx,amsmath,amssymb,color,placeins,tikz,pgflibraryarrows,braket,colortbl,xcolor,mathtools}
\usepackage[caption=false]{subfig}
\usetikzlibrary{decorations.pathreplacing}
\def\k{\kappa}
\def\a{\alpha}

\def\d{\delta}
\def\DD{\Delta}
\def\w{\omega}

\def\bk{{\bf k}}
\def\bK{{\bf K}}
\def\bq{{\bf q} }
\def\e{\epsilon}
\def\ve{\varepsilon}

\def\<{\langle}
\def\>{\rangle}

\def\D{\partial}
\def\Re{\hspace{0.5pt}{\rm Re}}
\def\Im{\hspace{0.5pt}{\rm Im}}
\let\hide\iffalse

\def\bk{{\bf k}}

\def\br{{\bf r}}

\def\bq{{\bf q}}

\def\a{{\alpha}}
\def\b{{\beta}}

\def\bR{{\bf R}}

\def\btau{{\bm \tau}}
\def\D{\partial}
\def\d{\delta}
\def\w{\omega}

\def\bb{{\bf b}}

\def\bG{{\bf G}}

\def\<{\langle}
\def\>{\rangle}
\def\k{\kappa}
\def\ve{\varepsilon}
\def\e{\epsilon}

\AtBeginDocument{%
    \newwrite\bibnotes
    \def\bibnotesext{Notes.bib}
    \immediate\openout\bibnotes=\jobname\bibnotesext
    \immediate\write\bibnotes{@CONTROL{REVTEX41Control}}
    \immediate\write\bibnotes{@CONTROL{%
    apsrev41Control,author="08",editor="1",pages="1",title="0",year="1"}}
     \if@filesw
     \immediate\write\@auxout{\string\citation{apsrev41Control}}%
    \fi
}%

\begin{document}

\title{Theory of the special displacement method for\\[1pt]electronic structure
calculations at finite temperature}
\author{Marios Zacharias$^{1}$}
\author{Feliciano Giustino$^{2,3}$\vspace{2pt}}
\email{fgiustino@oden.utexas.edu}
\affiliation{
$^1$Fritz-Haber-Institut der Max-Planck-Gesellschaft, Faradayweg 4–6, D-14195 Berlin, Germany\\
$^2$Oden Institute for Computational Engineering and Sciences, The University of Texas at Austin, 
Austin, Texas 78712, USA\\
$^3$Department of Physics, The University of Texas at Austin, Austin, Texas 78712, USA}

\date{\today}

\begin{abstract}
Calculations of electronic and optical properties of solids at finite temperature including 
electron-phonon interactions and quantum zero-point renormalization have enjoyed considerable 
progress during the past few years. Among the emerging methodologies in this area, we recently 
proposed a new approach to compute optical spectra at finite temperature including phonon-assisted 
quantum processes via a single supercell calculation [Zacharias and Giustino, Phys. Rev. B {\bf 94}, 
075125 (2016)]. In the present work we considerably expand the scope of our previous theory 
starting from a compact reciprocal space formulation, and we demonstrate that this improved
approach provides accurate temperature-dependent band structures in three-dimensional and 
two-dimensional materials, using a special set of atomic displacements in a single supercell 
calculation. We also demonstrate that our special displacement reproduces the thermal ellipsoids 
obtained from X-ray crystallography, and yields accurate thermal averages of the mean-square 
atomic displacements. At a more fundamental level, we show that the special displacement 
represents an exact single-point approximant of an imaginary-time Feynman's path integral 
for the lattice dynamics. This enhanced version of the special displacement method enables 
non-perturbative, robust, and straightforward {\it ab initio} calculations of the electronic 
and optical properties of solids at finite temperature, and can easily be used as a 
post-processing step to any electronic structure code. To illustrate the capabilities of 
this method, we investigate the temperature-dependent band structures and atomic displacement 
parameters of prototypical nonpolar and polar semiconductors and of a prototypical two-dimensional 
semiconductor, namely Si, GaAs, and monolayer MoS$_2$, and we obtain excellent agreement with 
previous calculations and experiments. Given its simplicity and numerical stability, the present 
development is suited for high-throughput calculations of band structures, quasiparticle corrections, 
optical spectra, and transport coefficients at finite temperature.
\end{abstract}

\maketitle

\section{Introduction}\label{sec.1}

The calculation of the electronic and optical properties of materials at finite temperature
is a long-standing challenge for {\it ab initio} electronic structure methods. In typical 
semiconductors, insulators, metals, and semiconductors, the key mechanism leading to temperature 
dependent properties is the thermal motion of the atoms in the crystal lattice, and the effect of 
this motion on the electronic structure of the system. Recent advances have made it possible to 
study these effects from first principles with predictive accuracy~\cite{Giustino_2017}.

In the case of semiconductors and insulators, one problem that attracted considerable attention 
during the past decade is the electron-phonon renormalization of band structures, including both 
quantum zero-point effects and temperature dependence~\cite{Capaz_2005,Marini_2008,Giustino_2010,
Cannuccia_2011,Gonze_2011,Cannuccia_2012,Ponce_2014_2,Kawai_2014,Antonius_2014,Ponce_2014,
Ponce_2015,Antonius_2015,Molina_2016,Bartomeu_2014,Engel_2015,Bartomeu_2015,Patrick_2015,
Bartomeu_2016,Bartomeu_2016_GW,Nery_2016,Allen_2017,Karsai_2018}. This problem is becoming 
increasingly important as we strive to achieve close quantitative agreement between {\it ab initio} 
calculations and experimental data. Furthermore this problem underpins calculations of many 
important properties, from temperature-dependent optical absorption~\cite{Noffsinger_2012,
Zacharias_2015,Zacharias_2016} and emission spectra~\cite{Paleari_2019} to 
temperature-dependent transport coefficients~\cite{Gunst_2017,Palsgaard_2018}.

The calculation of temperature-dependent electronic and optical properties has been demonstrated 
using both perturbative approaches based on density functional perturbation theory~\cite{Baroni_2001} 
(DFPT) calculations in the crystal unit cell, and non-perturbative approaches based on density 
functional theory (DFT) calculations in large supercells. In perturbative methods the key ingredient 
of the calculations is the electron-phonon matrix element, that is the scattering matrix element 
between two Kohn-Sham states which results from the variation of the self-consistent potential 
associated with a vibrational eigenmode. The matrix elements are employed to obtain temperature-dependent 
band structures in the Allen-Heine (AH) method~\cite{Allen_Heine_1976,Allen_Cardona_1981}, and 
to compute indirect optical absorption in the Hall, Bardeen, and Blatt theory~\cite{Hall_1954,
Parravicini_1975}. These approaches have enjoyed considerable success during the past decade 
across a broad range of materials~\cite{Marini_2008,Giustino_2010,Cannuccia_2011,Cannuccia_2012,
Noffsinger_2012,Antonius_2014,Ponce_2014,Ponce_2014_2,Kawai_2014,Antonius_2015,Ponce_2015,
Molina_2016,Saidi_2016,Menendez_2017,Yanfeng_2019_arxiV,Querales_2019}. 

Non-perturbative supercell-based methods offer an alternative approach to calculations of 
electronic and optical properties at finite temperature. The central idea of these methods 
is that the interactions between electrons and phonons can be described using large 
supercells where the atoms are displaced to capture thermal disorder~\cite{Monserrat_2018}. 
Supercell methods derive from earlier frozen-phonon approaches~\cite{Chang_1986,Capaz_2005}, 
and do not require the explicit evaluation of electron-phonon matrix elements. In order 
to sample thermal disorder, the first attempts in this area relied on the Monte 
Carlo sampling of quantum nuclear wavefunctions, either in the atomic configuration
space~\cite{Patrick_2013,Patrick_2014,Zacharias_2015} or in the space of normal vibrational 
modes~\cite{Bartomeu_2016,Bartomeu_2016_GW,Ivona_2019}. The formal basis for this approach 
is provided by a theory developed in the 1950s by Williams~\cite{Williams_1951} and Lax~\cite{Lax_1952} 
to study defects in solids, and recently extended by Zacharias {\it et al}. to finite-temperature 
optical spectra, phonon-assisted optical processes, and band gap renormalization~\cite{Zacharias_2015}.

The central idea behind Monte Carlo supercell methods is that the temperature dependence 
of the dielectric function $\e ( \omega;T)$ results from a multivariate Gaussian integral given by~\cite{Zacharias_2016}: 
 \begin{eqnarray}\label{eq.WL-epsilon_2}
  \e ( \omega;T) = \prod_\nu \int x_\nu \frac{\text{exp}(-x^2_\nu/2 \sigma^2_{\nu,T} )}
                     {\sqrt{2\pi}\sigma_{\nu,T}} \e ( \omega;x) 
 \end{eqnarray}
where the product runs over the normal coordinates $x_\nu$, and
each width $\sigma_{\nu,T}$ of the independent Gaussians represents the root
mean square displacement of the atoms at temperature $T$ along a phonon mode $\nu$.
The interpretation of Eq.~\eqref{eq.WL-epsilon_2} is that the temperature dependence of the spectrum
can directly be obtained as an ensemble average over the spectra calculated at fixed nuclear coordinates $x$, 
whose probability density function is a multivariate Gaussian.

The disadvantages of stochastic supercell approaches are that (i) it is difficult to control
the rate of convergence with the number of random samples, and (ii) the final outcome of the
calculations is a distribution of values at each temperature, e.g. band gaps, rather than 
a unique value as in perturbative methods. Furthermore, performing stochastic sampling on large
supercells can be prohibitively costly from the computational standpoint. In order to overcome 
these limitations, in Ref.~[\onlinecite{Zacharias_2016}] we demonstrated that it is possible to
replace the stochastic sampling of the nuclear wavefunctions by a {\it single} supercell 
calculation with a precise choice of atomic displacements. The possibility of studying electronic 
and optical spectra at finite temperature using a single supercell calculation enabled several 
interesting applications. For example, in Ref.~[\onlinecite{Zacharias_2016}] we reported 
temperature-dependent phonon-assisted optical spectra of Si, diamond, and GaAs based on this 
method; Refs.~[\onlinecite{Tathagata_2017}],~[\onlinecite{Zhang_2018}] and~[\onlinecite{Kang_2018}] 
applied this method to the dielectric function of Zn$_2$Mo$_3$O$_8$, metallic hydrogen, and 
BaSnO$_3$, respectively; Ref.~[\onlinecite{Karsai_2018}] used this technique to compute GW 
quasiparticle band gaps at finite temperature for diamond, BN, SiC, Si, AlP, ZnO, GaN, ZnS, 
GaP, AlAs, ZnSe, CdS, GaAs, Ge, AlSb, CdSe, ZnTe, and CdTe; Ref.~[\onlinecite{Karsai_2018_b}] 
calculated exciton-phonon couplings in hexagonal boron nitride; and Refs.~[\onlinecite{Gunst_2017},~\onlinecite{ 
Palsgaard_2018}] demonstrated calculations of finite-temperature carrier mobilities in silicon 
$n$-i-$n$ and $p$-$n$ junctions using this approach. In retrospect, these successes across 
a broad range of applications are not too surprising, since the method of Ref.~[\onlinecite{Zacharias_2016}] 
is designed to provide, in the limit of large supercell, the exact thermodynamic average 
of any property that can be expressed in the form of Fermi's golden rule: this includes in 
principle photoelectron spectra (hence band gaps and band structures), optical spectra, tunneling 
spectra, and transport coefficients.

One potential limitation of Ref.~[\onlinecite{Zacharias_2016}] is that the theory was developed 
using a $\Gamma$-point formalism, therefore phonon calculations to determine the vibrational 
eigenmodes in the supercell are demanding. Furthermore, by addressing only $\Gamma$-point 
properties, the calculations are limited to angle-integrated spectra, such as density-of-states 
(DOS) and optical absorption spectra. Finally, it has been pointed out that the choice of the 
special displacement employed in Ref.~[\onlinecite{Zacharias_2016}] might be improved to reduce 
the size of the supercell required to achieve convergence~\cite{Karsai_2018}.

In this manuscript, we significantly expand the scope of the methodology introduced in
Ref.~[\onlinecite{Zacharias_2016}] by formulating the theory within a compact reciprocal
space formulation, and exploiting translational invariance and time-reversal symmetry.
This upgrade allows us to determine the special supercell displacement using quantities 
that are computed in a crystal unit cell via DFPT. In a nutshell, we demonstrate that a 
supercell calculation where the atoms are displaced according to:
 \begin{eqnarray}\label{eq.realdtau_method00}
   \DD\btau_{ p\k} &=&  \!\sum_{\bq  \in \mathcal{B}, \nu}\! 
   S_{\bq  \nu} \left[\!\frac{\hbar}{2 N_p M_\k \w_{\bq\nu}} ( 2n_{\bq \nu,T} 
   + 1) \!\right]^{\!\frac{1}{2}} \nonumber \\ &\times& 2\,{\rm Re}
   \Big[ e^{i\bq \cdot {\bf R}_p} {\bf e}_{\k,\nu} (\bq ) \Big].
 \end{eqnarray}
yields the {\it exact} thermodynamic average of electronic and optical properties at the
temperature $T$ in the adiabatic and harmonic approximations. In the above expressions
$\DD\btau_{ p\k}$ indicates the displacement of the atom $\k$ with mass $M_\k$ in the unit 
cell with lattice vector ${\bf R}_p$, and $N_p$ is the number of unit cells in the supercell.
${\bf e}_{\k,\nu} (\bq )$ is the phonon polarization vector of the normal mode (normalized
within the unit cell) with wavevector $\bq$, branch index $\nu$, frequency $\w_{\bq\nu}$, 
and Bose-Einstein occupation $n_{\bq \nu,T}$. The quantities $S_{\bq  \nu}$ are signs, $+$
or $-$, which depend on the normal mode, as specified in Sec.~\ref{sec.ZG_theory}. 
The summation is restricted to phonon wavevectors that are not time-reversal partners.
 In particular, we define this group of phonons as set $\mathcal{B}$. The real
part in the equation arises from grouping together a phonon with wavevector $\bq\in \mathcal{B}$ 
and its partner $-\bq$.
Phonon wavevectors that coincide with their time-reversal partners are
grouped in a {\it finite} set $\mathcal{A}$, and their contribution to the atomic displacements vanishes in
the limit of dense Brillouin-zone sampling. This partitioning is
discussed in greater detail in Sec.~\ref{sec.WL_theory}. 

Equation~(\ref{eq.realdtau_method00}) reduces to our previous prescription provided in 
Ref.~[\onlinecite{Zacharias_2016}] if we perform $\Gamma$-point sampling, but is much more 
powerful since the construction of the special displacement relies on DFPT calculations in 
the unit cell. Therefore the only expensive step is {\it one calculation} of the desired 
property in a large supercell. To demonstrate the power of this approach, we report 
for the first time complete temperature-dependent {\it band structures} based on 
Eq.~(\ref{eq.realdtau_method00}) for both three-dimensional semiconductors, Si and GaAs, 
and two-dimensional semiconductors (MoS$_2$), and we show that this method delivers
an accuracy comparable to perturbative techniques, without requiring the so-called 
rigid-ion approximation of the Debye-Waller self-energy~\cite{Ponce_2015,Molina_2016}.
We also demonstrate that Eq.~(\ref{eq.realdtau_method00}) is an exact single-point
approximant to an imaginary-time path integral in the space of nuclear configurations,
and that it reproduces very accurately the thermal displacement ellipsoids measured
by X-ray diffraction (XRD).

The method of Ref.~[\onlinecite{Zacharias_2016}] was originally named `one-shot 
configuration', but subsequently it has variably been referred to as the `WL-HA' 
method~\cite{Zhang_2018} or the `STD' method~\cite{Gunst_2017}. In order to avoid a 
proliferation of acronyms, in this manuscript we will refer to the general theory as 
the {\it special displacement method} (SDM), and to Eq.~(\ref{eq.realdtau_method00}) 
as the {\it ZG displacement}.

The manuscript is organized as follows. In Sec.~\ref{sec.WL_theory} we describe the
Williams-Lax theory that underpins our method, and we write the key equations using 
a reciprocal-space formulation.  In Sec.~\ref{sec.DW_factor} we establish the connection 
between the ZG displacement and thermal ellipsoids.
In Sec.~\ref{sec.link_to_PI} we show that the Williams-Lax 
theory can be recast in the form of an imaginary-time Feynman's path integral, and that 
the ZG displacement enables an efficient single-point evaluation of this path integral.
Section~\ref{sec.ZG_theory} is devoted to the construction of the ZG displacement and 
the determination of the mode signs $S_{\bq \nu}$ in Eq.~\eqref{eq.realdtau_method00}.
In Sec.~\ref{sec.methods_details} we provide all computational details of the present 
calculations, including the unfolding of bands and spectral functions. 
In Sec.~\ref{sec.method-results} we briefly outline
the procedure that we follow to determine the ZG displacement, and we demonstrate 
our main computational results. 
We show the temperature-dependent thermal ellipsoids, band structures and 
band gaps of Si, GaAs, and monolayer MoS$_2$ using the SDM.  Here we also provide 
detailed convergence tests and show that accurate results can be obtained even with 
relatively small supercells. 
In Sec.~\ref{sec.conclusion} 
we summarize our key findings and discuss future directions. More technical aspects are left to the Appendices.

\section{The Williams-Lax theory}\label{sec.WL_theory}

\subsection{General remarks}

The theoretical framework underlying our present methodology was laid out in two works 
by Williams~\cite{Williams_1951} and Lax~\cite{Lax_1952} in the 1950s, which we refer 
to collectively as the Williams-Lax theory. This theory starts from the Herzberg-Teller 
rate~\cite{Herzberg_1966,Patrick_2014} that describes transitions between coupled 
electron-phonon states driven by an external field, and replaces the final quantum nuclear 
states by a semiclassical continuum. Formally this step corresponds to neglecting commutators
involving the nuclear kinetic energy operator, and is related (albeit not identical) to 
the adiabatic Born-Oppenheimer approximation~\cite{Patrick_2014}. This approach is also 
closely related to Feynman path integrals~\cite{Feynman_1965}, as we discuss in detail in 
Sec.~\ref{sec.link_to_PI}.

If we denote a Born-Oppenheimer quantum state using the ket $|\alpha n\>$, with the Greek 
letter referring to the electronic part and the integer to the nuclear part~\cite{Patrick_2014}, 
then the Williams-Lax theory provides a semiclassical approximation for the transition rate 
from an initial state $|\alpha n\>$ to all final states $|\beta m\>$ with the same $\beta$ 
and every possible vibrational state $m$, evaluated using Fermi's golden rule:
  \begin{equation}\label{eq.transition_rate}
   \Gamma_{\alpha n \rightarrow \beta} (\w) = \!\!\int \!d\tau |\chi_{\alpha n}(\{\tau\})|^2\, 
   \Gamma_{\alpha \rightarrow\beta}^{\{\tau\}}(\w).
  \end{equation}
In this expression $\w$ is the frequency of the driving external field, the $\chi_{\alpha n}
(\{\tau\})$ are the quantum nuclear wavefunctions for the potential energy surface associated 
with the electronic state $\alpha$, and $\{\tau\}$ denotes the set of all atomic coordinates. 
$\Gamma_{\alpha \rightarrow\beta}^{\{\tau\}}(\w)$ is the transition rate evaluated with the 
atoms clamped in the positions ${\{ \tau \}}$:
  \begin{equation}\label{eq.transition_rate_clamped}
   \Gamma_{\a\rightarrow \beta}^{\{ \tau \}}(\w) = \frac{2\pi}{\hbar} 
   |M^{\{ \tau \}}_{\a\rightarrow \beta}|^2 \delta\left(E^{\{ \tau \}}_{\beta} - 
   E^{\{ \tau \}}_{\a} - \hbar \w\right),
  \end{equation}
where $E^{\{ \tau \}}_\alpha$, $E^{\{ \tau \}}_\b$, and $M^{\{ \tau \}}_{\a\rightarrow \beta}$
denote the energies of the initial and final electronic states and the associated transition
matrix elements, respectively, all evaluated at clamped atoms. For example $\Gamma_{\alpha 
\rightarrow\beta}^{\{\tau\}}(\w)$ can represent the optical transition rates for light 
absorption, or the charge current in electron tunneling. Equation~(\ref{eq.transition_rate})
formalizes the intuitive concept that the transition rate including quantum nuclear effects 
can be obtained by averaging `static' clamped-ion rates over the nuclear probability distributions 
$|\chi_{\alpha n}(\{\tau\})|^2$ of the initial state.

The transition rate at a finite temperature $T$ is obtained from Eq.~(\ref{eq.transition_rate}) 
by carrying out a canonical average over the vibrational quantum numbers of the initial 
state~\cite{Lax_1952,Patrick_2014,Zacharias_2016}:
 \begin{equation}\label{eq.WL}
  \Gamma_{\alpha \rightarrow \beta} (\w,T) = \frac{1}{Z} \sum_n \exp(-E_{\alpha n}/k_{\rm B}T)\, 
  \Gamma_{\alpha n \rightarrow \beta} (\w),
 \end{equation}
where $k_{\rm B}$ stands for the Boltzmann constant and $Z = \sum_n \exp(-E_{\alpha n}/k_{\rm B}T)$ 
is the canonical partition function. Here $E_{\alpha n}$ denotes the vibrational energy of 
the state $|\alpha n\>$. This approach has been used in a number of investigations and is 
well established by now~\cite{Williams_1951,Lax_1952,Sala_2004,Patrick_2014,Zacharias_2015,
Zacharias_2016,Tathagata_2017,Monserrat_2018_b,Morris_2018,Karsai_2018}.

Since Eqs.~(\ref{eq.transition_rate})-(\ref{eq.WL}) involve the calculations of transition rates 
for displaced atomic configurations $\{ \tau \}$, the results automatically incorporate the effect 
of electron-phonon couplings, because the method probes the change in the electronic energies 
and wavefunctions upon displacing the atoms. We emphasize that Eqs.~(\ref{eq.transition_rate})-(\ref{eq.WL})
apply to any property that can be obtained from the Fermi golden rule, in particular optical 
spectra, photoemission spectra, transport coefficients, and tunneling spectra. Furthermore, these 
equations provide the formal basis for computing temperature-dependent electronic eigenvalues. 
Indeed, if we consider transitions from an electronic state $|\alpha n\>$ of a solid into free 
electron states, as in photoemission experiments, the final free electron states are independent 
of atomic positions, therefore the first frequency moment of the transition rates, $\int d\w \,\w\, 
\Gamma_{\alpha \rightarrow \beta} (\w,T)$, yields the temperature-dependent electronic energy of the
initial state: 
 \begin{equation}\label{eq.T_dep_ene}
  E_{\alpha}(T) = \frac{1}{Z} \sum_n \exp(-E_{\alpha n}/k_{\rm B}T) \int d\tau 
  |X_{\alpha n}(\{\tau\})|^2 E^{\{ \tau \}}_{\alpha}.
  \end{equation}
This relation states that, in the Williams-Lax theory, the temperature-dependent electronic energy 
$E_{\alpha}(T)$ can be obtained by averaging the position-dependent energy $E^{\{ \tau \}}_{\alpha}$ 
over thermal fluctuations. This observation forms the basis for all non-perturbative supercell 
calculations of temperature-dependent electronic properties, and is also at the basis of the 
adiabatic formulation of the Allen-Heine theory of temperature-dependent band structures~\cite{Allen_Heine_1976}.

An important implication of Eq.~(\ref{eq.T_dep_ene}) is that, if we describe ions classically, 
then $E_{\alpha}(T)$ corresponds to averaging electronic energies over the snapshots of a molecular 
dynamics trajectory. This concept has been used extensively in the literature, but to the best of 
our knowledge the connection to the Williams-Lax theory is still not fully appreciated. 
  
\subsection{Reciprocal space formulation}\label{sec.recsp}

In this section we recast Eqs.~(\ref{eq.transition_rate})-(\ref{eq.WL}) in a language appropriate 
for {\it ab initio} calculations in periodic crystals. From now on we assume the harmonic 
approximation, so that we can define phonons in the usual way. We employ the same standard 
notation as in Ref.~[\onlinecite{Giustino_2017}], and we choose the convention for the phase of 
vibrational eigenmodes as in Ref.~[\onlinecite{Maradudin_1968}]. 

We consider a Born-von K\'arm\'an supercell containing $N_p$ unit cells with $M$ atoms each. 
The position vector of the atom $\k$ in the unit cell $p$ at zero temperature is ${\boldsymbol 
\tau}_{\k p} = {\bf R}_p + {\boldsymbol \tau}_\k$, where ${\bf R}_p$ indicates a direct lattice 
vector, and $\btau_\k$ is the position vector in the primitive unit cell, with Cartesian components 
$\tau_{p\k\alpha}$. The atomic displacements from equilibrium are written as a linear combination 
of normal vibrational modes~\cite{Giustino_2017}:
 \begin{equation}\label{eq.dtau}
  \DD\tau_{ p\k \a} = N_p^{-1/2} \bigg(\frac{M_0}{M_\k}\bigg)^{1/2} \sum_{ \bq \nu} 
  e^{i\bq \cdot {\bf R}_p} e_{\k\a,\nu} (\bq ) \,z_{\bq  \nu}, 
 \end{equation}
where the $z_{\bq  \nu}$ are the complex-valued normal coordinates~\cite{Bruesch_book_1982}. 
The inverse relation is:
 \begin{equation}\label{eq.zeta}
   z_{\bq  \nu} = N_p^{-1/2}  \sum_{p\k\a}  \bigg(\frac{M_\k}{M_0}\bigg)^{1/2} e^{-i\bq 
   \cdot {\bf R}_p} e^*_{\k\a,\nu} (\bq ) \, \DD\tau_{p\k \a} \, .
 \end{equation}
In both equations $M_0$ is an arbitrary reference mass, usually chosen to be the proton mass, 
and the summations run over all the $3MN_p$ atomic degrees of freedom. For completeness, in 
Appendix~\ref{app.normal_modes} we give the standard textbook relations between normal modes. 
  
If we write $z_{\bq  \nu} = x_{\bq  \nu} + iy_{\bq  \nu}$, with $x_{\bq  \nu}$ and 
$y_{\bq  \nu}$ being the real normal coordinates, Eq.~\eqref{eq.zeta} and time-reversal 
symmetry imply $x_{-\bq \nu} = x_{\bq \nu}$ and $y_{-\bq \nu} = -y_{\bq \nu}$. Therefore only
half of the real normal coordinates are independent.  The wavevectors $\bq$ of these independent 
coordinates occupy half of the first Brillouin zone. This is illustrated in Fig.~\ref{fig_1}, 
where we consider a regular $\bq$-grids of size 8$\times$8. This grid can be partitioned in three sets, $\mathcal{A}$, $\mathcal{B}$, and 
$\mathcal{C}$, following Appendix~B of Ref.~[\onlinecite{Giustino_2017}]. 
Set $\mathcal{A}$ includes the $\bq$-points that are invariant under inversion, modulo a 
reciprocal lattice vector. Therefore the $\bq$-points in this set correspond to the center of 
the Brillouin zone, the centers of its faces, and the corners, as shown by the filled disks 
in Fig.~\ref{fig_1}. Set $\mathcal{B}$ includes all the $\bq$-points that are not inversion 
partners (modulo a reciprocal lattice vector), as shown by the empty circles in Fig.~\ref{fig_1}. 
Set $\mathcal{C}$ is obtained by changing the sign of all the points in $\mathcal{B}$, and is 
denoted by crossed filled circles in Fig.~\ref{fig_1}. Using this partitioning we can write 
the atomic displacements in terms of independent real normal coordinates as:  
 \begin{eqnarray}\label{eq.realdtau}
  \DD\tau_{ p\k \a} &=& N_p^{-1/2} \bigg(\frac{M_0}{M_\k}\bigg)^{1/2} \Bigg[ \sum_{\bq  
  \in \mathcal{A}, \nu} e_{\k\a,\nu} (\bq ) x_{\bq  \nu} \text{cos}(\bq \cdot {\bf R}_p) 
  \nonumber \\ &+& 2 \Re \!\!\sum_{\bq  \in \mathcal{B}, \nu} e^{i\bq \cdot {\bf R}_p} 
  e_{\k\a,\nu} (\bq ) (x_{\bq  \nu} + iy_{\bq \nu}) \Bigg].
 \end{eqnarray}
Equation~(\ref{eq.realdtau}) allows us to write the total quantum nuclear wavefunction of the 
state $|\alpha n\>$ in the harmonic approximation~\cite{Ziman_1960}:
 \begin{eqnarray}\label{eq.ind_harm_oscil}
   && \chi_n(\{\btau_{ p\k}+\DD\btau_{ p\k}\}) = \nonumber \\ && \hspace{40pt}\prod_{\bq  \in A, \nu}\!\!
   \chi_{n_{\bq \nu}}(x_{\bq \nu}) \!\!\prod_{\bq  \in \mathcal{B}, \nu}\!\!  
   \chi_{n_{\bq \nu}}(x_{\bq \nu}) \chi_{n_{\bq \nu}}(y_{\bq \nu}).
 \end{eqnarray}
Here we omitted the subscript $\alpha$ for notational simplicity. $\chi_{n_{\bq \nu}}(x)$ represents 
the wavefunction of a quantum harmonic oscillator with frequency $\w_{\bq\nu}$ and quantum number 
$n_{\bq \nu}$; in particular, for $\bq \in \mathcal{A}$ we have:
 \begin{equation}
   \chi_{n_{\bq \nu}}(x) = \frac{(4\pi l_{\bq\nu}^2)^{-1/4}}{\sqrt{2^{n_{\bq\nu}}
   n_{\bq\nu} !\ }}e^{-x^2/8l_{\bq\nu}^2} H_{n_{\bq \nu}}(x/2l_{\bq\nu}),
 \end{equation}
while for $\bq \in \mathcal{B}$ we have: 
  \begin{equation}
   \chi_{n_{\bq \nu}}(x) = \frac{(\pi l_{\bq\nu}^2)^{-1/4}}{\sqrt{2^{n_{\bq\nu}}
   n_{\bq\nu} !\ }}e^{-x_{\bq \nu}^2/2l_{\bq\nu}^2} H_{n_{\bq \nu}}(x/l_{\bq\nu}).
 \end{equation}
Here $l_{ \bq \nu}=  (\hbar/2 M_{0} \w_{\bq \nu})^{1/2} $ is the zero-point vibrational amplitude, 
and $H_m(x)$ denotes the Hermite polynomial of order $m$. The total energy of the state $\chi_n$ 
is given by the standard expression:
 \begin{equation}\label{eq.ind_harm_oscil_ene}
    E_n = \sum_{\bq \nu} (n_{ \bq \nu} +1/2) \hbar \w_{ \bq \nu}.
 \end{equation}
By substituting Eqs.~\eqref{eq.ind_harm_oscil}-\eqref{eq.ind_harm_oscil_ene} inside 
Eqs.~\eqref{eq.transition_rate} and \eqref{eq.WL}, and using Mehler's sum rule for Hermite 
functions~\cite{Watson_1933}, we obtain the compact expression~\cite{Patrick_2014}:
 \begin{eqnarray}\label{eq.WL3}
    && \hspace{-5pt}\Gamma_{\alpha\rightarrow \beta}(\w,T) = 
    \prod_{\bq  \in \mathcal{A}, \nu} \! \int \!\frac{dx_{\bq \nu}}{2
    \sqrt{ \pi} \sigma_{\bq \nu}} e^{-\frac{x^2_{\bq \nu}}{4 
    \sigma^2_{ \bq \nu}} }  \nonumber \\ &&\quad\times \!\!\!  \prod_{\bq  \in 
    \mathcal{B}, \nu} \!\int\! \frac{dx_{\bq \nu} dy_{\bq \nu} }{\pi \sigma^2_{\bq \nu}} 
    e^{-\frac{(x^2_{\bq \nu} + y^2_{\bq \nu} )}{\sigma^2_{ \bq \nu}} } \Gamma_{\alpha\rightarrow 
    \beta}^{\{x_{\bq \nu},y_{\bq \nu} \}}(\w), \hspace{10pt}
 \end{eqnarray}
 where $\sigma_{\bq \nu}$ is 
the mean square displacement of the normal mode at the temperature $T$ and is given by: 
 \begin{eqnarray}\label{eq.sigma}
  \sigma^2_{ \bq \nu} = l^2_{ \bq \nu} ( 2n_{\bq \nu,T} + 1).
 \end{eqnarray}
Here $n_{\bq \nu,T}=[{\rm exp}(\hbar \w_{\bq \nu}/k_{\rm B} T)-1]^{-1}$ is the Bose-Einstein 
occupation of the mode with frequency $\w_{\bq \nu}$.

Equation~\eqref{eq.WL3} states that the temperature-dependent transition rate $\Gamma_{\alpha\rightarrow 
\beta}(\w,T)$ is obtained by averaging the transition rates calculated at clamped atoms for a 
variety of atomic configurations specified by the normal coordinates $\{ x_{\bq  \nu},y_{\bq  \nu} \}$, 
and that the average is to be taken using a multidimensional Gaussian importance function.
In this context the temperature $T$ sets the width of each Gaussian via Eq.~\eqref{eq.sigma}.
As a sanity check, we note that if $\Gamma_{\alpha\rightarrow \beta}^{\{\tau \}}(\w)$ 
does not depend on the atomic coordinates, Eq.~\eqref{eq.WL3} yields the correct temperature-independent 
rate.

The core of the special displacement method described in this manuscript is to identify {\it one} 
set of atomic displacements so that a single evaluation of $\Gamma_{\alpha\rightarrow \beta}^{\{ 
x_{\bq\nu},y_{\bq\nu} \}}(\w)$ yields the same result as the multidimensional integral in 
Eq.~\eqref{eq.WL3}.  From this perspective, the task of finding the ZG displacement is similar 
to finding the mean value point of a definite integral; the only difference is that we are dealing 
with a multi-dimensional integral with hundreds to thousands of variables.

\section{ZG displacement and thermal ellipsoids} \label{sec.DW_factor}

In this section we discuss the physical meaning of the ZG displacement in 
Eq.~\eqref{eq.realdtau_method00}. In particular we prove rigorously that the ZG displacement 
reproduces the displacement autocorrelation function obtained from the quantum canonical 
average~\cite{Kittel_1976}, and therefore encodes information about the thermal ellipsoids 
measured via XRD. Furthermore, using explicit calculations, we demonstrate that the ZG 
displacement yields the correct thermal distribution of the atomic coordinates to all orders.

When a harmonic crystal is in thermodynamic equilibrium at the temperature $T$, 
the atomic displacements $\DD \tau_{ p \k \a}$ follow a normal distribution. This a direct 
consequence of the fact that the marginal distribution of a multivariate normal is also normal. 
In particular we have (cf.\ Eq.~7.2.21 of Ref.~[\onlinecite{Maradudin_Weiss_1963}]):
 \begin{equation}\label{eq.prob_distribution}
    P(\DD \tau_{ p \k \a};T) = \frac{1}{\sqrt{2 \pi} \sigma_{\k\a}(T)} 
    \text{exp}\left[-\frac{\DD \tau_{ p \k \a}^2}{2 \sigma^2_{\k\a}(T)}\right],
  \end{equation}
where the width $\sigma_{\k\a}(T)$ of the Gaussian is given by (cf.\ Eq.~7.2.5 of 
Ref.~[\onlinecite{Maradudin_Weiss_1963}]):
  \begin{equation}\label{eq.Debye_waller_factor}
   \sigma^2_{\k\a}(T) = \frac{2}{N_p} \sum_{\bq \in \mathcal{B}, \nu} 
   |e_{\k \a,\nu} (\bq)|^2 \frac{\hbar}{2M_\k \w_{\bq\nu}} (2n_{\bq\nu,T}+1). 
 \end{equation}
Here the limit of dense Brillouin-zone sampling is implied; in this limit the contribution
to the sum of phonons with $\bq \in \mathcal{A}$ vanishes. The width $\sigma_{\k\a}(T)$ is 
a particular case of the tensor of anisotropic displacement parameters (ADPs), defined 
as~\cite{Trueblood_1996}: 
  \begin{equation}\label{eq.adp}
   U_{\k,\a\b}(T) = \< \DD \tau_{ p \k \a} \DD \tau_{ p \k \b} \>_T,
 \end{equation}
where $\< \cdot \>_T$ denotes the canonical average over vibrational quantum states.
In fact, by combining Eqs.~\eqref{eq.adp} and~\eqref{eq.prob_distribution} it follows 
directly that $\sigma^2_{\k\a}(T) = U_{\k,\a\a}(T)$.  The ADPs of Eq.~\eqref{eq.adp} are 
the values employed to generate thermal ellipsoids when visualizing crystal structures at 
finite temperature.

{If we start from the ZG displacement, and we take the mean square displacement 
of the atom $\k$ over all the unit cells of the supercell, then
we obtain precisely the thermodynamic 
average given by Eq.~(\ref{eq.Debye_waller_factor}).  In fact, by replacing the ZG displacement 
of Eq.~\eqref{eq.realdtau_method00} inside the sum $\sum_{p} \DD\tau_{p \k \a}^2/N_p$, and 
using the sum rule in Eq.~\eqref{eqa.sum_rules}, we find immediately:
 \begin{eqnarray}\label{eq.mean_ampl_1C}
  &&\frac{1}{N_p} \!\sum_{p} \DD\tau_{p \k \a}^2 \! = \frac{2}{N_p}\! \sum_{\bq \in 
  \mathcal{B}, \nu} \!|e_{\k \a,\nu} (\bq)|^2 \frac{\hbar}{2M_\k \w_{\bq\nu}} (2n_{\bq\nu,T}+1)  
  \nonumber \\ &&\,\,+
  \frac{2}{N_p} \frac{M_0}{M_\k}  \sum_{\bq \in \mathcal{B}}^{\nu \ne\nu'} S_{\bq\nu}S_{\bq\nu'}
  \Re[e_{\k \a,\nu} (\bq) e^*_{\k \a,\nu'} (\bq)] \sigma_{\bq \nu} \sigma_{\bq \nu'} \nonumber  \\
  &&\,\,+\frac{1}{N_p}\sum_{\bq \in \mathcal{A}} \left[ \cdots \right].
 \end{eqnarray} 
The third line contains terms associated with $\bq \in \mathcal{A}$ phonons. As we prove below
in Sec.~\ref{sec.normcoord}, all contributions arising from $\bq \in \mathcal{A}$ phonons vanish
in the limit of large supercell. Furthermore, the choice of signs $S_{\bq\nu}$ in the ZG 
displacement guarantees that also the second line of Eq.~\eqref{eq.mean_ampl_1C} vanishes 
in the thermodynamic limit. Therefore in this limit we recover precisely the mean-square
displacements $\sigma^2_{\k\a}(T)$ of Eq.~\eqref{eq.Debye_waller_factor}. 

In Sec.~\ref{sec.displ} we show that these results are confirmed by explicit calculations on 
Si, GaAs, and MoS$_2$.

\section{The ZG displacement and Feynman's path integrals} \label{sec.link_to_PI}

An important property of the ZG displacement is that it can be understood as an exact 
single-point approximant to a Feynman's path integral~\cite{Feynman_1965}. 


In path integral approaches, such as path integral Monte Carlo simulations~\cite{Ruben_1998,
Sala_2004,Tuckerman_2010} and path integral molecular dynamics~\cite{Hernandez_2006,
Zhang_2018,Marx_2009}, the electronic and ionic degrees of freedom are decoupled
using the adiabatic approximation, and the quantum nature of the atomic nuclei is 
taken into account by averaging the electronic properties over all possible trajectories 
of the atomic nuclei. In order to obtain the correct quantum statistics, each trajectory
is weighted by $\exp(-S_{\rm E}/\hbar)$, where $S_{\rm E}$ is the Euclidean action associated
with a path~\cite{Feynman_1965}, and is defined in Eq.~\eqref{eq.action} below. 

We now show that the Williams-Lax rate can be recast
exactly in the language of path integrals. The equivalence between the two approaches 
was originally identified in Ref.~[\onlinecite{Sala_2004}].
By combining Eqs.~\eqref{eq.WL} and \eqref{eq.transition_rate} and using the relations 
$\chi_{\a n}(\{\tau\}) = \< \tau | \a n\>$, $\sum_n  \ket{\a n} \bra{ \a n} = 1$, and 
$\hat{H}_{\rm n} \ket{\a n} = E_{\a n} \ket{\a n}$, we find: 
 \begin{equation}\label{eq.WL_PI-2}
  \Gamma_{\a\rightarrow \beta} (\w;T)  = \frac{1}{Z} \int d\tau 
  \braket{\tau |  e^{-\hat{H}_{\rm n}/k_{\rm B}T}  |\tau} 
  \Gamma_{\a\rightarrow \beta}^{\{ \tau \}}(\w).
 \end{equation}
In these expressions $\ket{\tau}$ represents position eigenstates for all atomic coordinates, 
and $\hat{H}_{\rm n}$ is the nuclear Hamiltonian for the adiabatic potential energy surface 
$E_\a^{\{\tau\}}$. 

Now by combining together  Eqs.~\eqref{eq.WL_PI-2b}-\eqref{eq.WL_PI-5}, 
the Williams-Lax transition rate assumes the form of a thermodynamic Feynman 
path integral~\cite{Feynman_1965,Sala_2004,Tuckerman_2010,Zhang_2018}:
 \begin{eqnarray}\label{eq.WL_PI-11}
  \Gamma_{\a\rightarrow \beta} (\w;T)  &=& \frac{1}{Z_{\rm F}} \int d\tau  \,
   \Gamma_{\a\rightarrow \beta}^{\{ \tau \}}(\w)  \\ 
   &\times& \int_{\tau'(0) = \tau}^{\tau'(\hbar/k_{\rm B}T) = \tau}
    \mathcal{D} \tau' {\rm exp}\bigg(-\frac{1}{\hbar} S_E [\tau']\bigg), \nonumber 
  \end{eqnarray}
where the notation $\int_{\tau'(0) = \tau}^{\tau'(k_{\rm B}T) = \tau} \mathcal{D} \tau'$
indicates the sum over all paths that begin from the set of atomic coordinates $\{\tau\}$ 
at the time $t=0$ and go back to the same initial coordinates at the time $t=\hbar/k_{\rm B}T$.
$S_{\rm E}$ is the Euclidean action evaluated along each one of these paths:
  \begin{equation}\label{eq.action} 
    S_E = \int_0^{\hbar/k_{\rm B}T}\! 
     \left[\,\sum_{p\k\a}\frac{1}{2}M_\k \dot{\tau}_{p \k\a}^2 + U_\a(\{\tau\})\right] dt,
  \end{equation}
where $\dot{\tau}_{p \k\a}$ represents the classical velocities of the nuclei
and $U_\a$ is the potential energy of the nuclear Hamiltonian.
The partition function $Z_{\rm F}$ is equal to the second line of Eq.~\eqref{eq.WL_PI-11}.

Equation~\eqref{eq.WL_PI-11} states that the Williams-Lax transition rates can be obtained by 
averaging the electronic transitions at clamped ions using thermodynamic path integrals as 
weighting coefficients. For each atomic configuration `snapshot' $\{\tau\}$, we must evaluate 
all closed-loop path integrals that begin and terminate at $\{\tau\}$, see Fig.~\ref{fig_2}. 
At high temperature, $k_{\rm B}T \gg \hbar \w$ (with $\w$ being the highest vibrational frequency), 
Eq.~\eqref{eq.WL_PI-11} assigns the largest weights to paths that minimize the classical action, 
therefore the William-Lax theory reduces to a standard configurational average using classical 
Boltzmann weights.  Based on these observations, we can regard Eq.~\eqref{eq.WL_PI-11} as
the {\it link} between the Williams-Lax theory, path-integral molecular dynamics calculations 
of electronic and optical properties at finite temperature, and finite-temperature calculations 
using classical molecular dynamics simulations~\cite{Sala_2004}. We note that Eq.~\eqref{eq.WL_PI-11}
does not require the harmonic approximation, and maintains its validity even for strongly 
anharmonic systems.

Since in the case of {\it harmonic} systems the special displacement method presented here
yields the same result as the Williams-Lax theory (as we prove in Sec.~\ref{sec.ZG_theory}), 
and since we made no assumption about the form of the transition rates $\Gamma_{\a\rightarrow 
\beta}^{\{ \tau \}}(\w)$ in Eq.~\eqref{eq.WL_PI-11}, it follows that any calculation of 
finite-temperature properties using path integrals can be replaced altogether by a single
evaluation with the ZG displacement. In symbols, for a generic observable $O(\{\tau\})$ which 
is a parametric function of the atomic coordinates $\{\tau\}$, we have:
 \begin{equation}
  Z_{\rm F}^{-1} \int d\tau  \, O(\{\tau\}) \int \mathcal{D} \tau'
   {\rm exp}\bigg(-\frac{1}{\hbar} S_E [\tau']\bigg) = O(\{\tau^{\rm ZG}\}).
 \end{equation}
In other words, the ZG displacement provides an {\it exact single point} approximant to an
thermodynamic Feynman path integral, as is shown schematically in Fig.~\ref{fig_2}.  This 
result constitutes a significant computational simplification. For example, calculations of 
temperature-dependent band gaps using ring-polymer molecular dynamics require simulating hundreds 
of bids and averaging over thousands of snapshots~\cite{Hernandez_2006}, while the SDM can perform 
the same operation by performing a single supercell calculation.

\section{The ZG displacement}\label{sec.ZG_theory}

In this section, using a rigorous reciprocal space formulation, we prove that the ZG displacement 
given by Eq.~\eqref{eq.realdtau_method00} yields the correct Williams-Lax transition rates of 
Eq.~\eqref{eq.WL} in the limit of large supercell. The key ingredients of our derivation are 
(i) the translational invariance of the lattice, (ii) time-reversal symmetry, and (iii) a smooth 
connection between vibrational eigenmodes at nearby wavevectors.

The strategy that we follow is similar to our previous work~\cite{Zacharias_2016}, that is 
we choose a displacement defined by normal mode coordinates of magnitude 
$|x_{\bq  \nu}| = \sqrt{2}\,\sigma_{\bq \nu}$ for $\bq \in\mathcal{A}$,
and $|x_{\bq  \nu}| = |y_{\bq  \nu}| = \sigma_{\bq \nu}/\sqrt{2}$ for $\bq \in\mathcal{B}$.
These choices leave us with the freedom to select appropriate 
signs that define in which direction the normal coordinates are being displaced ($\pm$). 
In Ref.~[\onlinecite{Zacharias_2016}] we considered a $\Gamma$-point formalism, the vibrational 
modes were ordered by increasing energy, and the choice of signs was simply the sequence 
$+,-,+,\cdots$, see Eq.~(5) of Ref.~[\onlinecite{Zacharias_2016}].
In the present work the situation is more complicated, because we now have to deal
with phonon wavevectors $\bq$ and phonon branch indices $\nu$. As we discuss below,
in this case a simple sequence of alternating signs is not sufficient to recover
the correct Williams-Lax rate, and a more structured choice is necessary. The following
sections describe how we make this choice and why.

\subsection{The simplest case: one vibrational mode}\label{sec.simplest}

Before proceeding with the derivations, it is useful to examine the key idea of this
method using the simplest example: a hypothetical system with only one vibrational mode.
We rewrite Eq.~(\ref{eq.WL3}) by replacing the transition rate with a generic `observable'
$O^{\{x\}}$, and we call the WL thermal average of this quantity $O(T)$. For definiteness 
we consider the form of the integral corresponding to $\bq \in \mathcal{A}$ in Eq.~(\ref{eq.WL3}):
  \begin{equation} \label{eq.simple.0}
    O(T) = \frac{1}{2\sqrt{\pi}\sigma} \int dx\, e^{-x^2/4\sigma^2}\, O^{\{x\}}.
  \end{equation}
By expanding $O^{\{x\}}$ in powers of $x$ near the equilibrium coordinates $x=0$ and evaluating 
the integrals, we find:
  \begin{equation} \label{eq.simple.1}
    O(T) = O^{\{0\}}+\left.\frac{\D^2 O^{\{x\}}}{\D x^2}\right|_{x=0}\sigma^2,
  \end{equation}
which is correct to fourth order in $\sigma$. Alternatively, we can evaluate $O^{\{x\}}$ for 
the normal coordinate $x=\sqrt{2}\,\sigma$. This procedure yields: 
  \begin{equation} \label{eq.simple.2}
    O^{\{x=\sigma/\sqrt{2}\}}  = O^{\{0\}}+
    \left.\frac{\D O^{\{x\}}}{\D x}\right|_{0}\sqrt{2}\,\sigma+
    \left.\frac{\D^2 O^{\{x\}}}{\D x^2}\right|_{0}\sigma^2 + \cdots.
  \end{equation}
If we now average the expansions for $x=\pm\sqrt{2}\,\sigma$ we obtain:
  \begin{equation} \label{eq.simple.3}
    \frac{1}{2}\left[O^{\{x=+\sqrt{2}\,\sigma\}} +O^{\{x=-\sqrt{2}\,\sigma\}}\right] = 
    O^{\{0\}}+ \left.\frac{\D^2 O^{\{x\}}}{\D x^2}\right|_{0}\sigma^2,
  \end{equation}
up to fourth order in $\sigma$. By comparing Eq.~\eqref{eq.simple.1} and \eqref{eq.simple.3}
we see that {\it two} evaluations of the observable $O^{\{x\}}$ at clamped nuclei are sufficient to
reproduce the thermal average given by Eq.~\eqref{eq.simple.0}.

If we now consider the form of the integral corresponding to $\bq\in\mathcal{B}$ in Eq.~(\ref{eq.WL3}),
we arrive at a similar result. In this case the integral yielding the thermal average is to be 
replaced by two evaluations of the property $O^{\{x\}}$ at $x=\pm \sigma/\sqrt{2}$.

In the case of a real system with many vibrational modes, our strategy is to exploit the same 
mechanism leading to Eq.~\eqref{eq.simple.3}, by leveraging the cancellation between `similar' 
modes, i.e. modes of the same branch at adjacent $\bq$-points. In the case of many vibrational
modes, the displacements can be chosen so as to reproduce the thermal average with only 
{\it one} atomic configuration.

\subsection{Thermal average of an observable in the Williams-Lax formalism}\label{sec.second_order_exp}

In this section we derive the expression for the WL thermal average of an observable $O^{\{ \tau \}}$ 
which depends parametrically on the atomic positions $\{\tau\}$, starting from Eq.~(\ref{eq.WL3}) 
and following the same steps that lead from Eq.~\eqref{eq.simple.0} to Eq.~\eqref{eq.simple.1}.

To second order in the displacements $\DD\tau_{ p\k \a}$ from equilibrium, the observable
$O^{\{ \tau \}}$ reads:
  \begin{eqnarray}\label{eq.expansion_dtau}
        O^{\{ \tau \}} &=& O^{\{0\}} + \sum_{p\k\a} \frac{\D O^{\{ \tau \}}}
        {\D\tau_{p\k \a }} \DD\tau_{p\k \a} \\
       &+&  \frac{1}{2}\sum_{\substack {p\k\a \\ p'\k'\a'}} 
        \frac{\D ^2 O^{\{ \tau \}}}{{\D\tau_{ p\k \a}}
        {\D\tau_{p' \k' \a'}}} \DD\tau_{p\k \a} \DD\tau_{p'\k' \a'} \, , \nonumber
  \end{eqnarray}
where $O^{\{0\}}$ indicates the observable evaluated at the equilibrium configuration. We can 
express the displacements in term of the real normal coordinates using Eq.~(\ref{eq.realdtau}) 
and the chain rule. The result for the linear variation is:
  \begin{eqnarray}\label{eq.linear_dtau}
    \sum_{p\k\a} \frac{\D O^{\{ \tau \}}}
        {\D\tau_{p\k \a }} \DD\tau_{p\k \a}  &=& 
       \sum_{\bq  \in \mathcal{A}, \nu}  \frac{\D O^{\{ \tau \}}} {\D x_{\bq  \nu}} x_{\bq  \nu}  \\
               &+&  \sum_{\bq  \in \mathcal{B}, \nu} \bigg[ \frac{\D O^{\{ \tau \}}} 
           {\D x_{\bq  \nu}} x_{\bq  \nu} 
           + \frac{\D O^{\{ \tau \}}} {\D y_{\bq  \nu}}  y_{\bq  \nu} \bigg], \nonumber
  \end{eqnarray}
while the result for the quadratic variation is:
 \begin{eqnarray}\label{eq.quadratic_dtau}
   && \sum_{\substack {p\k\a \\ p'\k'\a'}} \frac{\D ^2 O^{\{ \tau \}}}{{\D\tau_{ p\k \a}}
   {\D\tau_{p' \k' \a'}}} \DD\tau_{p\k \a} \DD\tau_{p'\k' \a'} = \nonumber \\
   && \hspace{0.3cm}  \sum_{\substack { \bq  \in \mathcal{A}, \nu \\ \bq ' \in \mathcal{A}, \nu'}} 
   \frac{\D^2 O^{\{ \tau \}} } {\D x_{\bq  \nu} \D x_{\bq ' \nu'}} x_{\bq  \nu} x_{\bq ' \nu'} 
   \nonumber \\ && \hspace{0.2cm}+ 2 \!\!\!\!\sum_{\substack { \bq  \in \mathcal{A}, \nu \\ 
   \bq ' \in \mathcal{B}, \nu'}} \bigg[ \frac{\D^2 O^{\{ \tau \}}} {\D x_{\bq  \nu} 
   \D x_{\bq ' \nu'}} x_{\bq  \nu} x_{\bq ' \nu'}   + \frac{\D^2 O^{\{ \tau \}} } {\D x_{\bq \nu} 
   \D y_{\bq' \nu'}} x_{\bq  \nu} y_{\bq ' \nu'} \bigg] \nonumber \\ && \hspace{0.2cm}+ \!\!\! 
   \sum_{\substack { \bq  \in \mathcal{B}, \nu \\ \bq ' \in \mathcal{B}, \nu'}} \bigg[  
   \frac{\D^2 O^{\{ \tau \}} } {\D x_{\bq  \nu} \D x_{\bq ' \nu'}} x_{\bq  \nu} x_{\bq ' \nu'}
   +  \frac{\D^2 O^{\{ \tau \}} } {\D y_{\bq  \nu} \D y_{\bq ' \nu'}} y_{\bq  \nu} y_{\bq ' \nu'} 
   \nonumber \\ && \hspace{1.5cm}+  2\frac{\D^2 O^{\{ \tau \}} } {\D x_{\bq  \nu} \D y_{\bq ' \nu'}} 
   x_{\bq  \nu} y_{\bq' \nu'} \bigg]. 
 \end{eqnarray}
The derivation of Eqs.~\eqref{eq.linear_dtau} and~\eqref{eq.quadratic_dtau} is lengthy but 
straightforward. 

We now replace the transition rate $\Gamma_{\a\rightarrow \beta}^{\{ \tau \}}(\w)$ in Eq.~\eqref{eq.WL3} 
with the generic observable $O^{\{ \tau \}}$, we use Eqs.~\eqref{eq.linear_dtau}-\eqref{eq.quadratic_dtau}, 
and we evaluate the integrals in the real normal coordinates $x_{\bq\nu}$ and $y_{\bq\nu}$. There 
are only two types of integrals, those that are odd in $x_{\bq\nu}$ or $y_{\bq\nu}$, which vanish 
identically, and those that involve the powers $x_{\bq\nu}^2$ or $y_{\bq\nu}^2$. The latter are standard 
Gaussian integrals of the form $\int\! dx \,x^2e^{-x^2} = \sqrt{\pi}/2$. The resulting expression 
for the WL thermal average of $O^{\{ \tau \}}$, correct to third order in $\sigma_{\bq\nu}$, is:
  \begin{eqnarray}\label{eq.WL_exp}
   O(T) &=&  O^0 + \sum_{\bq \in \mathcal{A}, \nu}  
   \frac{\D^2 O^{\{ \tau \}}} {\partial x^2_{\bq \nu}} \sigma^2_{\bq  \nu} \\ 
   &+& \frac{1}{4} \sum_{\bq  \in \mathcal{B}, \nu} \bigg[ \frac{\D^2 O^{\{ \tau \}}} 
   {\partial x^2_{\bq  \nu}}  + \frac{\D^2 O^{\{ \tau \}}} {\partial y^2_{\bq  \nu}} \bigg] 
   \sigma^2_{\bq  \nu} \nonumber.
  \end{eqnarray}
This result constitutes the generalization to many $\bq$-points and many phonon branches
of the result in Eq.~\eqref{eq.simple.1} for the single-mode model.

Here we see that all odd powers in the normal coordinates and all second order cross-terms 
with $\bq \neq \bq'$ or $\nu\ne \nu'$ vanish upon integration.  

\subsection{Taylor expansion of an observable in normal coordinates}\label{sec.Transl_inv}

\subsubsection{Linear variations}\label{sec.linvar}

Now we perform a step similar to Eq.~\eqref{eq.simple.2}, but for the general case
of many $\bq$-points and many phonon branches. Although  we already derived the required 
Taylor expansion in Eqs.~\eqref{eq.expansion_dtau}-\eqref{eq.quadratic_dtau}, it is convenient 
to find equivalent expressions which are easier to handle and where the translational invariance 
and time-reversal symmetry are built in from the outset.

Translational invariance implies that the derivative of $O^{\{ \tau \}}$ with respect to 
an atomic displacement be the same for every unit cell:
 \begin{equation}\label{eq.transl1}
  \frac{\D O^{\{ \tau \}} } {\D \tau_{ p\k \a} } = \frac{\D O^{\{ \tau \}} } {\D \tau_{ 0\k \a} }.
 \end{equation}
This property can be used to prove that:
 \begin{equation}\label{eq.new1}
    \frac{\D O^{\{ \tau \}} } {\D x_{\bq \nu} } =
    \frac{\D O^{\{ \tau \}} } {\D y_{\bq \nu} } = 0 \mbox{ if } \bq \in \mathcal{B}.
 \end{equation}
To see this we write the derivatives using the chain rule and we employ Eq.~\eqref{eq.realdtau}. 
For $\bq \in \mathcal{B}$ we find:
  \begin{equation}
   \frac{\D O^{\{ \tau \}} } {\D x_{\bq \nu} } = 
    \sum_{p\k\a} \frac{\D O^{\{ \tau \}} } {\D \tau_{ p\k \a} }
   \bigg(\frac{M_0}{N_p M_\k}\bigg)^{1/2} 
   2 \Re \left[ e_{\k\a,\nu} (\bq ) e^{i\bq \cdot {\bf R}_p} \right].
 \end{equation}
Now using Eq.~\eqref{eq.transl1} we have:
  \begin{equation}
   \frac{\D O^{\{ \tau \}} } {\D x_{\bq \nu} } = 
    \sum_{\k\a} \frac{\D O^{\{ \tau \}} } {\D \tau_{ 0\k \a} }
   \bigg(\frac{M_0}{N_p M_\k}\bigg)^{\!\!1/2} 
   \!\!\!\!2 \Re \left[ e_{\k\a,\nu} (\bq ) \sum_p e^{i\bq \cdot {\bf R}_p} \right].
 \end{equation}
The sum rule in Eq.~\eqref{eqa.sum_rules} requires $\bq=0$ for the sum over $p$ to be nonzero, 
but this condition cannot be fulfilled when $\bq \in \mathcal{B}$. The same argument applies 
to derivatives with respect to $y_{\bq \nu}$. This proves the result in Eq.~\eqref{eq.new1}.
The main consequence of this result is that the linear variation of the observable $O^{\{ \tau \}}$ 
in Eq.~\eqref{eq.linear_dtau} takes the simple form:
 \begin{equation}\label{eq.linear_terms_A}
   \sum_{p\k\a} \frac{\D O^{\{ \tau \}}} {\D\tau_{p\k \a }} \DD\tau_{p\k \a}
   = \sum_{\bq \in \mathcal{A}, \nu} \frac{\D O^{\{ \tau \}}} {\D x_{\bq \nu} } x_{\bq \nu},
 \end{equation} 
where the right-hand side contains only phonons with $\bq \in \mathcal{A}$. 
This result indicates that, if we perform a supercell calculation
with atoms displaced away from equilibrium, then the contribution to the observable $O^{\{ \tau \}}$
which is linear in the displacements comes entirely from phonons 
with wavevectors $\bq \in \mathcal{A}$.

\subsubsection{Quadratic variations}

Here we employ again translational invariance to simplify the expression for the quadratic 
variations of the observable $O^{\{ \tau \}}$ appearing in Eq.~\eqref{eq.quadratic_dtau}.
In this case translational invariance dictates:
  \begin{equation}\label{eq.transl}
    \frac{\D ^2 O^{\{ \tau \}}}{{\partial \tau_{p\k\a}} {\D \tau_{p'\k'\a'} }} =
    \frac{\D ^2 O^{\{ \tau \}}}{{\partial \tau_{0\k\a}} {\D \tau_{p'-p,\k'\a'} }},
  \end{equation}
where the unit cell $p'-p$ corresponds to the lattice vector $\bR_{p'}\!-\!\bR_p$.
We now rewrite the second derivatives in Eq.~\eqref{eq.quadratic_dtau} in terms of
the real normal coordinates $x_{\bq\nu}$ and $y_{\bq\nu}$ using the chain rule and
Eq.~\eqref{eq.realdtau}. After some lengthy but straightforward algebra we obtain:
  \begin{eqnarray}\label{eq.trinv.1}
   \frac{\D^2 O^{\{ \tau \}} } {\D x_{\bq \nu}\D x_{\bq'\nu'} } &=&
      \d_{\bq,\bq'} \sum_{\k\a,\k'\a'} \bigg(\frac{M_0^2}{M_\k M_{\k'}}\bigg)^{\!\!1/2} 
      \! \sum_p \frac{\D^2 O^{\{ \tau \}} } {\D \tau_{0\k\a}\D \tau_{p,\k'\a'} }
      \nonumber \\ & \times& \Big\{ \d_{\bq \in \mathcal{A}}\, 
         e_{\k\a,\nu} (\bq ) e_{\k'\a',\nu'} (\bq ) \text{cos}(\bq \cdot {\bf R}_p) 
        \nonumber \\ && \hspace{-0.1cm}+ \d_{\bq \in \mathcal{B}} \, 2 \Re 
   \big[ e^{i\bq \cdot {\bf R}_p} e^*_{\k\a,\nu} (\bq ) e_{\k'\a',\nu'} (\bq ) \big] \! \Big\},
 \end{eqnarray}
 \begin{eqnarray}\label{eq.trinv.2}
  \frac{\D^2 O^{\{ \tau \}} } {\D x_{\bq \nu}\D y_{\bq'\nu'} } &= &
    -\d_{\bq,\bq'} \!\!
   \sum_{\k\a,\k'\a'} 
        \bigg(\frac{M_0^2}{M_\k M_{\k'}}\bigg)^{\!\!1/2} 
   \!\sum_{p}\frac{\D^2 O^{\{ \tau \}} } {\D \tau_{0\k\a}\D \tau_{p\k'\a'} } \nonumber \\ 
   &\times& \d_{\bq \in \mathcal{B}} \,2 \Im \left[ e^{i\bq \cdot {\bf R}_p} e^*_{\k\a,\nu} 
  (\bq ) e_{\k'\a',\nu'} (\bq) \right],
 \end{eqnarray}
  \begin{eqnarray}\label{eq.trinv.3}
   \frac{\D^2 O^{\{ \tau \}} } {\D y_{\bq \nu}\D y_{\bq'\nu'} } & = &
  \d_{\bq,\bq'} \sum_{\k\a,\k'\a'} \bigg(\frac{M_0^2}{M_\k M_{\k'}}\bigg)^{1/2} 
     \sum_p \frac{\D^2 O^{\{ \tau \}} } {\D \tau_{0\k\a}\D \tau_{p\k'\a'} } \nonumber \\
    &\times& \d_{\bq\in \mathcal{B}} \,
   2 \Re \left[ e^{i\bq \cdot {\bf R}_{p}} e^*_{\k\a,\nu} (\bq ) e_{\k'\a',\nu'} (\bq ) \right],
 \end{eqnarray}
where $\d_{\bq \in \mathcal{A}}=1$ for $\bq\in\mathcal{A}$ and 0 otherwise, and similarly
for $\d_{\bq \in \mathcal{B}}$. From these expressions we see that, as a consequence of
translational invariance, all derivatives with $\bq\ne\bq'$ vanish identically.
By comparing Eqs.~\eqref{eq.trinv.1} and \eqref{eq.trinv.3} we have:
  \begin{equation}\label{eq.xx-yy}
     \frac{\D^2 O^{\{ \tau \}} } {\D y_{\bq \nu}\D y_{\bq'\nu'} } = 
        \frac{\D^2 O^{\{ \tau \}} } {\D x_{\bq \nu}\D x_{\bq'\nu'} }\, \mbox{ if } \bq,\bq' \in \mathcal{B}.
  \end{equation}
Furthermore, by using translational invariance in Eq.~\eqref{eq.trinv.2} it can be verified 
that: 
 \begin{equation}\label{eq.trinv.2b}
  \frac{\D^2 O^{\{ \tau \}} } {\D x_{\bq \nu'}\D y_{\bq'\nu} } =
  -\frac{\D^2 O^{\{ \tau \}} } {\D x_{\bq \nu}\D y_{\bq'\nu'} }.
 \end{equation}
These properties allow us to simplify Eq.~\eqref{eq.quadratic_dtau} as follows:
 \begin{eqnarray}\label{eq.quadratic_dtau2}
   && \frac{1}{2}\sum_{\substack {p\k\a \\ p'\k'\a'}} \!\!
   \frac{\D ^2 O^{\{ \tau \}}}{{\D\tau_{ p\k \a}}
  {\D\tau_{p' \k' \a'}}} \DD\tau_{p\k \a} \DD\tau_{p'\k' \a'} =   \nonumber \\
   && \hspace{0.9cm}
   \frac{1}{2}\sum_{\bq \in \mathcal{A},\nu\nu' } \!
   \frac{\D^2 O^{\{ \tau \}} } {\D x_{\bq  \nu} \D x_{\bq \nu'}} x_{\bq  \nu} x_{\bq \nu'} 
  \nonumber \\ && \hspace{0.6cm}+ \frac{1}{2}\sum_{ \bq  \in \mathcal{B}, \nu\nu'} 
  \frac{\D^2 O^{\{ \tau \}} } {\D x_{\bq  \nu} \D x_{\bq \nu'}}
  (   x_{\bq  \nu} x_{\bq \nu'}
  + y_{\bq  \nu} y_{\bq \nu'} )\nonumber \\ &&
  \hspace{0.6cm}
  +  \frac{1}{2}\sum_{ \bq  \in \mathcal{B}, \nu\ne\nu'} 
     2\frac{\D^2 O^{\{ \tau \}} } {\D x_{\bq  \nu} \D y_{\bq\nu'}} x_{\bq  \nu} y_{\bq \nu'}. 
 \end{eqnarray}
As in the case of the linear variations in Sec.~\ref{sec.linvar}, also for the quadratic
variations the contribution from phonons with $\bq \in\mathcal{A}$ vanishes in the
limit of dense Brillouin zone sampling. 

\subsection{Choice of normal coordinates for the ZG displacement}\label{sec.normcoord}

The last step of the procedure outlined in Sec.~\ref{sec.simplest} is to choose the values 
of the normal coordinates $x_{\bq\nu}$ and $y_{\bq\nu}$ so that Eqs.~\eqref{eq.linear_terms_A} 
and \eqref{eq.quadratic_dtau2} reproduce the WL average given by Eq.~\eqref{eq.WL_exp} in 
the limit of dense Brillouin-zone sampling.

By comparing Eq.~\eqref{eq.WL_exp} with Eq.~\eqref{eq.quadratic_dtau2} it is evident that 
the magnitude of the normal coordinates must be:  $|x_{\bq\nu}|=\sqrt{2}\,
\sigma_{\bq\nu}$ for ${\bq \in \mathcal{A}}$,  and $|x_{\bq\nu}|=|y_{\bq\nu}|=\sigma_{\bq\nu}/\sqrt{2}$ 
for ${\bq \in \mathcal{B}}$. The remaining degrees of freedom are the signs of the normal 
coordinates, say $S_{\bq\nu,x}=\pm 1$ and $S_{\bq\nu,y}=\pm 1$, which are to be determined.
If we evaluate the observable $O^{\{\tau\}}$ at the atomic positions specified by the normal 
coordinates $x_{\bq\nu}= \sqrt{2}\,S_{\bq\nu,x}\sigma_{\bq\nu}$ for 
${\bq \in \mathcal{A}}$, and $x_{\bq\nu}=S_{\bq\nu,x}\sigma_{\bq\nu}/\sqrt{2}$, 
$y_{\bq\nu}=S_{\bq\nu,y}\sigma_{\bq\nu}/\sqrt{2}$ for ${\bq \in \mathcal{B}}$, then the 
combination of Eq.~\eqref{eq.expansion_dtau} with Eqs.~\eqref{eq.linear_terms_A} and 
\eqref{eq.quadratic_dtau2} yields, to second order in the displacements:
  \begin{eqnarray}\label{eq.expansion.zg}
     && O^{\{ \tau \}} = O^{\{0\}} + \sqrt{2}
     \sum_{\bq \in \mathcal{A}, \nu} \frac{\D O^{\{ \tau \}}} {\D x_{\bq \nu} } 
     \sigma_{\bq \nu} S_{\bq\nu,x}  \nonumber \\ &&
     + \sum_{\bq \in \mathcal{A},\nu\nu'} \!
     \frac{\D^2 O^{\{ \tau \}} } {\D x_{\bq  \nu} \D x_{\bq \nu'}} 
     \sigma_{\bq  \nu} \sigma_{\bq \nu'} S_{\bq\nu,x}S_{\bq\nu',x}\nonumber \\ &&  
     + \frac{1}{4} \sum_{ \substack{\bq  \in \mathcal{B}\\ \nu \nu'}} 
     \frac{\D^2 O^{\{ \tau \}} } {\D x_{\bq  \nu} \D x_{\bq \nu'}}
     \sigma_{\bq  \nu} \sigma_{\bq \nu'} ( S_{\bq\nu,x}S_{\bq\nu',x} 
     +  S_{\bq\nu,y}S_{\bq\nu',y}) \nonumber\\
     && + \frac{1}{4} \sum_{ \bq  \in \mathcal{B}, \nu\ne \nu'} 
     2\frac{\D^2 O^{\{ \tau \}} } {\D x_{\bq  \nu} \D y_{\bq \nu'}}
     \sigma_{\bq  \nu} \sigma_{\bq \nu'} S_{\bq\nu,x}S_{\bq\nu',y}.
  \end{eqnarray}
Using the WL average in Eq.~\eqref{eq.WL_exp}, the last expression takes the form:
  \begin{equation}
    O^{\{ \tau \}} = O(T) + \Delta_\mathcal{A} + \Delta_\mathcal{B},
  \end{equation}
where $\Delta_\mathcal{A}$ and $\Delta_\mathcal{B}$ represent the deviation of $O^{\{ \tau \}}$ 
with respect to the WL average $O(T)$, and are given by:
  \begin{eqnarray}\label{eq.delta.a}
   \Delta_\mathcal{A} &=& \sqrt{2}\sum_{\bq \in \mathcal{A}, \nu} \frac{\D O^{\{ \tau \}}}
   {\D x_{\bq \nu} } \sigma_{\bq \nu} S_{\bq\nu,x}  \nonumber \\ 
   &+& \sum_{\substack{\bq \in \mathcal{A}\\\nu\ne \nu'}} \!
   \frac{\D^2 O^{\{ \tau \}} } {\D x_{\bq  \nu} \D x_{\bq \nu'}} 
   \sigma_{\bq  \nu} \sigma_{\bq \nu'} S_{\bq\nu,x}S_{\bq\nu',x},
  \end{eqnarray}
  \begin{eqnarray}\label{eq.delta.b}
  \Delta_\mathcal{B} &=& \frac{1}{4}\sum_{ \substack{\bq  \in \mathcal{B}\\ \nu\ne\nu'}} 
   \sigma_{\bq  \nu} \sigma_{\bq \nu'} \nonumber \\ &\times& \Bigg\{ 
  \frac{\D^2 O^{\{ \tau \}} } {\D x_{\bq  \nu} \D x_{\bq \nu'}}
  ( S_{\bq\nu,x}S_{\bq\nu',x} +  S_{\bq\nu,y}S_{\bq\nu',y}) \nonumber\\
  &&+ 2\frac{\D^2 O^{\{ \tau \}} } {\D x_{\bq  \nu} \D y_{\bq \nu'}}
  S_{\bq\nu,x}S_{\bq\nu',y} \Bigg\}.\hspace{0.6cm}
  \end{eqnarray}
The term $\Delta_\mathcal{A}$ only contains contributions from phonons with $\bq \in 
\mathcal{A}$. In the limit of dense sampling of the Brillouin zone, the number of elements 
of $\mathcal{A}$ remains finite, while the number of elements in $\mathcal{B}$ goes to 
infinity. Therefore the Lesbegue measure of set $\mathcal{A}$ vanishes in this limit, 
and we have the result:
  \begin{equation}\label{eq.limA}
   {\lim}_{N_p \rightarrow \infty}\, \Delta_\mathcal{A} = 0.
  \end{equation}
Given that phonons with $\bq \in \mathcal{A}$ do not contribute for large supercells,
to reproduce the WL average using a single atomic configuration, we only need to determine 
the signs $S_{\bq\nu,x}$ and $S_{\bq\nu,y}$ so that the term $\Delta_\mathcal{B}$ in 
Eq.~\eqref{eq.delta.b} vanishes.  In order to simplify the task, we choose to assign the 
{\it same} sign to $x_{\bq\nu}$ and $y_{\bq\nu}$: $S_{\bq\nu,x}=S_{\bq\nu,y}=S_{\bq\nu}$. 
If we replace this choice of normal coordinates inside Eq.~\eqref{eq.dtau} 
and ignore the contributions from $\bq\in \mathcal{A}$ [which vanish in the limit of large 
supercell according to Eq.~\eqref{eq.limA}], then we obtain precisely the ZG displacement 
in Eq.~\eqref{eq.realdtau_method00}. Strictly speaking the result carries an additional 
phase $e^{i\pi/4}$, but this can be absorbed in the eigenmodes (adding this extra phase 
does not pose any problem because the set $\bq \in \mathcal{B}$ does not contain time-reversal 
partners).

Using the choice $S_{\bq\nu,x}=S_{\bq\nu,y}=S_{\bq\nu}$ inside Eqs.~\eqref{eq.trinv.1} 
and \eqref{eq.trinv.2}, we can rewrite $\Delta_\mathcal{B}$ as follows:
  \begin{eqnarray}\label{eq.delta.b.3}
  \Delta_\mathcal{B} &=& \sum_{ \substack{\bq  \in \mathcal{B}\\ \nu < \nu'}}
     S_{\bq\nu}S_{\bq\nu'} A_{\nu\nu'}(\bq),
  \end{eqnarray}
having defined:
  \begin{eqnarray}\label{eq.delta.b.4}
  && A_{\nu\nu'}(\bq) =
    2\!\!\sum_{\k\a,\k'\a'} \bigg(\frac{M_0^2}{M_\k M_{\k'}}\bigg)^{\!\!1/2}
      \! \sum_p \frac{\D^2 O^{\{ \tau \}} } {\D \tau_{0\k\a}\D \tau_{p,\k'\a'} } \nonumber\\
  &&\hspace{0.7cm}\times 
     \Re \big[ (1+i)
     e^{i\bq \cdot {\bf R}_p} e^*_{\k\a,\nu} (\bq ) e_{\k'\a',\nu'} (\bq ) \big] 
     \sigma_{\bq  \nu} \sigma_{\bq \nu'}.\hspace{0.5cm}
  \end{eqnarray}
Clearly the quantity $A_{\nu\nu'}(\bq)$ in the last expression is a relative of the 
lattice dynamical matrix, and the second derivatives $\D^2 O^{\{ \tau \}}/\D \tau_{0\k\a}
\D \tau_{p,\k'\a'}$ are relatives of the interatomic force constants, but for the observable 
$O^{\{\tau\}}$ instead of the total energy.
We note that the construction of the ZG displacement does not require the explicit
evaluation of the second derivatives $\D^2 O^{\{ \tau \}}/\D \tau_{0\k\a}\D \tau_{p,\k'\a'}$ to
achieve the minimization of $\Delta_\mathcal{B}$.

In order to determine the ZG displacement, we need to choose the signs
$S_{\bq\nu}$ in such a way that $\Delta_{\mathcal{B}}$ in Eq.~\eqref{eq.delta.b.3}
vanishes in the limit of large supercells. To this aim we note that in the limit of 
dense Brillouin-zone sampling, the vibrational eigenmodes $e_{\k\a,\nu} (\bq )$ and 
eigenfrequencies $\w_{\bq\nu}$ can be chosen to be nearly the same between adjacent $\bq$-points:
\begin{equation}\label{eq.samew}
   \lim_{\Delta \bq\rightarrow 0}\w_{\bq+\Delta \bq,\nu} = \w_{\bq\nu}, \quad
 \lim_{\Delta \bq\rightarrow 0} e_{\k\a,\nu} (\bq+\Delta\bq )= e_{\k\a,\nu} (\bq ).
\end{equation}
These conditions are always true for the frequencies, but are usually not fulfilled by 
the eigenmodes obtained from the diagonalization of the dynamical matrix. In fact nondegenerate 
eigenmodes may carry an arbitrary complex phase, and in case of degeneracy any rotation in
the degenerate subspace is admissible. In order to {\it enforce} Eq.~\eqref{eq.samew},
we set up a smooth Bloch gauge in the Brillouin zone by performing a unitary rotation
of all eigenmodes. This is described in detail in Sec.~\ref{synch_eigenv}. Then, by 
combining Eqs.~\eqref{eq.samew} and Eq.~(\ref{eq.delta.b.4}) we have:
  \begin{equation}\label{eq.samew2}
    \lim_{\Delta \bq\rightarrow 0} A_{\nu\nu'}(\bq+\Delta\bq)= A_{\nu\nu'}(\bq).
 \end{equation}
Owing to this relation, if we select a set $\mathcal{D}$ of adjacent $\bq$-points in the 
Brillouin zone, in the limit of dense sampling we can rewrite Eq.~\eqref{eq.delta.b.3} as:
   \begin{equation}\label{eq.sumE.2}
     \!\lim_{\Delta \bq\rightarrow 0} 
     \!\sum_{\substack{\nu<\nu'\\\bq  \in \mathcal{D}}} \!
      S_{\bq\nu}S_{\bq\nu'} A_{\nu\nu'}(\bq) \!=\!
      \!\sum_{\nu<\nu'}\!\! A_{\nu\nu'}(\bar{\bq})\!\sum_{\bq  \in \mathcal{D}} 
      \!S_{\bq\nu}S_{\bq\nu'},
 \end{equation}
where $\bar{\bq}$ is the centroid of $\mathcal{D}$. For the sum on the right-hand side 
to vanish, we need to determine a set of signs $S_{\bq\nu}$ so that half of the products 
$S_{\bq\nu}S_{\bq\nu'}$ are positive and the other half is negative. 

If we denote by $n$ the number of phonon branches, we can assign $2^n$ unique combinations
of signs to these modes. Let us call such combinations $S_{\bq\,1}^{(i)}, S_{\bq\,2}^{(i)},
\cdots, S_{\bq\,n}^{(i)}$, with $i=1,2,\cdots,2^n$. It is easy to show that, for $\nu\ne\nu'$,
$\sum_{i=1}^{2^n} S_{\bq\nu}^{(i)}S_{\bq\nu'}^{(i)} = 0$. The proof proceeds by induction:
the equality is trivially verified for $n=2$; for $n>2$ assume that the result holds for a given $n$;
when considering $n+1$ all the possible $2^{n+1}$ combinations of signs are obtained by
duplicating the previous $2^n$ combinations of $n$ signs, and appending an extra sign at 
the end of each duplicate sequence. By construction, when $\nu,\nu'\le n$ we have
$\sum_{i=1}^{2^{n+1}} S_{\bq\nu}^{(i)}S_{\bq\nu'}^{(i)} = 2 \sum_{i=1}^{2^n} S_{\bq\nu}^{(i)}
S_{\bq\nu'}^{(i)} =0$; when $\nu=n+1$ or $\nu' = n+1$ every term of the sequence for $n$ 
modes appears twice and with opposite signs, therefore also in this case the sum vanishes. 

At this point we can perform a limiting procedure and consider a dense sampling of
the Brillouin zone. If we partition the set $\mathcal{B}$ of $\bq$-points
into disjoint subsets containing $2^n$ elements each, and we attach to each of 
these elements one of the $2^n$ unique sequences of signs $S_{\bq\nu}^{(i)}$, then the
right-hand side of Eq.~\eqref{eq.sumE.2} vanishes identically, and so does the 
error $\Delta_{{\mathcal B}}$ in Eq.~\eqref{eq.delta.b.3}:
  \begin{equation}\label{eq.limB}
   \lim_{N_p\rightarrow \infty} \Delta_{{\mathcal B}} = 0.
  \end{equation}
Taken together, Eqs.~\eqref{eq.limA} and \eqref{eq.limB} constitute the formal proof 
that the ZG displacement provided by Eq.~\eqref{eq.realdtau_method00} yields exactly the 
Williams-Lax thermal average, to second order in the mean-square displacements. 

While we have not proven formally that the equivalence between the ZG displacement and the 
Williams-Lax theory extends beyond second order in the displacements, 
in Sec.~\ref{sec.displ} we show
numerically that Eqs.~\eqref{eq.limA} and \eqref{eq.limB} also hold for higher orders.
A proof of equivalence of the two formalisms to all orders in the displacements was provided in 
Ref.~[\onlinecite{Zacharias_2016}] for the simpler case of $\Gamma$-point calculations.
Therefore the present method does not rely only on the quadratic expansion of the relevant operator, 
and the equivalence between the ZG displacement and the multivariate Gaussian integral 
in Eq.~\eqref{eq.WL-epsilon_2} holds to all orders in the atomic displacements.

We also note that the same proof leading to Eq.~\eqref{eq.limB} can be employed to demonstrate 
the equivalence between the ZG displacement and the mean-square displacements in 
Eq.~\eqref{eq.mean_ampl_1C}. To this aim it is sufficient to replace the first line of 
Eq.~\eqref{eq.delta.b.4} by a constant, absorb the imaginary factor $e^{i\pi/4}$ inside the 
eigenmodes, and observe that with these changes $\Delta_\mathcal{B}$ of Eq.~\eqref{eq.delta.b.3} 
reduces precisely to the second line of Eq.~\eqref{eq.mean_ampl_1C}.

\subsection{Additional considerations for calculations using small supercells}

The proof outlined in the previous section considers the limit of very large supercells. 
For practical calculations it is important to ensure that the ZG displacement delivers good 
accuracy also in the case of computationally tractable, smaller supercells. This improvement 
can be achieved as follows.

The combinations of signs necessary to achieve the cancellation in Eq.~\eqref{eq.limB}
carries some redundancy: for every set $S_{\bq\,1}^{(i)}, S_{\bq\,2}^{(i)}, \cdots, 
S_{\bq\,n}^{(i)}$ there is also the set $-S_{\bq\,1}^{(i)}, -S_{\bq\,2}^{(i)}, \cdots, 
-S_{\bq\,n}^{(i)}$.  Obviously the latter combination yields the same products as the former,
$-S_{\bq\nu}^{(i)}\times -S_{\bq\nu'}^{(i)} = S_{\bq\nu}^{(i)}S_{\bq\nu'}^{(i)}$,
therefore it is not useful to achieve the cancellation in Eq.~\eqref{eq.sumE.2}.
We can call the latter combination `antithetic' to the former. In order to reduce
the size of the supercell required to achieve convergence, we can eliminate the
antithetic combinations.

This reasoning implies that the minimum number of $\bq$-points required to achieve 
exact cancellation of $\Delta_{\mathcal{B}}$ [in the limit where Eq.~\eqref{eq.sumE.2} 
is satisfied] is precisely $2^{n-1}$. For example, in a tetrahedral semiconductor like 
Si or GaAs we would need at least $2^{6-1}=32$ $\bq$-points and therefore supercells
of size at least $4\times 4\times 4$ are necessary to enable such cancellation. 

To see how the choice of signs without antithetic sets works in a simple example, let us 
consider a system with $n=3$ phonon branches. In this case we have $2^3=8$ distinct 
combinations of signs as follows:
\begin{eqnarray}
  &&
   \hspace{0.0cm}   \begin{matrix}
        \nu\,\, & & \\
        1\,\,\, & 2\,\,\, & 3\,\,\,\\
      \end{matrix} \nonumber \\
  &&
     \begin{matrix} 
        + & + & + \\
        + & - & - \\
        - & + & - \\
        - & - & + \\
        - & - & - \\
        - & + & + \\
        + & - & + \\
        + & + & - 
      \end{matrix}
      \hspace{0.3cm}
     \begin{matrix} 
        1 & i\\
        2 & \\
        3 & \\
        4 & \\
        5 & \\
        6 & \\
        7 & \\
        8 &
      \end{matrix},\label{eq.signs2}
\end{eqnarray}
Here the combinations $i=5,\cdots,8$ are antithetic to $i=1,\cdots,4$ and we can discard them.
If we now consider a set $\mathcal{D}$ of $2^{3-1}=4$ adjacent $\bq$-points,
say $\bq_1,\cdots,\bq_4$, we can choose the `sign matrix' as follows:
\begin{eqnarray}
  &&
   \hspace{1.3cm}   \begin{matrix}
        \nu\,\, & & \\
        1\,\,\, & 2\,\,\, & 3\,\,\,\\
      \end{matrix} \nonumber \\
  &&
  S_{\bq\nu} = 
     \begin{bmatrix} 
        + & + & + \\
        + & - & - \\
        - & + & - \\
        - & - & + 
      \end{bmatrix}
     \begin{matrix} 
        1 & \bq\\
        2 & \\
        3 & \\
        4 &
      \end{matrix}.\label{eq.signs}
\end{eqnarray}
With this choice the summation in Eq.~\eqref{eq.sumE.2} becomes:
\begin{eqnarray}\label{eq.example}
+ A_{1,2}(\overline{\bq})  
+ A_{1,3}(\overline{\bq})    
+ A_{2,3}(\overline{\bq})\phantom{,}   \nonumber \\
- A_{1,2}(\overline{\bq})  
- A_{1,3}(\overline{\bq}) 
+ A_{2,3}(\overline{\bq})\phantom{,}   \nonumber \\
- A_{1,2}(\overline{\bq}) 
+ A_{1,3}(\overline{\bq})
- A_{2,3}(\overline{\bq})\phantom{,}   \nonumber \\
+ A_{1,2}(\overline{\bq}) 
- A_{1,3}(\overline{\bq}) 
- A_{1,3}(\overline{\bq}).
\end{eqnarray}
By inspecting each column we see that terms with the same superscript $\nu,\nu'$ appear 
the same number of times with positive and negative signs, therefore the sum vanishes. 

In the above example we selected 4 rows from the complete set of combinations in 
Eq.~\eqref{eq.signs2}. How to perform this selection and in which order is not specified 
by the theory leading to Eq.~\eqref{eq.limB}. For example we could have selected rows 
7, 8, 1, 2, and this choice would have led again to the same cancellation of Eq.~\eqref{eq.example}. 
In practice, however, since we sample the Brillouin zone on a discrete grid of $\bq$-points,
the property $A_{\nu\nu'}(\bq)\simeq A_{\nu\nu'}(\overline{\bq})$ used in Eq.~\eqref{eq.sumE.2} 
does not hold for small supercells, and the cancellation is incomplete. This observation
can be exploited to identify an optimal choice of signs from the complete set in Eq.~\eqref{eq.signs2}.
Indeed, we can select $2^{n-1}$ inequivalent rows in such a way as to numerically {\it minimize} 
the error $\Delta_\mathcal{B}$.

A global optimization would require us to evaluate the second derivatives of the observable 
$O^{\{\tau\}}$ for every atomic displacement $\Delta\tau_{p\k\a}$, as it can be seen from 
Eq.~\eqref{eq.delta.b.4}. However, this step would be computationally costly and would make 
the special displacement method equivalent to standard frozen-phonon 
techniques~\cite{Chang_1986,Capaz_2005}.

Instead we make the observation that, as for the matrix of interatomic force constants,
the second derivatives $\D^2 O^{\{ \tau \}}/\D \tau_{0\k\a}\D \tau_{p,\k'\a'}$ must vanish 
for $|\bR_p|\rightarrow \infty$. Therefore the leading components of $A_{\nu\nu'}(\bq)$ 
in Eq.~\eqref{eq.delta.b.4} must be those for small $|\bR_p|$. This suggests to minimize 
Eq.~\eqref{eq.delta.b.3} by retaining only the second line of Eq.~\eqref{eq.delta.b.4}, 
multiplied by the respective signs, for every $\k\a$ and $\k'\a'$ when $\bR_p=0$. In
practice we chose to minimize the following normalized function of the signs~$S_{\bq\nu}$:
  \begin{eqnarray}
  && E(\{S_{\bq\nu}\}) =\nonumber \\ &&\sum_{\substack{\k\a\\\k'\a'}}
  \frac{\Big|\displaystyle\sum_{\substack{\bq\in\mathcal{B}\\\nu<\nu'}}
  \Re [ e^*_{\k\a,\nu} (\bq ) e_{\k'\a',\nu'} (\bq ) ]
  \sigma_{\bq  \nu} \sigma_{\bq \nu'} S_{\bq\nu}S_{\bq\nu'}\Big|}
  {\Big|\displaystyle\sum_{\substack{\bq\in\mathcal{B}\\\nu}}
  \Re [ e^*_{\k\a,\nu} (\bq ) e_{\k'\a',\nu} (\bq ) ]
  \sigma_{\bq  \nu}^2 \Big|}.\quad\label{eq.minimize}
  \end{eqnarray}
The minimization of $E(\{S_{\bq\nu}\})$ with respect to $\{S_{\bq\nu}\}$ only involves the 
phonon eigenmodes and eigenfrequencies, and does not require any evaluation of the observable 
$O^{\{\tau\}}$. Hence the resulting ZG displacement is agnostic to the specific temperature-dependent 
property that is being computed, and can be constructed from quantities that are easily evaluated 
via standard DFPT in the crystal unit cell. When the signs are obtained by minimizing Eq.~\eqref{eq.minimize},
the ZG displacements of a given crystal becomes {\it uniquely} defined for each temperature and
each supercell.

The physical meaning of selecting the signs in such a way as to minimize Eq.~\eqref{eq.minimize}
is that this choice leads precisely to the ZG displacements that best approximates the exact
thermal mean-square displacements given by Eq.~\eqref{eq.Debye_waller_factor}. Therefore this
choice constitutes a way to construct displacements that reproduce XRD spectra as accurately 
as possible, as we show in Sec.~\ref{sec.displ}.

\section{Implementation and computational details}\label{sec.methods_details}

\subsection{Computational setup}

All calculations were performed using the local density approximation~\cite{Ceperley_1980,Perdew_1981} 
to DFT for Si and GaAs, and the PBE generalized gradient approximation~\cite{GGA_Pedrew_1996}
for monolayer MoS$_2$. We employed planewaves basis sets, norm-conserving 
pseudopotentials~\cite{Fuchs_1999} for Si and GaAs, and ultrasoft 
pseudopotentials~\cite{Vanderbilt_1990} for monolayer MoS$_2$, as implemented in the 
{\tt Quantum ESPRESSO} package~\cite{QE}. The planewaves kinetic energy cutoff was 
set to 40~Ry for Si and GaAs, and 50~Ry for MoS$_2$. To minimize interactions between 
periodic images of the monolayer, we used an interlayer separation of 15~\AA.  
The ZG displacement was constructed using vibrational eigenmodes and eigenfrequencies 
obtained from DFPT~\cite{Baroni_2001}, starting from Brillouin zone grids of 
8$\times$8$\times$8 points (Si), 8$\times$8$\times$8 points (GaAs), 8$\times$8$\times$1 
points (monolayer MoS$_2$), and then using standard Fourier interpolation to generate 
dynamical matrices for coarser or denser grids.

In the case of polar materials (GaAs and monolayer Mo$_2$) our calculations correctly 
include the long-range component of the interatomic force constants via the non-analytic 
correction to the dynamical matrix~\cite{Gonze_1997}.

The algorithm used to construct the ZG displacement, the generation of a smooth 
connection between vibrational eigenmodes in the Brillouin zone, and the unfolding 
of spectral functions and band structures from the supercell to the unit cell are 
described in the following three sections, respectively.

\subsection{Generation of the ZG displacement}\label{sec.zg-generate}

To compute the ZG displacement in Eq.~\eqref{eq.realdtau_method00} we proceed as follows. 
First we perform a DFPT calculation of phonons in the crystal unit cell, using a coarse 
uniform grid of $\bq$-points in the Brillouin zone. Then we decide the size of the 
supercell for the ZG displacement, say $N_1\times N_2\times N_3$ unit cells, so that 
$N_p = N_1 N_2 N_3$. This choice sets the grid of $\bq$-points that we need to use in 
Eq.~\eqref{eq.realdtau_method00}, namely $\bq = \sum_{i=1}^{3} \bb_i (n_i-1)/N_i$ 
with $0\le n_i \le N_i$. From this grid we extract the sets $\mathcal{A}$ and $\mathcal{B}$
as illustrated in Fig.~\ref{fig_1}, and discard all remaining points. Using the 
DPFT results from the coarse Brillouin-zone grid, we perform standard Fourier interpolation 
to obtain the eigenmodes $e_{\k\a,\nu}(\bq)$ and eigenfrequencies $\w_{\bq \nu}$ for 
$\bq$-points in sets $\mathcal{A}$ and $\mathcal{B}$. This operation is identical to the 
procedure for calculating phonon dispersion relations using the matrix of interatomic 
force constants.

In order to enforce a smooth Berry connection between phonon eigenmodes as dictated by 
Eq.~\eqref{eq.samew}, we determine unitary rotations for adjacent $\bq$-points using 
a singular-value decomposition (SVD), as described in Sec.~\ref{synch_eigenv}. This 
calculation only requires evaluating scalar products between eigenmodes at adjacent 
$\bq$-points.

At this point we are left with the determination of the signs $S_{\bq\nu}$. Here we 
could proceed with a global optimization of all the signs, using Eq.~\eqref{eq.minimize}. 
However, in order to keep the algorithm as simple as possible, we proceed with a 
partial optimization as follows. We order the $\bq$-points along a path that
is designed to span the entire set $\mathcal{B}$ in the Brillouin zone. A simple
representative path in two dimensions is shown in Fig.~\ref{fig_3}. More complex
choices such as the Peano-Hilbert space-filling curve are possible~\cite{Sagan_book_1994},
but we have not explored these alternatives. The only requirement for the construction
of this path is that it should exhibit `locality', in the sense that pairs of $\bq$-points
which are close in the Brillouin zone should also be close along the path, so that
Eq.~\eqref{eq.sumE.2} is fulfilled.
We then group the $\bq$-points along the path in sets of neighbors, with 
$2^{n-1}$ $\bq$-points per set. The signs in each set are determined by extracting 
$2^{n-1}$ sequences from the $2^n$ possible combinations, as explained in 
Sec.~\ref{sec.normcoord}, excluding antithetic sets. Then we consider $2^{n-1}$
consecutive sets like $\mathcal{D}$, and we choose the signs as cyclic permutations
of those from the first set. This procedure allows us to assign $S_{\bq\nu}$
for a total of $2^{2(n-1)}$ $\bq$-points. At this stage we evaluate the error 
$E(\{S_{\bq\nu}\})$ of Eq.~\eqref{eq.minimize} for these $\bq$-points. If the error 
is above a pre-defined threshold (defined as an external parameter, say $\delta = 5$\%) 
then we restart the procedure by performing a new selection of the $2^{n-1}$ sign 
sequences and their order in the first $\mathcal{D}$ set. 

We emphasize that the above optimization is unnecessary for 
large supercells, and many other choices of signs will lead to essentially the same results. 
This procedure is only advantageous when trying to obtain temperature-dependent properties 
using small supercells (e.g. 4$\times$4$\times$4 supercells). If the set $\mathcal{B}$ 
contains more than $2^{2(n-1)}$ $\bq$-points, then we continue the sequence by restarting 
from the beginning. We note that, while this procedure does not necessarily respect the 
periodic gauge across the Brillouin-zone boundaries, this does not constitute a limitation 
since we are only interested in minimizing $E(\{S_{\bq\nu}\})$.

Having established the signs $S_{\bq\nu}$, we finally compute the ZG displacement
using Eq.~\eqref{eq.realdtau_method00}. In this expression the temperature $T$ 
is an external parameter and enters via the Bose-Einstein factors $n_{\bq\nu}$.

\subsection{Smooth gauge of phonon eigenmodes along a path in reciprocal space} \label{synch_eigenv}

In order to satisfy Eq.~\eqref{eq.samew}, we perform unitary transformations
of phonon eigenmodes at adjacent $\bq$-points on the space-filling curve described
in Sec.~\ref{sec.zg-generate}. The transformation is defined in such a way that 
similar eigenmodes at different $\bq$-points have a similar complex phase, and 
the ordering of eigenmodes is preserved in the case of branch crossing. This
is equivalent to enforcing a smooth Berry connection across the Brillouin zone~\cite{Vanderbilt_Book_2018}. 
These ideas are related to standard concepts used in the theory of maximally-localized
Wannier functions~\cite{Marzari_1997}.

Given two adjacent reciprocal-space vectors $\bq$ and $\bq+\Delta\bq$, we seek for 
a transformation such that $e_{\k\a,\nu}(\bq)$ and $e_{\k\a,\nu}(\bq+\DD\bq)$ be 
as similar as possible. We can define similarity between eigenmodes using the 
overlap matrix:
\begin{eqnarray} \label{eqa.overlap_mat}
   M_{\nu \nu'} = \sum_{\k \a} e_{\k\a,\nu}(\bq+\DD\bq) e^*_{\k\a,\nu'}(\bq).
\end{eqnarray}
Using this definition and using the orthonormality relations in Eq.~\eqref{eqa.orth_norm} 
we have:
\begin{eqnarray} \label{eqa.overlap_mat2}
  e_{\k\a,\nu}(\bq+\DD\bq) = \sum_{\nu'}  M_{\nu \nu'}  e_{\k\a,\nu'}(\bq). 
\end{eqnarray}
For the modes $e_{\k\a,\nu}(\bq)$ and $e_{\k\a,\nu}(\bq+\DD\bq)$ to be as similar 
as possible, the overlap matrix $M_{\nu\nu'}$ needs to be as close as possible to 
the identity matrix. Generally this is not the case since the diagonalization of 
the dynamical matrix at different $\bq$-points introduces an arbitrary gauge in 
the normal modes. To address this problem we perform a transformation of 
$e_{\k\a,\nu}(\bq+\DD\bq)$ as follows:
\begin{equation} \label{eqa.overlap_mat3}
  e^\prime_{\k\a,\nu}(\bq+\DD\bq) = \sum_{\nu'}  U_{\nu \nu'}  e_{\k\a,\nu'}(\bq+\DD\bq),
\end{equation}
where $U_{\nu\nu'}$ is a unitary rotation among the vibrational eigenmodes. After 
this transformation, the new overlap matrix reads:
\begin{equation} \label{eqa.overlap_mat4}
   M^\prime_{\nu \nu'} = \sum_{\k \a} e^\prime_{\k\a,\nu}(\bq+\DD\bq) 
   e^*_{\k\a,\nu'}(\bq) = \sum_{\mu} U_{\nu \mu} M_{\mu \nu'}.
\end{equation}
We want to find the unitary matrix $U$ such that $M^\prime = UM$ is as close as 
possible to the identity $I$. To this aim we need to minimize the quantity 
$|| M^\prime - I ||^2$, where $||A||$ represents the Frobenius norm of the matrix $A$ 
and is given by $||A||^2={\rm Tr}(A^\dagger A)$. Using $M^\prime = UM$ we can write: 
 \begin{equation} \label{eqa.maximize}
  || M^\prime - I ||^2 = {\rm Tr}( M ^\dagger M +I) - {\rm Tr}(M ^\dagger U^\dagger + UM). 
 \end{equation}
Minimizing the left-hand side with respect to $U$ is equivalent to mazimizing 
${\rm Tr}(M ^\dagger U^\dagger + UM)$. Using the properties of the matrix trace 
this is the same as maximizing $\Re\, {\rm Tr}\, (U M)$.

The matrix $M$ is not Hermitian in general, but it can be decomposed via SVD as:
 \begin{eqnarray} \label{eqa.SLR}
  M = LSR^\dagger,
 \end{eqnarray}
where $L$ and $R$ are unitary matrices and $S$ is a diagonal matrix with non-negative 
real values on the diagonal. With these definitions, it can be shown that the matrix 
$U$ that maximizes $\Re\, {\rm Tr}\, (U M)$ is precisely $U= R L^\dagger$~\cite{Matlab}.

In order to use this strategy in practical calculations, we determine unitary
transformations for each $\bq$-point along a space-filling curve, by evaluating 
overlap matrices $M$ between each pair of successive $\bq$-points, say $\bq_1$ and 
$\bq_2$. Then we apply the transformation $U$ to the modes in $\bq_2$, and repeat 
the procedure for $\bq_2$ and $\bq_3$, and so on for all wavevectors. 

Figure~\ref{fig_4} shows the eigenmodes $e_{\k\a,\nu}(\bq)$ of silicon before and 
after the procedure just described. We can see that the resulting eigenmodes
vary continuously between adjacent $\bq$-points, as desired.

\subsection{Generation of temperature-dependent band structure by Brillouin-zone unfolding}\label{sec.unfolding}

In this section we present the recipe for calculating spectral functions that 
represent momentum-resolved density of states of the electrons at a given temperature.
In order to obtain the spectral functions in the Brillouin zone of the primitive
unit cell we employ the following unfolding procedure.

The spectral function is defined as~\cite{Popescu_2012,Medeiros_2014}:
 \begin{equation}\label{eq.sprtl_fn_T}
  A_\bk(\ve;T) =  \sum_{m\bK} P_{m\bK,\bk}(T) \,\delta[\ve - \ve_{m\bK}(T)],
 \end{equation}
where $P_{m\bK,\bk}(T)$ are temperature-dependent spectral weights given by: 
 \begin{equation}\label{eq.sprtl_weight_T}
  P_{m\bK,\bk}(T) = \sum_n |\<\psi^{\rm SC}_{m\bK}(T)| \psi^{\rm PC}_{n\bk}\>|^2. 
 \end{equation}
In these expressions $\ve_{m\bK}(T)$ and $\psi^{\rm SC}_{m\bK}(T)$ represent
eigenvalue and wavefunction of a Kohn-Sham state in the supercell, respectively, 
obtained from a calculation with the ZG displacement at the temperature $T$; 
$\psi^{\rm PC}_{n\bk}$ denotes a state in the primitive unit cell. We employ 
lower (upper) case bold fonts to indicate wavevectors in the primitive cell 
(supercell). The integral is over a volume that encompasses the
supercell and is commensurate both with $\bK$ and $\bk$. The equivalence between 
Eq.~\eqref{eq.sprtl_fn_T} and the standard definition of the electron spectral 
function using the Lehmann representation is demonstrated in Ref.~[\onlinecite{Allen_2013}].

We now expand Kohn-Sham states in a planewaves basis set as follows: 
$\psi^{\rm PC}_{n\bk}(\br) = V^{-1/2} \sum_{\bf g} c^{\rm PC}_{n \bk} 
({\bf g}) \exp[i(\bk+{\bf g})\cdot \br]$, $\psi^{\rm SC}_{n\bK}(\br;T) = 
V^{-1/2} \sum_{\bf G} c^{\rm SC}_{n \bK} ({\bf G};T) 
\exp[i(\bK+{\bf G})\cdot \br]$, where $V$ is the volume where the 
wavefunctions are normalized. By replacing these expansions inside 
Eq.~\eqref{eq.sprtl_weight_T} and using the resolution of identity
$\sum_n c^{\rm PC}_{n \bk} ({\bf g})c^{\rm PC,*}_{n \bk} ({\bf g}')=
\delta_{{\bf g}{\bf g}'}$ one obtains:
 \begin{equation}\label{eq.sprtl_weight_T2}
  P_{m\bK,\bk}(T) = \sum_{\bf g}|c^{{\rm SC}}_{m \bK} ({\bf g} + \bk - \bK; T) |^2. 
 \end{equation}
We note that in the last expression only the planewaves coefficients of the supercell 
state appear; therefore the calculation of the spectral function at finite temperature 
does not require explicit projections onto the states of the primitive cell. In actual 
calculations we select a $\bk$-path in the Brillouin zone of the primitive cell, 
and for each $\bk$-point we proceed as follows: we identify all the supercell $\bK$-points
that unfold into $\bk$ via a reciprocal lattice vector $\bG$ of the supercell;
for each of these points we evaluate the weights $P_{m\bK,\bk}(T)$ using 
Eq.~\eqref{eq.sprtl_weight_T2}; then we use the calculated weights inside
Eq.~\eqref{eq.sprtl_fn_T}. In the case of ultrasoft pseudopotentials, which we 
employed for calculations on monolayer MoS$_2$ in Sec.~\ref{sec.mos2}, we use a 
slightly modified version of Eq.~\eqref{eq.sprtl_weight_T2} to account for the 
augmentation charge~\cite{Vanderbilt_1990}.  Starting from the spectral function 
$A_\bk(\ve;T)$, we extract the renormalized band structure by numerically identifying 
the quasiparticle peaks along the energy axis.

In principle one could also evaluate a momentum- and band- resolved spectral function,
by considering scalar products like in Eq.~\eqref{eq.sprtl_weight_T} but without
the summation over the states $n$ of the primitive unit cell. In this case 
Eq.~\eqref{eq.sprtl_weight_T2} must be replaced by a more complicated expression
which requires an explicit evaluation of wavefunctions in the primitive unit cell.
We explored this possibility, but we found that in order to achieve numerical 
convergence one would need an impractically large number of bands in the supercell.

\section{Summary of procedure and numerical results}\label{sec.method-results}

In this section we summarize the procedure for the calculation of temperature-dependent 
band structures using the ZG displacement, and we discuss our results for silicon,
gallium arsenide, and monolayer molybdenum disulfide. The special displacement method consists
of the following steps:
\begin{itemize}
  \item[(i)]  We compute the vibrational eigenmodes $e_{\k\a,\nu}(\bq)$ and eigenfrequencies 
              $\w_{\bq\nu}$ in the unit cell using DFPT on a coarse uniform grid of ${\bq}$-points.
  \item[(ii)] In preparation for calculations on a supercell of size $N_1\! \times\! N_2\! \times N_3$,
              we interpolate the vibrational eigenmodes and frequencies on a finer ${\bq}$-points grid
              with the same size as the supercell, $N_1\! \times\! N_2\! \times N_3$. We
              partition this grid into the sets $\mathcal{A}$, $\mathcal{B}$, and $\mathcal{C}$. 
 \item[(iii)] We enforce a smooth Berry connection for the vibrational eigenmodes. To this aim we perform 
              unitary transformations that makes eigenmodes at nearby $\bq$-points as similar 
              as possible, as described in Sec.~\ref{synch_eigenv}.
  \item[(iv)] We build an $N_1\! \times\! N_2\! \times N_3$ supercell with the atoms
              displaced by $\DD\btau_{ p\k}$ given in Eq.~(\ref{eq.realdtau_method00}).
              The signs $S_{\bq  \nu}$ in Eq.~(\ref{eq.realdtau_method00}) are determined 
              using the procedure described in Sec.~\ref{sec.normcoord}.
  \item[(v)] To compute band structures, we first set up the $\bk$-point path in the 
             Brillouin zone of the primitive cell, then we obtain the folded $\bk$-points in the 
             Brillouin zone of the supercell. We perform a DFT band structure calculation in the
             supercell, and we unfold the result using the method of Ref.~[\onlinecite{Popescu_2012}].    
             This procedure is discussed in Sec.~\ref{sec.methods_details}. 
\end{itemize}
In this manuscript we focus on temperature-dependent band structures to limit the length of 
the presentation. For calculations of temperature-dependent optical spectra, photoelectron 
spectra, tunneling spectra, or transport coefficients, steps (i)-(iv) above remain the same, 
while step (v) shall be replaced by the calculation of the required property.

We emphasize that the ZG displacement in Eq.~(\ref{eq.realdtau_method00}) does not contain
eigenmodes with $\bq \in \mathcal{A}$. These eigenmodes correspond to stationary lattice
waves, and break the crystal symmetry. In the following sections we show how this observation
can be used to analyze the convergence and the accuracy of the calculations as a function
of supercell size.

The main difference between Eq.~(\ref{eq.realdtau_method00}) and the displacement provided
in our previous work [Eq.~(5) of Ref.~[\onlinecite{Zacharias_2016}]] is that here a more structured
choice of signs allow us to better control the convergence rate and to perform
accurate calculations using relatively small supercells.

\subsection{Thermal mean-square displacements of Si, GaAs, and MoS$_2$}\label{sec.displ}

In Fig.~\ref{fig_5}(a) we compare the probability distribution $P(\DD \tau_{ p \k \a};T)$
in Eq.~\eqref{eq.prob_distribution} evaluated for silicon at $T=0$~K with an histogram of the 
displacements obtained from the ZG formula in Eq.~\eqref{eq.realdtau_method00}. 
The histogram is obtained numerically by binning the values of $\Delta \tau_{p\k\a}$ for all 
atoms along the [100] direction. It is remarkable that the distribution provided by the ZG 
displacement follows the exact thermodynamic average with very high precision. The choice of 
the Cartesian direction is not important in this case, since silicon is isotropic. 
In fact the inset of Fig.~\ref{fig_5}(a) shows that the distribution in the (100) plane is also
isotropic. For completeness we also show in Fig.~\ref{fig_5}(b) how the ZG displacement
appears in a three-dimensional rendering.

In Fig.~\ref{fig_6}(a)-(c) we show
the thermal mean-square displacements of Si, GaAs, and MoS$_2$ calculated using the ZG displacements
(colored disks), the mean-square displacements evaluated using the exact expression in
Eq.~(\ref{eq.Debye_waller_factor}) (grey disks), and experimental data from XRD where available
(triangles) \cite{Aldred_1973,Stern_1980,Matsushita_1974,Schonfeld_1983}. We can see that the
ZG displacement yields mean-square displacements in excellent agreement with
Eq.~(\ref{eq.Debye_waller_factor}), and that the agreement between our calculations and experiments
is also very good. These successful comparisons reinforce the notion that the ZG displacement
provides a very accurate classical representation of a thermodynamic average over the quantum
fluctuations of the atomic positions.

The ZG displacement can also be employed to obtain the thermal displacement ellipsoids and
compare the complete ADP tensor $U_{\k,\a\b}(T)$ of Eq.~\eqref{eq.adp} with experiments.
In Fig.~\ref{fig_6}(d)-(f) we show the computed thermal ellipsoids of Si, GaAs, and MoS$_2$
at $T=300$~K. The ellipsoids of Si, Ga, and As are found to be spheres with radius
0.49~\AA$^2$, 0.59~\AA$^2$, and 0.50~\AA$^2$, respectively. These findings are consistent
with the cubic symmetry of the Si and GaAs lattices. In the case of MoS$_2$, the ellipsoids
reflect the two-dimensional nature of the material: the in-plane parameters $U_\parallel$
are 0.22~\AA$^2$ and 0.42~\AA$^2$ for Mo and S atoms, respectively; the out-of-plane parameters
$U_\perp$ are 0.82~\AA$^2$ and 0.86~\AA$^2$ for Mo ans S, respectively. In all cases the ADPs
obtained from the ZG displacements are within 25\% of the corresponding experimental values
\cite{Aldred_1973,Stern_1980,Matsushita_1974,Schonfeld_1983}.

\subsection{Temperature dependent band structure of Si} \label{sec.silicon}

Figure~\ref{fig_7} shows our results and analysis for the band structure of Si at finite 
temperature. Full computational details, including the evaluation of the spectral function
and band energies in the primitive unit cell, are provided in Sec.~\ref{sec.methods_details};
here we only mention that the calculations are based on DFT in the local density approximation 
(LDA). In Fig.~\ref{fig_7}(a),  we show the spectral function $A_{\bk}(\ve;T)$ obtained from the
ZG displacement in a 8$\times$8$\times$8 supercell. We consider $T=0$~K to focus on the
effect of zero-point renormalization. The spectral function provides the momentum-resolved
electronic density of states~\cite{Abrikosov_1963}, and it is shown as a colormap.
In Fig.~\ref{fig_7}(b) we compare the bands $\ve_{n\bk}(T)$ extracted from the spectral function
$A_\bk(\ve;T)$ with the usual DFT band structure evaluated at clamped ions; in particular
we consider the dispersions along the L$\Gamma$X path of the Brillouin zone for $T=0$~K 
(blue) and $T=300$~K (green); the DFT bands at clamped ions are in black. This calculation 
indicates that, as a result of zero-point motion, the energy of the valence band maximum 
(VBM) at $\Gamma$ increases by 35~meV, while the energy of the indirect conduction band minimum 
(CBM) at $\simeq 0.87\,\Gamma X$ decreases by 22~meV. These values are in excellent agreement with 
Ref.~[\onlinecite{Ponce_2015}], which obtained 35~meV and 21~meV, respectively,
when using the perturbative Allen-Heine approach and the adiabatic approximation.

From the band structure in Fig.~\ref{fig_7}(b) we can also extract the phonon-induced
mass-enhancement. For simplicity we focus on the longitudinal electron mass of silicon, 
and we define the coupling strength $\lambda_T$ such that $m^*_l(T) = (1+\lambda_T)m^*_l$ 
where $m_l^* = {0.95}\,m_e$ is the DFT mass at clamped ions. From the calculated bands
we obtain $\lambda_T =0.03$ and $0.04$ for $T=0$~K and $T=300$~K, respectively. This 
finding is in agreement with the mass renormalization reported in Ref.~[\onlinecite{Ponce_2018}].

Our calculated band gap narrowing due to zero-point effects is 57~meV.  This value is in very good
agreement with previous calculations based on non-perturbative adiabatic approaches, yielding
56-65~meV~\cite{Bartomeu_2014,Bartomeu_2016,Zacharias_2015,Zacharias_2016,Karsai_2018} as well as
with experimental values, in the range of 62-64~meV~\cite{Cardona_2001,Cardona_2005}.
Ref.~[\onlinecite{Ponce_2015}] showed that non-adiabatic corrections within the Allen-Heine theory 
increase the zero-point renormalization by 8~meV~\cite{Ponce_2015}. This effect is not captured
by the present special displacement method, which is in essence an adiabatic approach.
We also point out that the most recent GW calculations using an earlier version of the
present approach~\cite{Karsai_2018} yield a slightly larger gap renormalization of 66~meV. 
This is expected given that the electron-phonon interaction is overscreened in DFT/LDA due
to the band gap problem.

In order to analyze the convergence behavior of the SDM, in Fig.~\ref{fig_7}(c) we plot
the dependence of the zero-point band gap renormalization on the supercell size.
Two sets of data are shown. The green lines and datapoints correspond to calculations
performed using Eq.~(\ref{eq.realdtau_method00}). The black lines and grey datapoints were obtained
after modifying Eq.~(\ref{eq.realdtau_method00}) to include the contributions of phonons with
$\bq \in \mathcal{A}$. Phonons with $\bq \in \mathcal{A}$ correspond to stationary waves
in the primitive unit cell (e.g. $\bq=0$ phonons). As we demonstrate in Sec.~\ref{sec.normcoord},
in the thermodynamic limit of infinite supercell the contribution of these modes vanishes identically.
Therefore, the calculations performed by including or excluding phonons with $\bq \in \mathcal{A}$ 
in the ZG displacement given by Eq.~(\ref{eq.realdtau_method00}) converge to the same limit.
However, by considering only $\bq \in \mathcal{B}$ phonons as in Eq.~(\ref{eq.realdtau_method00})
we reach convergence from the bottom (in terms of magnitude); while by including phonons
with $\bq \in \mathcal{B}$  and $\bq \in \mathcal{A}$ we reach convergence from the top.
By construction the converged answer must lie in between these two limits, therefore 
the analysis presented in Fig.~\ref{fig_7}(c) can be used as a way to quantify the convergence
error of the calculations. For example the data obtained for a 10$\times$10$\times$10 supercell 
in Fig.~\ref{fig_7}(c) show that the fully converged result for an infinitely-large supercell 
will be in the interval 55-65~meV. This observation carries general validity for all the 
systems considered in this work. 
  
The inset of Fig.~\ref{fig_7}(c) shows the spectral function near the threefold degenerate
VBM of silicon, as computed using the ZG displacement for a 4$\times$4$\times$4 supercell
at $T=0$~K. If we include $\bq \in \mathcal{A}$ points in the ZG displacement, then the
crystal symmetry is broken, and by consequence we observe a band splitting (black line). 
In contrast, when we use the pure ZG displacement from Eq.~(\ref{eq.realdtau_method00}), i.e. 
without phonons with $\bq \in \mathcal{A}$, the band degeneracy is correctly preserved. 
This analysis indicates that, when performing electron-phonon calculations using non-perturbative 
supercell approaches, $\bq=0$ phonons as well as all phonons with $\bq \in \mathcal{A}$ are 
the least representative since they break crystal symmetry, which may lead to calculation 
artifacts. This issue is resolved by the present formulation of the ZG displacement as 
provided by Eq.~(\ref{eq.realdtau_method00}).

In Fig.~\ref{fig_7}(d) we compare our calculations of the indirect band gap of silicon 
using the SDM with experiments~\cite{Alex_1996}, up to $T=500$~K. To facilitate comparison 
we scissor-shifted our DFT/LDA results by 0.73~eV, which is close to the GW correction reported 
in Ref.~[\onlinecite{Lambert_2013}]. The agreement between our calculation and experiment 
is very good, except that we underestimate slightly the temperature slope. This effect 
is a well-known consequence of the fact that the strength of the electron-phonon interaction 
is underestimated by DFT/LDA; the slope can be improved by using GW calculations in combination 
with the SDM, as demonstrated in Ref.~[\onlinecite{Karsai_2018}].

\subsection{Temperature dependent band structure of GaAs}\label{sec.gaas}

Figure~\ref{fig_8} shows our calculated band structure of GaAs at finite temperature. 
Also in this case we employ DFT and the LDA, as described in Sec.~\ref{sec.methods_details}. 
In Fig.~\ref{fig_8}(a) we have the spectral function $A_{\bk}(\ve;T)$ at $T=0$~K as a color map, 
and in Fig.~\ref{fig_8}(b) we have the bands extracted from the spectral function at $T=0$~K 
(blue) and $T=300$~K (green). The bands at clamped ions are shown in 
black for comparison.  For simplicity we are not including spin-orbit coupling in the calculations,
therefore we do not see the spin-orbit splitting of the split-off holes in the valence 
bands~\cite{Chelikowsky_1976,Jancu_1998,Soline_2004}. Since the holes at the top of the 
valence band have the same orbital character, we expect a similar zero-point renormalization 
for the split-off holes.

From the data in Fig.~\ref{fig_8}(b) we obtain zero-point corrections to the valence and conduction
band edges at $\Gamma$ of $+21$~meV and $-11$~meV, respectively. The resulting band gap narrowing, 
$\Delta E_{\rm g} = 32$~meV, lies within the experimental range of $57\pm29$~meV 
\cite{Lautenschlager_1987}. The calculations in Fig.~\ref{fig_8}(b) are performed using an 
8$\times$8$\times$8 supercell. To better compare with the finite-differences results of 
Ref.~[\onlinecite{Antonius_2014}], we repeated the calculations with a smaller, 4$\times$4$\times$4 
supercell. In this case we obtain $\Delta E_{\rm g} = 25$~meV, which matches 
the value of 25~meV reported in Ref.~[\onlinecite{Antonius_2014}].
It is well established that GW quasiparticle corrections change these results
by strengthening the electron-phonon coupling~\cite{Antonius_2014,Karsai_2018}: in 
order to incorporate this effect it will be sufficient to perform a GW calculation 
using the ZG displacement.

From the band structure in Fig.~\ref{fig_8}(b) we can determine the phonon-induced
mass-enhancement. Since we are not including spin-orbit effects, we only report the value 
for the CBM at $\Gamma$. By denoting $m_e^* = 0.056\,m_e$ the conduction band mass at clamped 
ions, we find $m^*_e(T) = (1+\lambda_T)m^*_e$ with $\lambda_T = 0.005$ and $0.007$ for $T=0$~K 
and $T=300$~K, respectively. We are unaware of previous calculations of the mass renormalization
in GaAs as a function of temperature.

In Fig.~\ref{fig_8}(c) we show the convergence of the zero-point renormalization of the 
direct gap of GaAs with respect to the supercell size. The converged value obtained with
Eq.~(\ref{eq.realdtau_method00}) is $\Delta E_{\rm g}=32$~meV and is obtained using an 
8$\times$8$\times$8 supercell. As for the case of Si in Fig.~\ref{fig_7}(c), also for GaAs 
the threefold degeneracy of the VBT is preserved when using the ZG displacement. And also 
in this case, if we include phonons with $\bq \in \mathcal{A}$ in Eq.~(\ref{eq.realdtau_method00}), 
the degeneracy is lifted due to symmetry breaking, as it can be seen in the inset of Fig.~\ref{fig_8}(c).

In previous work it has been suggested that in the case of polar materials it might 
be impossible to achieve convergence within the adiabatic approximation~\cite{Ponce_2015}.
Here we wish to emphasize that the divergence of perturbative approaches for polar materials 
is {\it not} a consequence of the adiabatic approximation, but it results from taking the limit 
$\omega_{\bq\nu}\rightarrow 0$ before performing the integration over phonon wavevectors 
to obtain the Fan-Migdal 
self-energy \cite{Giustino_2017}. When the limit $\omega_{\bq\nu} \rightarrow 0$ is 
correctly performed {\it after} evaluating the integral over $\bq$, there is no divergence. 
The correct procedure can be found in Sec.~IV of the early work by Fan~\cite{Fan_1951} and 
is summarized in Appendix~\ref{app.adiab_polar}. Figure~\ref{fig_8}(c) shows indeed that 
adiabatic calculations using the SDM converge smoothly as a function of supercell size, and 
that the fully converged band gap renormalization lies in a very narrow bracket between 
32~meV (ZG displacement without $\bq\in \mathcal{A}$ phonons) and 36~meV 
(ZG displacement with $\bq\in \mathcal{A}$ phonons) already for a 8$\times$8$\times$8 supercell.

Generally speaking, the smooth and fast convergence of the SDM can be ascribed to the fact that 
the formalism relies on a standard DFT calculation for a supercell with displaced atoms. This 
calculation is intrinsically easier to converge than perturbative approaches. In fact perturbative 
methods require stringent $\bq$-point sampling to evaluate principal-value integrals of first-order 
poles that appear in the Fan-Migdal and Debye-Waller self-energies; furthermore these integrals
yield large and opposite contributions, so the final result is obtained by subtracting two large 
numbers. The SDM method circumvents these numerical challenges.

In Fig.~\ref{fig_8}(d) we show our results for the direct gap of GaAs using the ZG displacement, 
in the temperature range 0-500~K, and we compare with experimental data from 
Ref.~[\onlinecite{Lautenschlager_1987}]. In order to take into account the non-negligible expansion 
of the GaAs lattice with temperature, we employ the quasi-harmonic approximation. We use a 
scissor-shift of 0.53~eV to mimic GW corrections as reported in Ref.~[\onlinecite{Remediakis_1999}]. 
The slight underestimation of the experimental slope with temperature can be corrected
by performing GW calculations instead of standard DFT/LDA~\cite{Antonius_2014,Karsai_2018},
but overall the agreement between the present calculations and experiments is very good.

For completeness, we also mention that the results shown in Fig.~\ref{fig_8}(d)
and obtained from the analysis of temperature-dependent band structures
are in excellent agreement with the values that we obtained in an earlier work
by analyzing the joint density of states~\cite{Zacharias_2016}.

\subsection{Temperature dependent band structure of monolayer MoS$_2$}\label{sec.mos2}

Figure~\ref{fig_9} shows our calculated temperature-dependent band structure
of monolayer MoS$_2$, as an example of application of the SDM to two-dimensional materials.
In this case we employed the PBE exchange and correlation functional~\cite{GGA_Pedrew_1996},
and fully relativistic ultrasoft pseudopotentials~\cite{Vanderbilt_1990} to take spin-orbit 
coupling into account (Sec.~\ref{sec.methods_details}).

In Fig.~\ref{fig_9}(a) we have the spectral function of MoS$_2$ at $T=0$~K obtained from the
ZG displacement in a 10$\times$10$\times$1 supercell. By extracting the corresponding
temperature-dependent bands we obtain a spin-orbit splitting of 136~meV for the valence band 
states at $K$, to be compared to the clamped-ion value of 142~meV. Our calculation is in agreement 
with the spliting of 135~meV obtained in Ref.~[\onlinecite{Molina_2016}] within the Allen-Heine theory.
We also determined the electron  effective mass renormalization at the $K$ point, and found 
$\lambda_T \simeq 0.04$ and $\simeq 0.06$ for $T=0$~K and $T=300$~K, respectively. These latter two
values are not fully converged since we used finite differences with a coarse $\bk$-point spacing 
of $6\cdot 10^{-3}\, 2\pi/a$.

In Fig.~\ref{fig_9}(b) and (c) we show convergence tests for the gap renormalization.
In Fig.~\ref{fig_9}(b) we have the renormalization as a function of interlayer separation 
by keeping the in-plane supercell size fixed; in (c)  we vary the supercell size, keeping 
the interlayer separation constant. As in the case of the three-dimensional
materials Si and GaAs, the band gap narrowing converges relatively quickly with the
size of the cell; furthermore the calculations are not very sensitive to the size
of the vacuum region provided periodic replica are separated by more than 10~\AA.

For a 10$\times$10$\times$1 supercell we obtain a zero-point gap renormalization
of 64~meV using the ZG displacement. By including also $\bq \in \mathcal{A}$ phonons
in Eq.~(\ref{eq.realdtau_method00}) the value changes only slightly to 65~meV,
therefore we estimate an error with respect to the fully converged result of less than
1~meV. Our result is in good agreement with the perturbative
calculations of Ref.~[\onlinecite{Molina_2016}], which reported 75~meV. The residual
difference may be due to the `rigid-ion' approximation used in the Allen-Heine
approach of Ref.~[\onlinecite{Molina_2016}],
and to differences in the DFT exchange and correlation functionals employed.

In Fig.~\ref{fig_9}(d) we compare our calculated temperature-dependent band gap of
MoS$_2$ (green) with experimental data from Ref.~[\onlinecite{Park_2018}] (black).
To facilitate comparison we introduced a scissor correction of 0.34~eV to match the
measured band gap at 4~K. This correction is similar to that obtained using GW and 
the Bethe-Salpeter equation (BSE)~\cite{Hongliang_2013}. Unlike in Si and GaAs, here 
the calculated data follow the experimental trend very closely. This is unexpected 
since we are computing electron-phonon effects at the DFT/PBE level, and suggests
that quasiparticle corrections of the electron-phonon coupling~\cite{Zhenglu_2019} 
are not significant in MoS$_2$.

\subsection{General remarks on the SDM results}

The applications to Si, GaAs, and MoS$_2$ described in the previous sections show that 
the SDM yields temperature-dependent bands of the same quality as perturbative approaches 
based on the Allen-Heine theory. In particular, the capability to access momentum-resolved 
quantities such as temperature-renormalization at specific $\bk$-points and the phonon-induced 
enhancement of the band mass are entirely novel, and considerably expand the range of 
applicability of the ZG displacement.

One important point of note is that the present reciprocal-space formulation of the ZG 
displacement allowed us to identify and remove the contributions of vibrational modes that 
break crystal symmetry. This improvement is important in order to avoid spurious and 
uncontrolled lifting of electronic degeneracy.  Since the symmetry-breaking modes are 
those with wavevector $\bq \in \mathcal{A}$, that is phonons corresponding to standing 
Bloch waves, the present analysis and results demonstrate that these modes (especially 
$\bq=0$ phonons) are the {\it least} representative in a thermodynamic average, and 
should be avoided for accurate calculations.

Another interesting point is that we can bracket the fully converged results by performing 
two calculations: one with the pure ZG displacement of Eq.~(\ref{eq.realdtau_method00}), 
and one with the modified version including $\bq \in \mathcal{A}$ phonons. This provides 
a simple and effective strategy to quantify the convergence error as a function of supercell 
size.

We also found that, when used in conjunction with the ZG displacement, the adiabatic 
approximation does not suffer from the convergence problems that are encountered in 
perturbative approaches for polar materials. This advantage arises from the fact that 
the special displacement method does not require integrating over poles as in perturbative 
approaches, and the calculation is as easy as a standard calculation at clamped ions.

For the systems considered in this work, the supercells required to achieve relatively 
accurate results are surprisingly small. For example, the ZG displacement in a 
4$\times$4$\times$4 Si supercell (128 atoms) yields a zero-point renormalization which 
differs by less than 10\% from the fully converged value; for GaAs we need a 6$\times$6$\times$6 
supercell (432 atoms) to achieve similar accuracy; for MoS$_2$ a 6$\times$6$\times$1 supercell 
(72 atoms) is enough to converge the results within 10\%. This indicates that the ZG 
displacement can be used to perform calculations with relatively small supercells, and this 
may open the way to post-DFT calculations at finite temperature, including GW quasiparticle 
calculations and exciton calculations via the BSE method.

\section{Conclusions and outlook} \label{sec.conclusion}

In this manuscript we develop a new methodology for performing electronic structure
calculations at finite temperature. In a nutshell, this method consists of performing
a single calculation in a supercell where the atoms have been displaced according to
Eq.~\eqref{eq.realdtau_method00}. We refer to this displacement as the ZG displacement,
and to the methodology as the special displacement method.

This work follows our earlier study in Ref.~[\onlinecite{Zacharias_2016}], where the
original idea was first proposed. The key novelty of the present study is that we
reformulate the entire theory using a compact and rigorous reciprocal space formulation,
building on density functional perturbation theory. This new formulation allows us to 
go beyond angle-integrated electronic and optical spectra, and to compute complete 
{\it band structures} at finite temperature. We demonstrate this concept for three-dimensional
bulk semiconductors, silicon and gallium arsenide, and for a two-dimensional semiconductor, 
monolayer molybdenum disulfide. In all cases our results match the accuracy of established 
perturbative techniques based on the Allen-Heine theory. The added advantage of the present 
approach over perturbative methods is that it does not require the evaluation of 
self-energy energy poles, and it does not require the rigid-ion approximation for the 
Debye-Waller self-energy. As a consequence, the method is robust, reliable, and simple to use.

Beyond demonstrating calculations of band structures at finite temperature, we show that 
the ZG displacement accurately reproduces the characteristic anisotropic displacement 
parameters measured by X-ray crystallography, and can be used to determine thermal 
ellipsoids as a function of temperature. Therefore the ZG displacement represents an 
accurate classical snapshot of thermal disorder in solids, and eliminates the need 
for the stochastic sampling of the nuclear wavefunctions.

More fundamentally, we show that the ZG displacement can be understood as the best 
single-point approximant to a thermodynamic Feynman path-integral. This link may
open new avenues in the study of path integrals using the quantum-classical analogy
and special displacements.

In the present approach, the choice of the atomic displacements is carefully performed by 
first establishing a smooth Berry connection among vibrational eigenmodes in the Brillouin 
zone, and then by choosing the phase of each eigenmode using rigorous sum rules. This 
approach allows us to prove that the ZG displacement yields the exact Williams-Lax average 
of an electronic observable in the thermodynamic limit of infinite supercell. We also provide
procedures for accelerating the convergence of the calculations for those cases where 
only small supercells are within reach.

For reasons of space we only discussed applications to finite-temperature band structures, 
but we emphasize that this methodology is much more general. For example the earlier version 
of this method~\cite{Zacharias_2016} has already been applied to compute phonon-assisted 
optical properties, dielectric functions, GW quasiparticle corrections, zero-point 
renormalization of band gaps, exciton-phonon couplings, and charge transport. This broad 
range of applications is possible because the Williams-Lax theory can be employed to 
compute any property which can be expressed as a Fermi's golden rule.

The present extension of the special displacement method to compute entire band structures 
is particularly appealing for testing quasiparticle corrections at finite temperature. 
Since only one supercell calculation is required, and since we developed a new algorithm 
to accelerate convergence with respect to the supercell size, this approach opens the way 
to systematic many-body calculations of band structures at finite temperature.

Up to this point the SDM relied on the harmonic approximation. It would be interesting
to consider extensions to anharmonic systems. We expect that the method will work seamlessly 
in conjunction with the quasi-harmonic approximation~\cite{Fleszar1990,Baroni2010} and 
with the self-consistent harmonic approximation~\cite{Hellman2011,Errea2014}. In fact, in 
both cases the original anharmonic potential is replaced by its `best' harmonic approximation; 
after this replacement the SDM method can be used without changes. In the case of strongly 
anharmonic systems, for example in the presence of double-well potential energy surfaces, 
it should be possible to modify the present method by treating all the harmonic modes as 
described in this manuscript, and by adapting the ZG configuration to describe averages 
over double-well quantum nuclear wavefunctions. The feasibility and accuracy of this approach will 
have to be demonstrated in future work.

\acknowledgments
Work by F.G. was supported by the U.S. Department of Energy (DOE), Office of
Science, Basic Energy Sciences (BES) under Award DE-SC0020129.
 The authors acknowledge the use of the ARCHER UK National
Supercomputing Service under the `T-DOPS' Leadership project. 

\appendix

\section{Dynamical matrix and Born-von K\'arm\'an boundary 
conditions}\label{app.normal_modes}

Here we summarize the relations between normal modes $e_{\k\a,\nu}(\bq)$ 
obtained from the diagonalization of the dynamical matrix at each $\bq$ wavevector,
and the standard sum rules resulting from the Born-von K\'arm\'an boundary conditions.  

In the harmonic approximation the dynamical matrix $D_{\k \a,\k'\a'}(\bq)$ at 
each $\bq$-point and the vibrational eigenmodes and frequencies satisfy
the equation:~\cite{Giustino_2017}
\begin{eqnarray}\label{eqa.dyn_mat}
\sum_{\k' \a'} D_{\k \a,\k'\a'}(\bq) e_{\k'\a',\nu'} (\bq ) = \w_{\bq \nu} 
e_{\k\a,\nu} (\bq ).
\end{eqnarray}
Since the dynamical matrix is Hermitian, the eigenmodes form an orthonormal basis:
\begin{eqnarray}\label{eqa.orth_norm}
   \sum_{\nu} e_{\k\a,\nu} (\bq ) e^{*}_{\k'\a',\nu} (\bq ) &=& \delta_{\k\k'} \delta_{\a\a'}, 
  \\ \sum_{\k\a} e^{*}_{\k\a,\nu} (\bq ) e_{\k\a,\nu'} (\bq )&=& \delta_{\nu\nu'}.
\end{eqnarray}
Equation~\eqref{eqa.dyn_mat} implies the relations:~\cite{Giustino_2017}
\begin{eqnarray}\label{eqa.orth_norm2}
    e_{\k\a,\nu} ( -\bq) = e^{*}_{\k\a,\nu} (\bq), \,\,\,\,\,\,\,\,\, \w_{\bq \nu} = 
\w_{-\bq \nu},
\end{eqnarray}
where we followed the convention of Ref.~[\onlinecite{Maradudin_1968}].

We denote the lattice vector pointing to the $p$-th unit cell as
${\bf R}_p= n_{p_1} {\bf a}_1 + n_{p_2} {\bf a}_2 + n_{p_3} {\bf a}_3$,
where the ${\bf a}_i$ represent the primitive lattice vectors and $n_{p_i}$  
are integers between 0 and $N_i-1$. The Born-von K\'arm\'an  supercell contains 
$N_p = N_1 \times N_2 \times N_3$ unit cells. For the uniform grid of phonon 
wavevectors $\bq$ in the Brillouin zone we choose $\bq = (m_1/N_1) {\bf b}_1 
+ (m_2/N_2) {\bf b}_2 + (m_3/N_3) {\bf b}_3$, where the ${\bf b}_i$ represent
the primitive reciprocal lattice vectors, and $m_i$  are integers between 0 
and $N_i-1$. The vectors ${\bf b}_i$ are such that ${\bf a}_i \cdot {\bf b}_j 
= 2\pi \delta_{ij}$. With these conventions we have the sum rules:
 \begin{eqnarray}\label{eqa.sum_rules}
  \sum_\bq  e^{i ({\bf R}_p - {\bf R}_{p'}) \cdot \bq }  &=& N_p \delta_{pp'} \\
  \sum_p e^{i (\bq  - \bq ') \cdot {\bf R}_p} &=& N_p \delta_{\bq \bq '}.
 \end{eqnarray}

\section{Link between Williams-Lax theory and Feynman's path 
integrals}\label{app.Feynman_path_integrals}
In this appendix we provide some of the mathematical background required
to link the Williams-Lax theory with thermodynamic path integrals. The following
equations are used in Sec.~\ref{sec.link_to_PI}.

Starting from Eq.~\eqref{eq.WL_PI-2}, using the Trotter 
decomposition $\exp(-\hat{H}_{\rm n}/k_{\rm B}T) = 
[\exp(-\hat{H}_{\rm n}/N k_{\rm B}T)]^N = \prod_{i=1,N} \exp(-\hat{H}_{\rm n}/N 
k_{\rm B}T)$,\cite{Ceperley_1995} and inserting the resolutions of identity $\int 
d\tau_i |\tau_i\> \< \tau_i| = 1$ with $i=1,\cdots,N\!-\!1$ in between each product, 
Eq.~\eqref{eq.WL_PI-2} becomes:
  \begin{eqnarray}\label{eq.WL_PI-2b}
  \Gamma_{\a\rightarrow \beta} (\w;T) & = & \frac{1}{Z} 
  \int d\tau_0 \cdots d\tau_{N-1} \nonumber \\ &\times&\!\! \prod_{i=0}^{N-1} 
  \< \tau_i |  e^{-\hat{H}_{\rm n}/N k_{\rm B}T} |\tau_{i+1}\> \Gamma_{\a\rightarrow \beta}^{\{ 
  \tau \}}(\w),\,\,
 \end{eqnarray}
where we introduced $|\tau_0\> = |\tau_N\> = |\tau \>$ to make the notation more compact.

The nuclear Hamiltonian can be expressed as a sum of kinetic ($\hat{T}$) 
and potential ($\hat{U}_\a$) energies, and the exponential in Eq.~\eqref{eq.WL_PI-2b} is 
rewritten for large $N$ using the Baker-Campbell-Hausdorff formula~\cite{Hagen_2009}:
  \begin{eqnarray}\label{eq.WL_PI-3}
   \lim_{N\rightarrow \infty} e^{-\hat{H}_{\rm n}/N k_{\rm B}T} = 
   e^{-\hat{T}/N k_{\rm B}T} e^{-\hat{U}_\a/N k_{\rm B}T}.
 \end{eqnarray}
In this limit the matrix elements appearing in Eq.~\eqref{eq.WL_PI-2b} take the form:
  \begin{equation}\label{eq.WL_PI-4}
    \< \tau_i |  e^{-\hat{T} \Delta t/\hbar} |\tau_{i+1}\>\, e^{-U_\a(\{\tau_{i+1}\})
    \Delta t/\hbar},
  \end{equation}
having defined $\Delta t = \hbar/N k_{\rm B}T$. The matrix element with the kinetic energy 
is simplified by inserting the resolution of identity for the eigenstates of the nuclear 
momentum operators, $\int d k |k\>\<k| = 1$, where $\{k\}$ represent a complete set of 
planewaves associated with the atomic positions $\{\tau\}$. After evaluating terms like 
$\<k|\tau_i\>$ and solving the resulting Gaussian integral, this procedure leads to:
  \begin{equation}\label{eq.WL_PI-5}
     \< \tau_i |  e^{-\hat{T}\Delta t/\hbar} |\tau_{i+1}\>
      \!= c \prod_{p\k\a} \!e^{-[M_\k(\tau_{p \k \a, i+1}-\tau_{p\k\a,i})^2/2\Delta t^2]\Delta t/\hbar},
  \end{equation}
where $c$ is a normalization prefactor. By taking the limit of large $N$, which is equivalent 
to taking $\Delta t \rightarrow 0$, the terms $(\tau_{p \k \a, i+1}-\tau_{p\k\a,i})/\Delta t$ 
become the classical velocities of the nuclei, and the exponent of Eq.~\eqref{eq.WL_PI-5}
yields the classical kinetic energy of the nuclei. 

The combination of Eqs.~\eqref{eq.WL_PI-2b}-\eqref{eq.WL_PI-5} with Eq.~\eqref{eq.WL_PI-2}
yields the link of the Wiliams-Lax transition rate to a thermodynamic Feynman path integral, 
as shown in Eq.~\eqref{eq.WL_PI-11}. 

\section{Adiabatic approximation for polar materials}\label{app.adiab_polar}

In recent work it has been proposed that the adiabatic approximation leads
to an ill-defined expression~\cite{Ponce2015} when computing the band structures 
of polar semiconductors at finite temperatures.

In this Appendix we show that the divergence noted in Ref.~[\onlinecite{Ponce2015}] is not
related to the adiabatic approximation, but to the procedure employed for evaluating
the principal-value integrals appearing in the self-energy.

For simplicity we write the Fan-Migdal self-energy used in Ref.~[\onlinecite{Ponce2015}]
within the Fr\"olich approximation.
In particular we consider an electron in an otherwise empty conduction band on mass $m^*$,
interacting with a single, dispersionless polar longitudinal-optical phonon of frequency
$\w_{\rm LO}$. At zero temperature the self-energy reads~\cite{Giustino_2017}:
  \begin{equation}\label{eq.b1}
   \Sigma^{\rm FM}_{\bk}(\w) = 
     \!\int\!\! \frac{d\bq}{\Omega_{\rm BZ}} 
   \frac{|g(\bq)|^2}{\hbar\w\!-\!\ve_{\bk+\bq}-\hbar\w_{\rm LO}+i\hbar\eta}.
  \end{equation}
Here $\ve_{\bk}=\hbar^2|\bk|^2/2m^*$, $\eta$ is a positive infinitesimal, $\Omega_{\rm BZ}$
is the volume of the Brillouin zone, and the Fr\"olich electron-phonon matrix element 
is given by~\cite{Sio2019b}:
 \begin{equation}\label{eq.b2}
  |g(\bq)|^2 = \frac{e^2}{4\pi\e_0}\frac{4\pi}{\Omega}\frac{\hbar\w_{\text{LO}}}{2}\frac{1}{\k\,|\bq|^2},
 \end{equation}
where $\Omega$ is the volume of the unit cell. The quantity $\kappa$ is defined as 
$1/\kappa = 1/\epsilon_0-1/\epsilon_\infty$, with $\epsilon_0$ and $\epsilon_\infty$
being the static and the high-frequency dielectric constants, respectively.

After replacing Eq.~\eqref{eq.b2} inside \eqref{eq.b1} and carrying out the algebra we
reach the standard expression for the self-energy at the band bottom ($\bk=0$, $\w=0$,
$\eta\rightarrow 0$):
  \begin{equation}\label{eq.b3}
   \Sigma^{\rm FM} = 
        -\alpha \hbar\w_{\rm LO}\, \frac{q_{\rm LO}}{\pi}\int_{-\infty}^\infty \frac{dq}
      {q^2+q_{\rm LO}^2},
  \end{equation}
having defined $q_{\rm LO}^2 = 2m^* \hbar\w_{\rm LO}/\hbar^2$. The integration was extended
to the infinite crystal since this does not alter the result~\cite{Fan_1951}, and we
used the definition of the Fr\"ohlich coupling constant~\cite{Sio2019b}:
   \begin{equation}\label{eq.b4}
    \alpha = \frac{e^2}{4\pi\e_0} \frac{1}{\hbar}\sqrt{\frac{m^*}{2\hbar\w_{\rm LO}}}\frac{1}{\kappa}.
   \end{equation}
The evaluation of the integral in Eq.~\eqref{eq.b3} yields $\pi/q_{\rm LO}$, therefore the
self-energy is:
  \begin{equation}\label{eq.b5}
   \Sigma^{\rm FM} = 
        -\alpha \,\hbar\w_{\rm LO}.
  \end{equation}
This is a standard result which is well known in the literature~\cite{Nery2016,Mahan1993}.

In Ref.~[\onlinecite{Ponce2015}] it was noted that, in the presence of Fr\"ohlich coupling
as in the above example, the adiabatic approximation to the self-energy diverges 
in the limit of dense Brillouin zone sampling. This result was reached by taking the limit 
$\w_{\rm LO} \rightarrow 0$ in Eq.~\eqref{eq.b1} {\it before} evaluating the integral. 
As a consequence of this limiting operation, the integral appearing in Eq.~\eqref{eq.b3} 
is replaced by $\int_{-\infty}^{+\infty} q^{-2}dq$, which diverges indeed.

However, one can alternatively perform the adiabatic approximation by taking the limit 
$\w_{\rm LO} \rightarrow 0$ {\it after} evaluating the integral. This alternative
approach corresponds to taking the limit $\w_{\rm LO} \rightarrow 0$ of Eqs.~\eqref{eq.b4} and 
\eqref{eq.b5}. In this case the limit is {\it finite} and the result is $\Sigma^{\rm FM} =0$.

This analysis shows that there is no fundamental flaw in the adiabatic approximation,
and that the spurious divergence of the Fan-Migdal self-energy is an artifact of the
integration procedure. Mathematically the divergence arises from collapsing the two
imaginary poles at $\pm i q_{\rm LO}$ into a double pole at the origin. A similar problem
arises in the textbook Fourier transform of the Coulomb potential.

In practical {\it ab initio} calculations the adiabatic approximation can be retained
without incurring into a divergence as follows. First we perform calculations where all the phonon
frequencies are set to a small constant, say $\w_0$. After converging the summation
over the Brillouin zone, we repeat the calculations for smaller values of
$\w_0$, so as to take the adiabatic limit $\w_0\rightarrow 0$. This procedure 
will yield a finite self-energy.
\bibliography{references} 

\clearpage
\newpage

\begin{figure}
\includegraphics[width=0.42\textwidth]{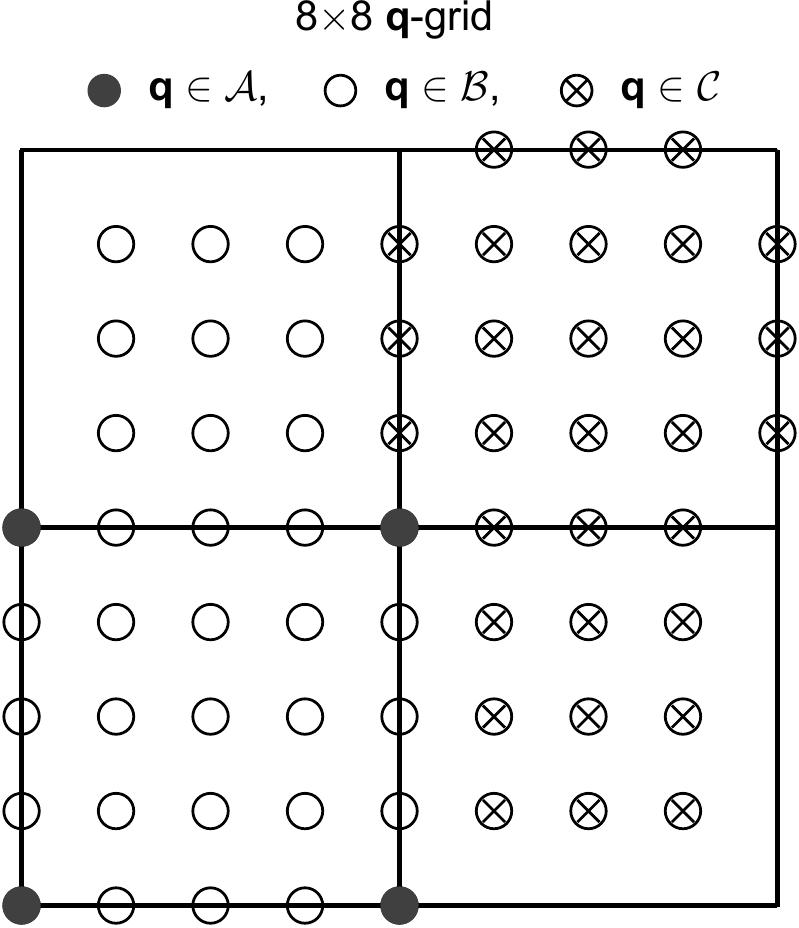}
  \caption{\label{fig_1}
  Partition of the Brillouin zone grid in sets $\mathcal{A}$, $\mathcal{B}$, and $\mathcal{C}$.
  Shown is an 8$\times$8 two-dimensional $\Gamma$-centered grid. Wavevectors which are invariant
  with respect to time-reversal $-\bq = \bq+\bG$ belong to set $\mathcal{A}$ (filled disks),
  The other wavevectors are separated into a set $\mathcal{B}$ which does not contain the
  time-reversal partner of any element (empty circles), and set $\mathcal{C}$ obtained by
  reversing the sign of the wavevectors in set $\mathcal{B}$ (crossed empty circles).
  }
\end{figure}

\clearpage
\newpage

\begin{figure}
\includegraphics[width=0.45\textwidth]{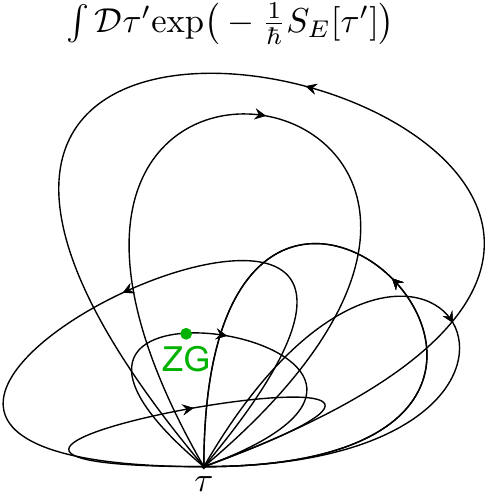}
    \caption{\label{fig_2}
    Schematic representation of closed-loop paths required to evaluate the finite-temperature
    electronic and optical properties using thermodynamic Feynman's path integrals.
    The Williams-Lax theory is equivalent to averaging the property of interest over many
    atomic configuration $\{\tau\}$. The weight of each configuration is given by the
    the sum of all possible path integrals starting and ending with the configuration $\{\tau\}$.
    The special displacement method replaces the average over paths by a single evaluation of 
    the property using the ZG displacement (green disk).
    } 
\end{figure}

\clearpage
\newpage

\begin{figure}
\vspace*{1cm}
\includegraphics[width=0.42\textwidth]{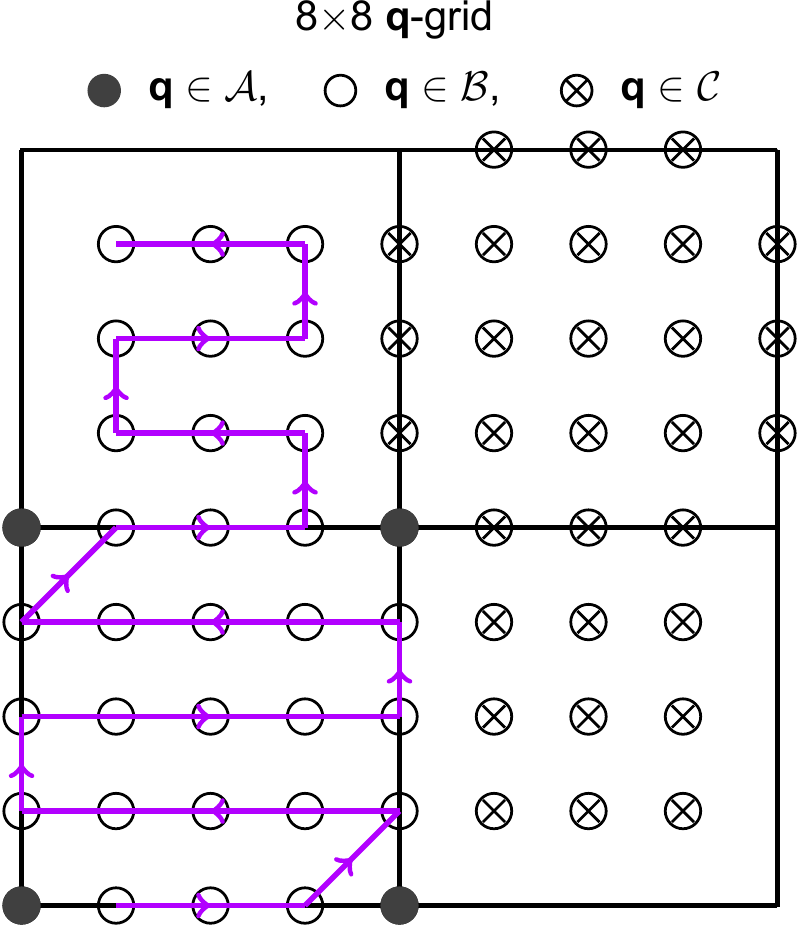}
   \caption{\label{fig_3} 
      Example of space-filling curve employed to order the phonon wavevectors
      along a one-dimensional array. For illustration purposes we show the
      curve passing through the $\bq$-points in set $\mathcal{B}$ of the Brillouin
      zone grid shown in Fig.~\ref{fig_1}.
    }
\end{figure}

\clearpage
\newpage

\begin{figure*}
\vspace*{5cm}
        \begin{tikzpicture}
        \node[inner sep=0pt] (russell) at (-9,0) 
{\includegraphics[width=0.48\textwidth]{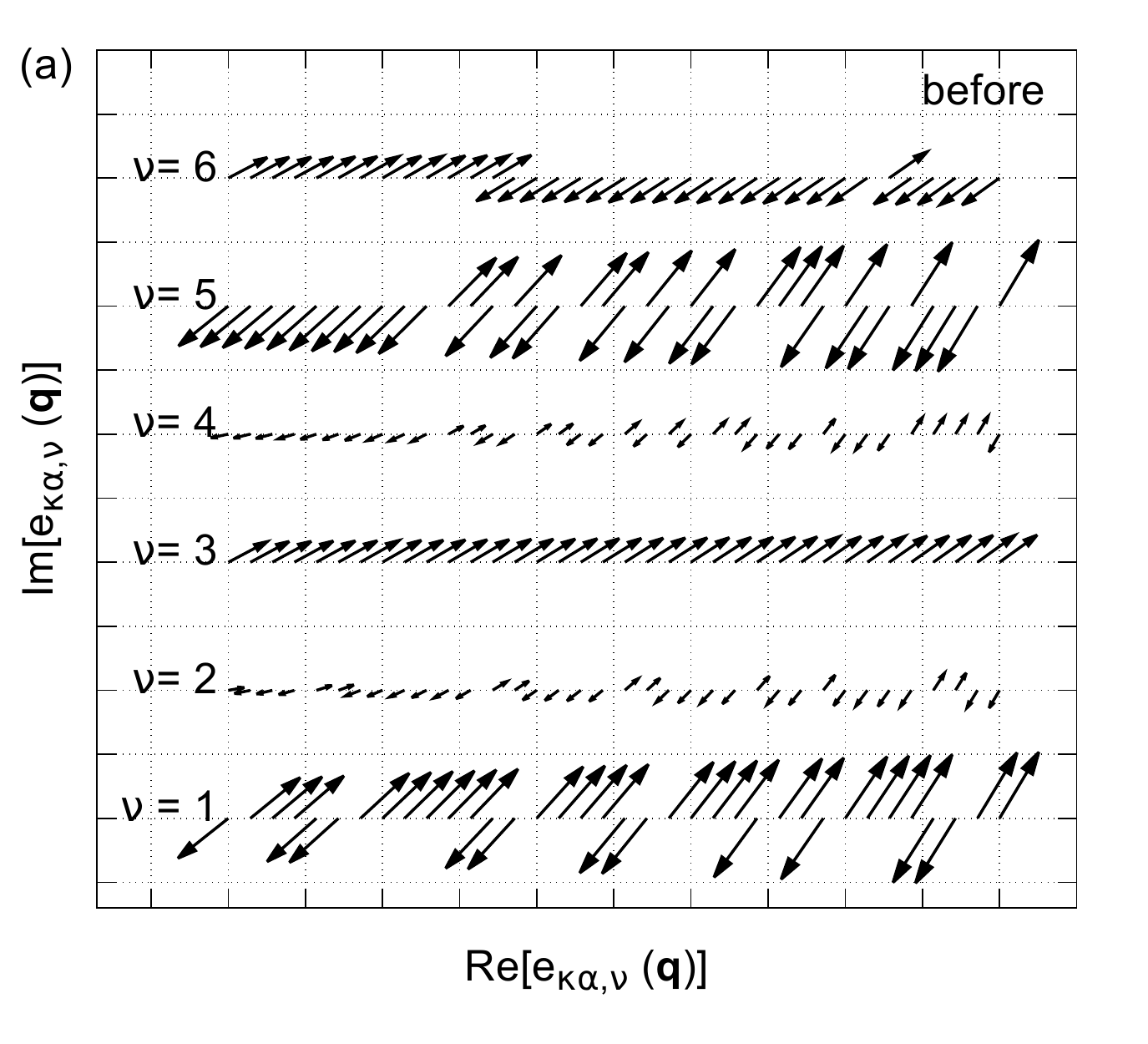}};
        \node[inner sep=0pt] (russell) at (0,0) 
{\includegraphics[width=0.48\textwidth]{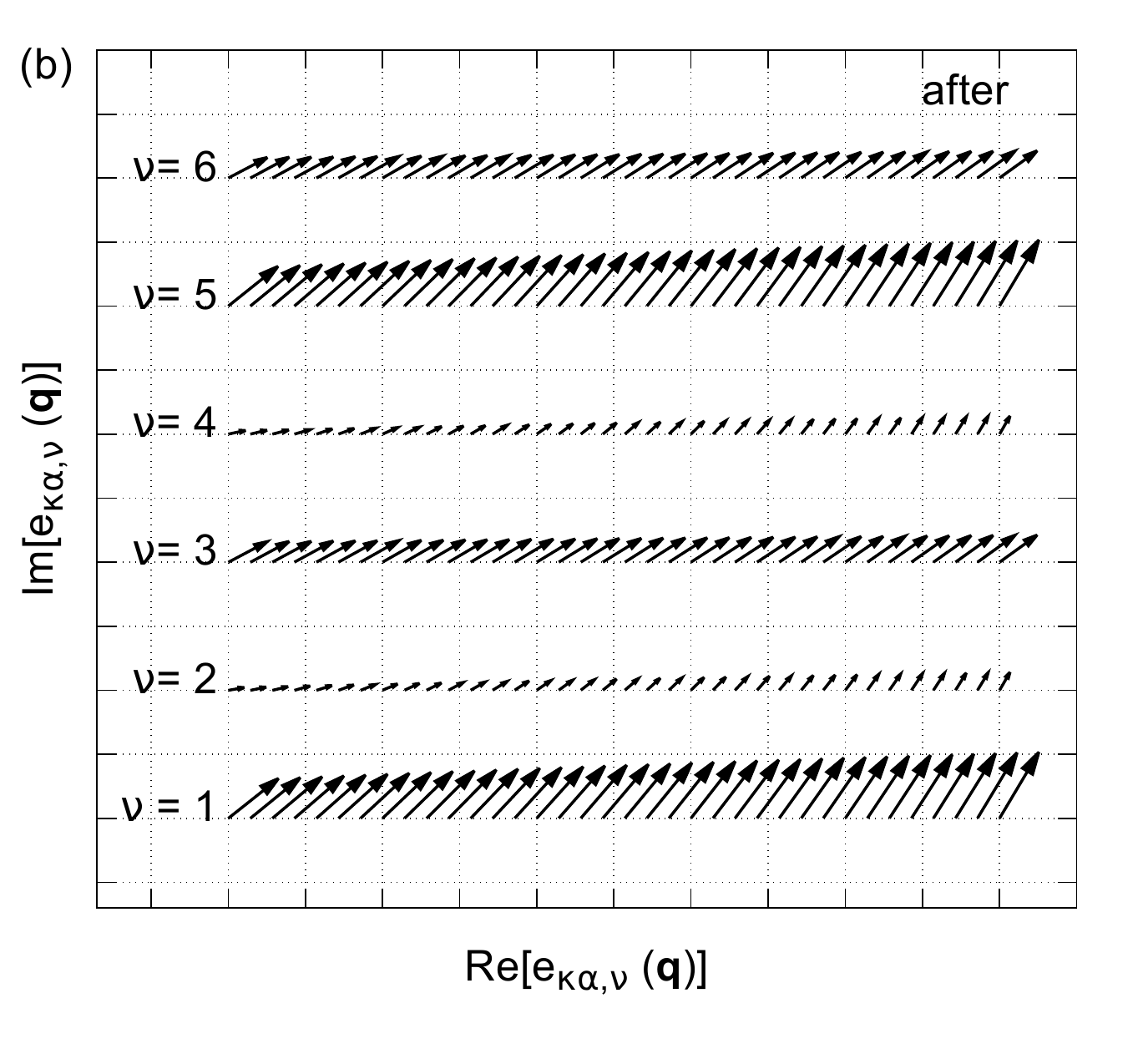}};
        \end{tikzpicture}
 \caption{\label{fig_4}
  Setting up a smooth Berry connection for the vibrational eigenmodes of 
  silicon.  (a) Eigenmodes as obtained from the diagonalization of the dynamical matrix, 
  shown as complex vectors. In this plot we set $\k = 1$, $\a=1$, and we follow
  a line along the [100] direction of a 100$\times$100$\times$100 Brillouin
  zone grid. The eigenmodes jump discontinuously
  between adjacent $\bq$-points. (b) The same eigenmodes as in (a), this time after using the 
  algorithm described in Sec.~\ref{synch_eigenv} to enforce a smooth Berry connection between 
  eigenmodes at adjacent $\bq$-points. In this case the eigenmodes vary smoothly along the 
  Brillouin-zone path.
 }
\end{figure*}

\newpage

\begin{figure*}
\begin{tikzpicture}
         \node[inner sep=0pt] (russell) at (-12,-0.0) 
{\includegraphics[width=0.5\textwidth]{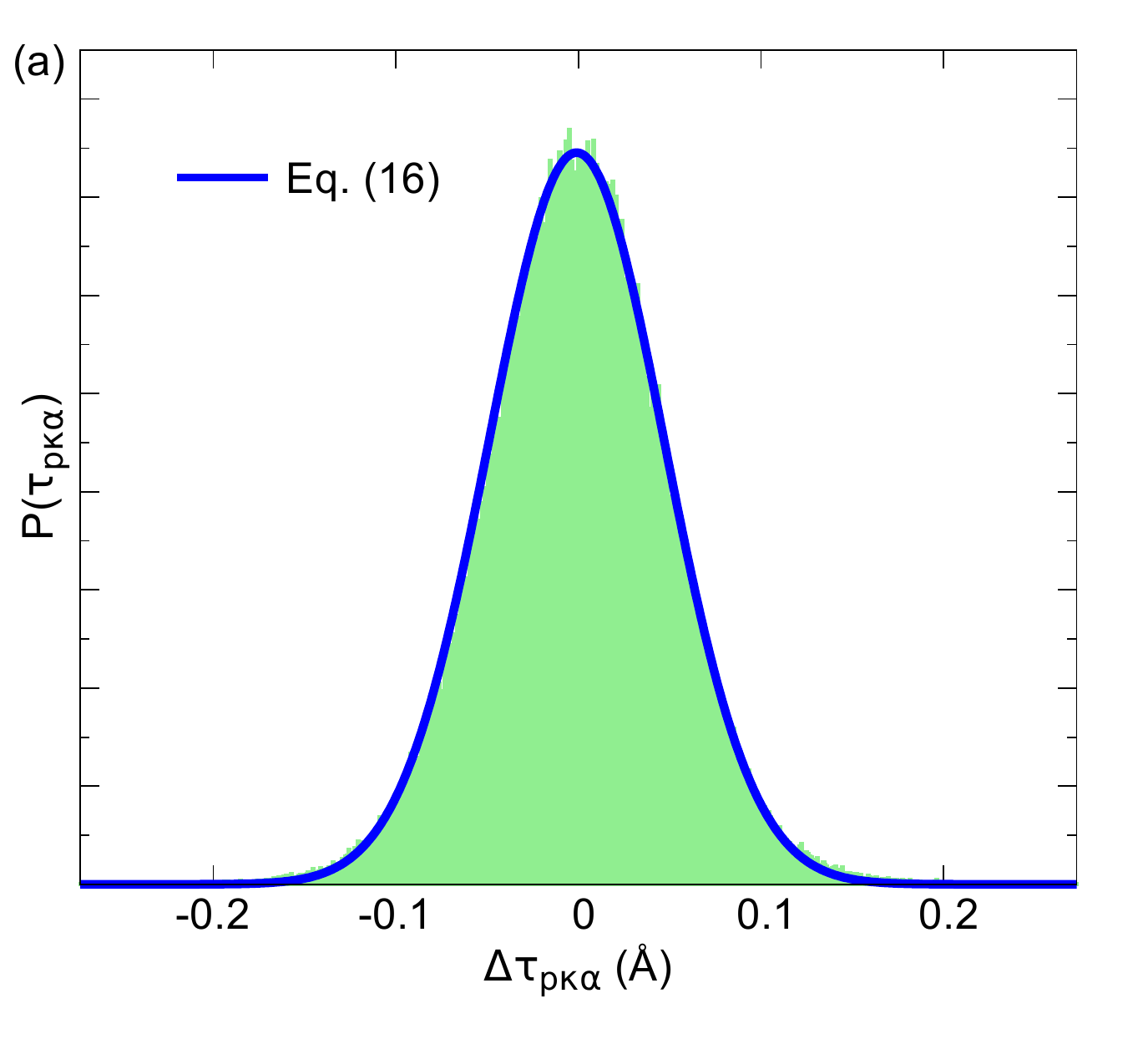}};
         \node[inner sep=0pt] (russell) at (-3,-0.0) 
{\includegraphics[width=0.465\textwidth]{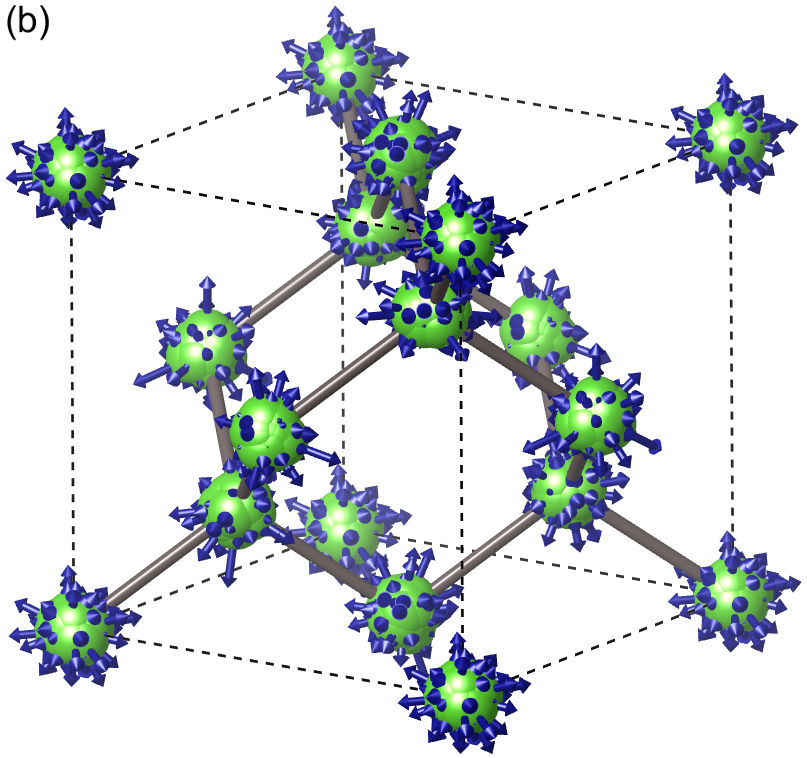}};
         \node[inner sep=0pt] (russell) at (-9.10,1.50) 
{\includegraphics[width=0.21\textwidth]{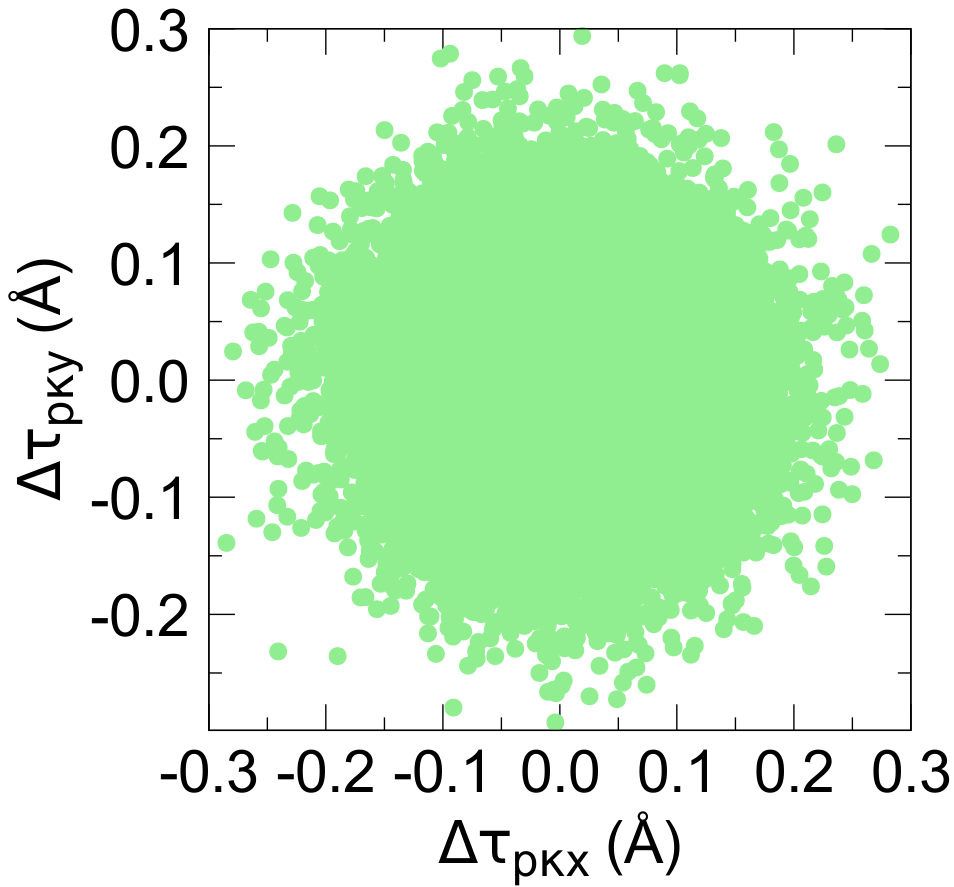}};
\end{tikzpicture}
 \caption{\label{fig_5} 
   (a) Normalized probability distribution $P(\DD \tau_{ p \k \a})$ of the atomic 
   displacements $\DD \tau_{ p \k \a}$ around a silicon atom, along the [100] direction . 
   The blue line represents the exact distribution
   from Eq.~\eqref{eq.prob_distribution}, and the filled green area represents a normalized 
   histogram of the ZG displacements, evaluated for a 50$\times$50$\times$50 supercell of
   silicon at $T=0$~K. In both cases the standard deviation is 0.05~\AA.
   The inset represents a scatter plot of the atomic displacements in the (001) plane, 
   all referred to the same Si atom. (b) Ball-stick model of silicon with the atoms 
   folded back in the unit cell  after the  ZG-displacement in a 8$\times$8$\times$8 supercell. 
   The ZG displacements are for $T=0$~K and are shown as arrows. The
   displacements have been scaled $\times$4.5 for clarity.
 }
\end{figure*}

\newpage

\begin{figure*}
\begin{tikzpicture}
         \node[inner sep=0pt] (russell) at (-14.7,-7.0) 
{\includegraphics[width=0.45\textwidth]{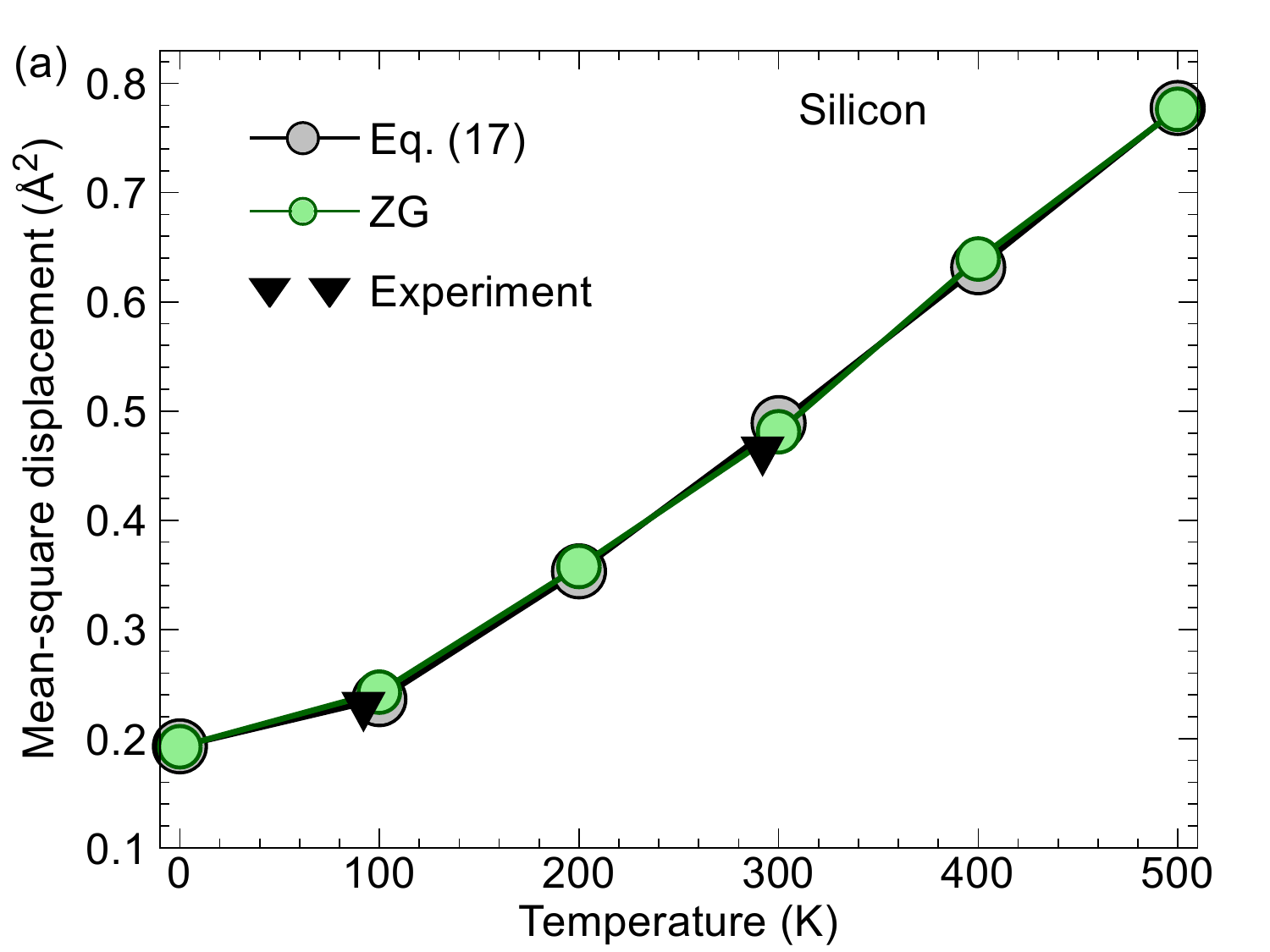}};
        \node[inner sep=0pt] (russell) at (-5.8,-7.15) 
{\includegraphics[width=0.34\textwidth]{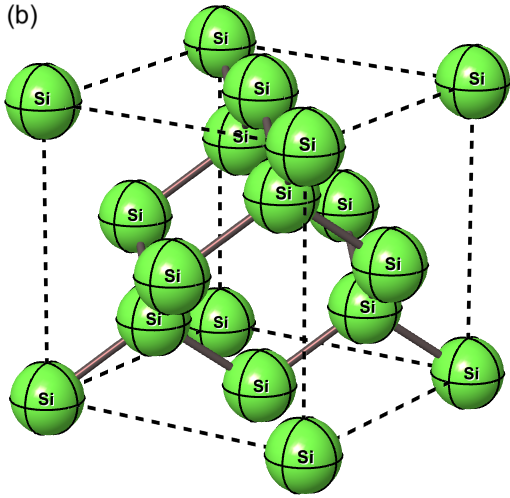}};
    \node[inner sep=0pt] (russell) at (-14.7,-13.0) 
{\includegraphics[width=0.45\textwidth]{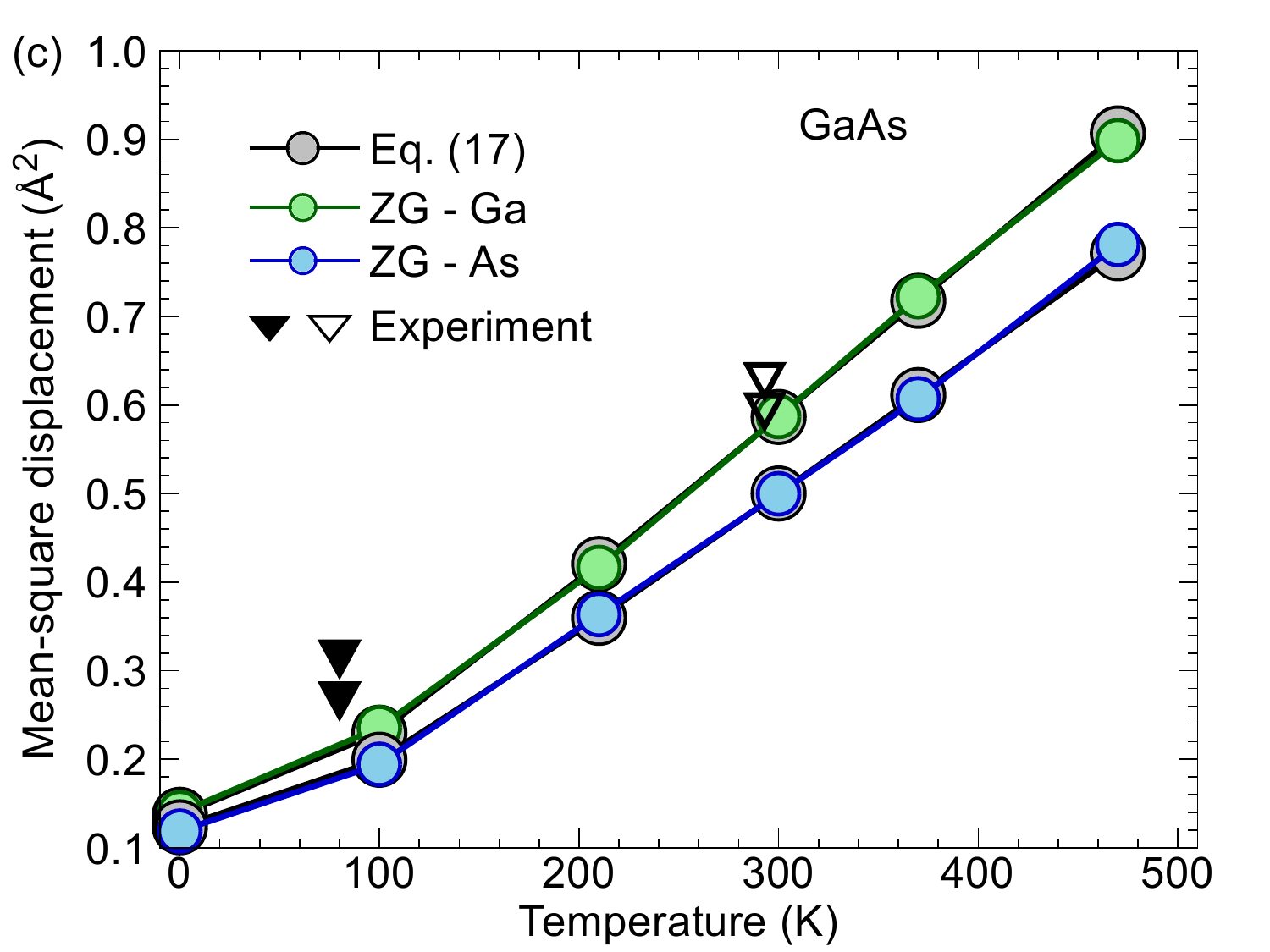}};
         \node[inner sep=0pt] (russell) at (-5.7,-13.15) 
{\includegraphics[width=0.34\textwidth]{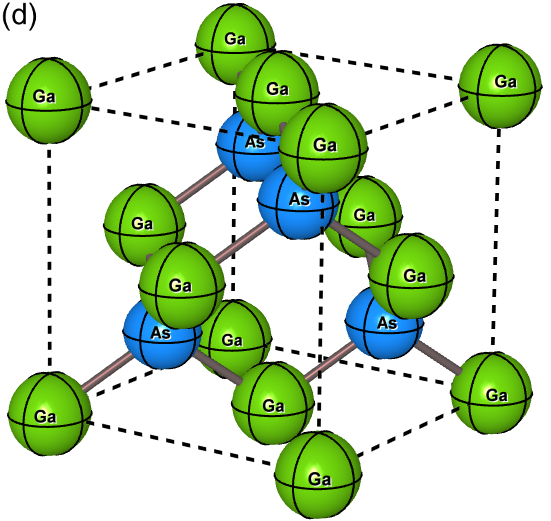}};
        \node[inner sep=0pt] (russell) at (-14.7,-19) 
{\includegraphics[width=0.45\textwidth]{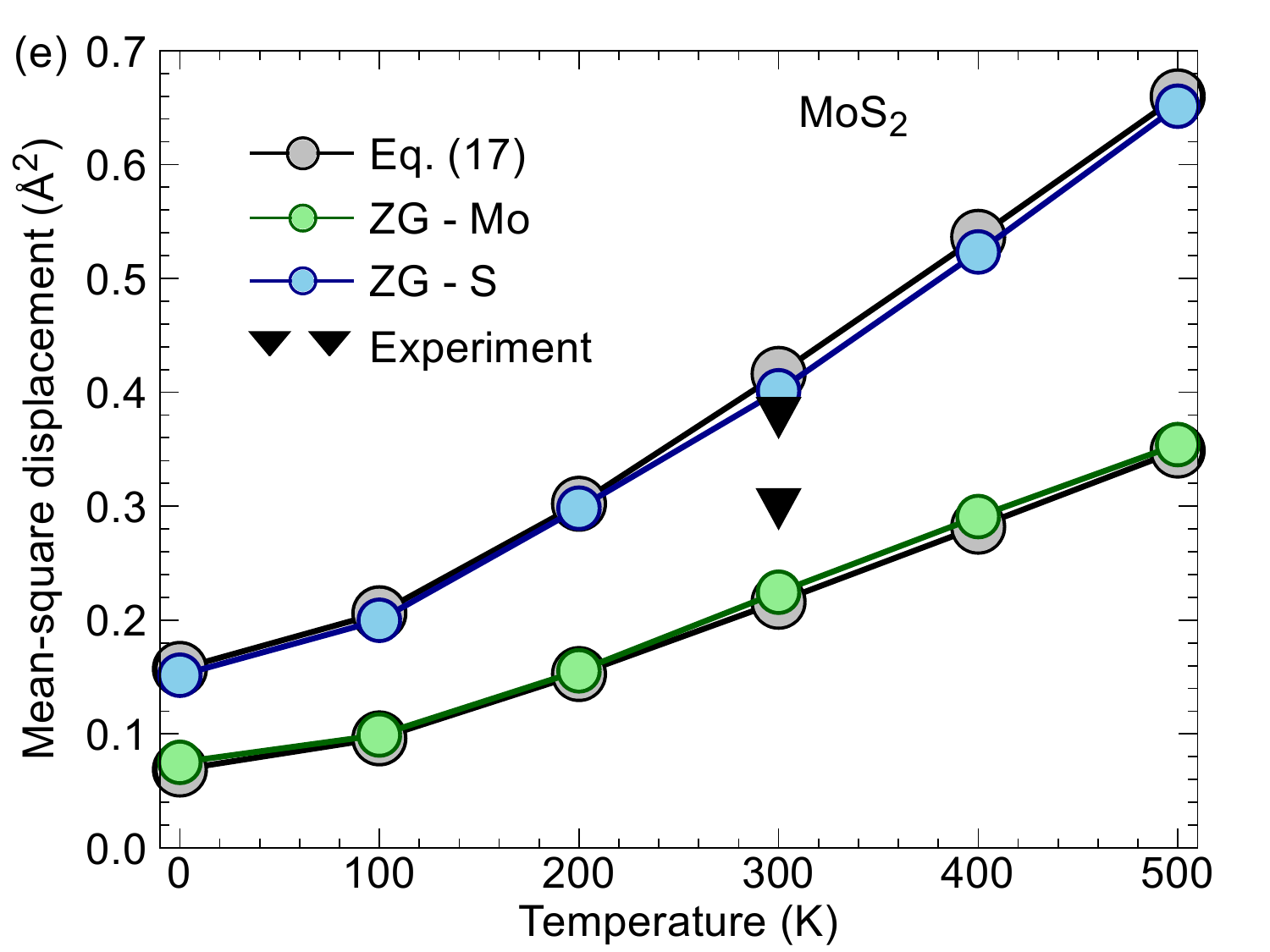}};
        \node[inner sep=0pt] (russell) at (-5.8,-19) 
{\includegraphics[width=0.5\textwidth]{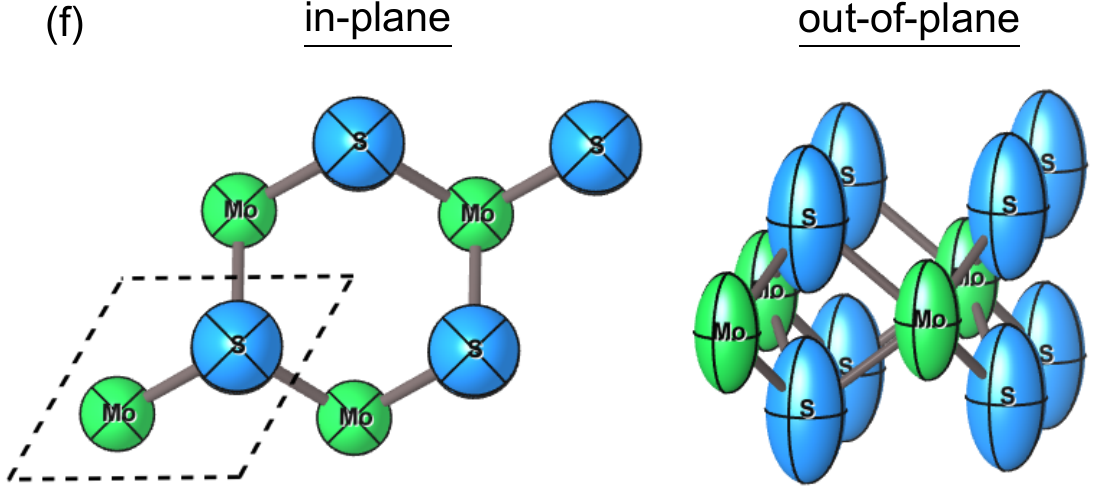}};
\end{tikzpicture}
 \caption{\label{fig_6} 
   (a) Mean-square thermal displacements of silicon as a function of temperature,
   evaluated using Eq.~\eqref{eq.Debye_waller_factor} (gray disks) and the ZG displacement
   (green disks). Experimental data from Ref.~[\onlinecite{Aldred_1973}] are shown as 
   black triangles. (b) Thermal ellipsoids of silicon at $T=300$~K, as obtained 
   from the ZG displacement. (c) Mean-square thermal displacements of GaAs as a function 
   of temperature, evaluated using Eq.~\eqref{eq.Debye_waller_factor} (gray disks) and the 
   ZG displacement (green and blue disks). We also report experimental data from 
   Ref.~[\onlinecite{Stern_1980}] (filled triangles) and Ref.~[\onlinecite{Matsushita_1974}]. 
   These data correspond to weighted averages of the displacements of Ga and As. 
   (d) Thermal ellipsoids of GaAs at $T=300$~K, as obtained from the ZG displacement.
   (e) In-plane mean-square thermal displacements of monolayer MoS$_2$ as a function of temperature,
   evaluated using Eq.~\eqref{eq.Debye_waller_factor} (gray disks) and the ZG displacement
   (green and blue disks). We also report experimental data from Ref.~[\onlinecite{Schonfeld_1983}].
   (f) Thermal ellipsoids of monolayer MoS$_2$ at $T=300$~K, as obtained from the ZG displacement.
 }
\end{figure*}

\newpage

\begin{figure*}
\begin{tikzpicture}
        \node[inner sep=0pt] (russell) at (-8,0) 
{\includegraphics[width=0.43\textwidth]{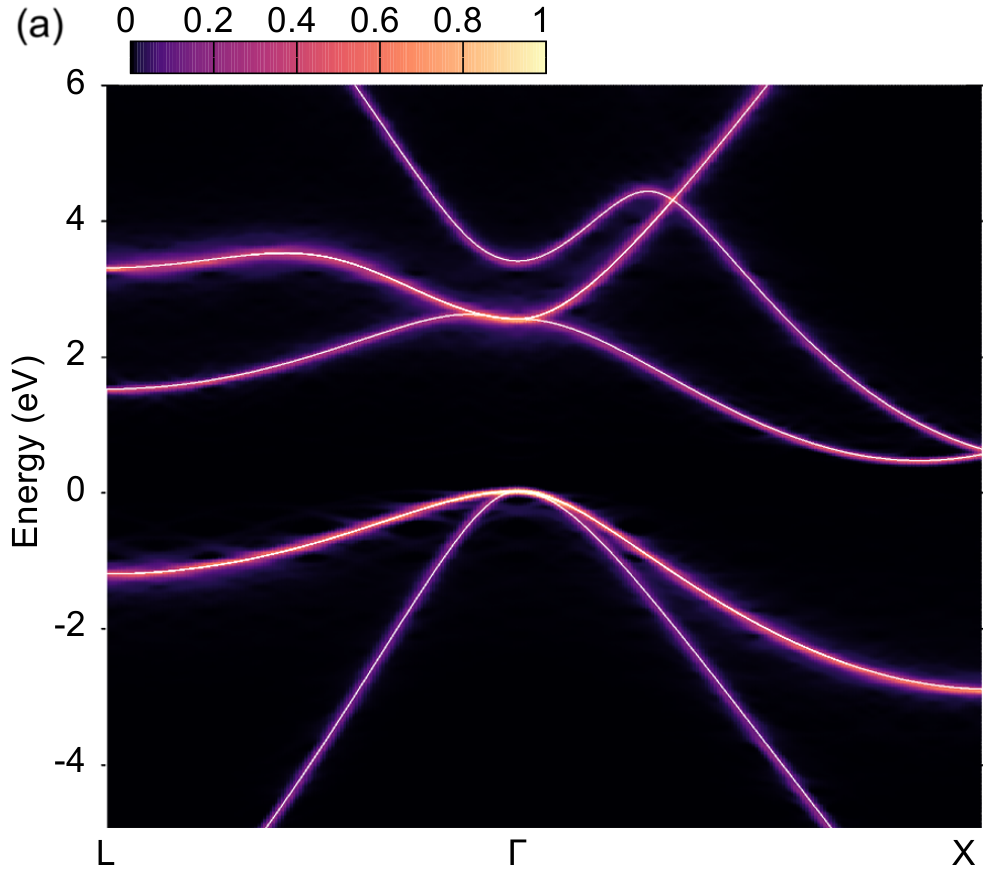}};         
        \node[inner sep=0pt] (russell) at (2.0,-0.95) 
{\includegraphics[width=0.22\textwidth]{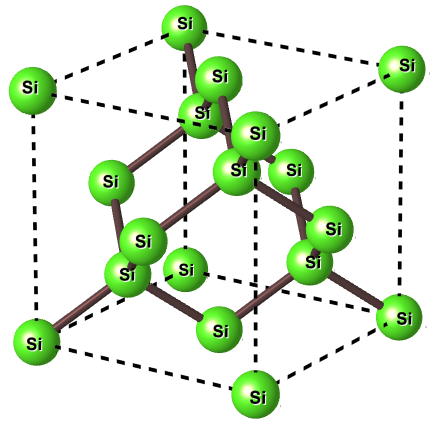}}; 
        \node[inner sep=0pt] (russell) at (0.4,0) 
{\includegraphics[width=0.45\textwidth]{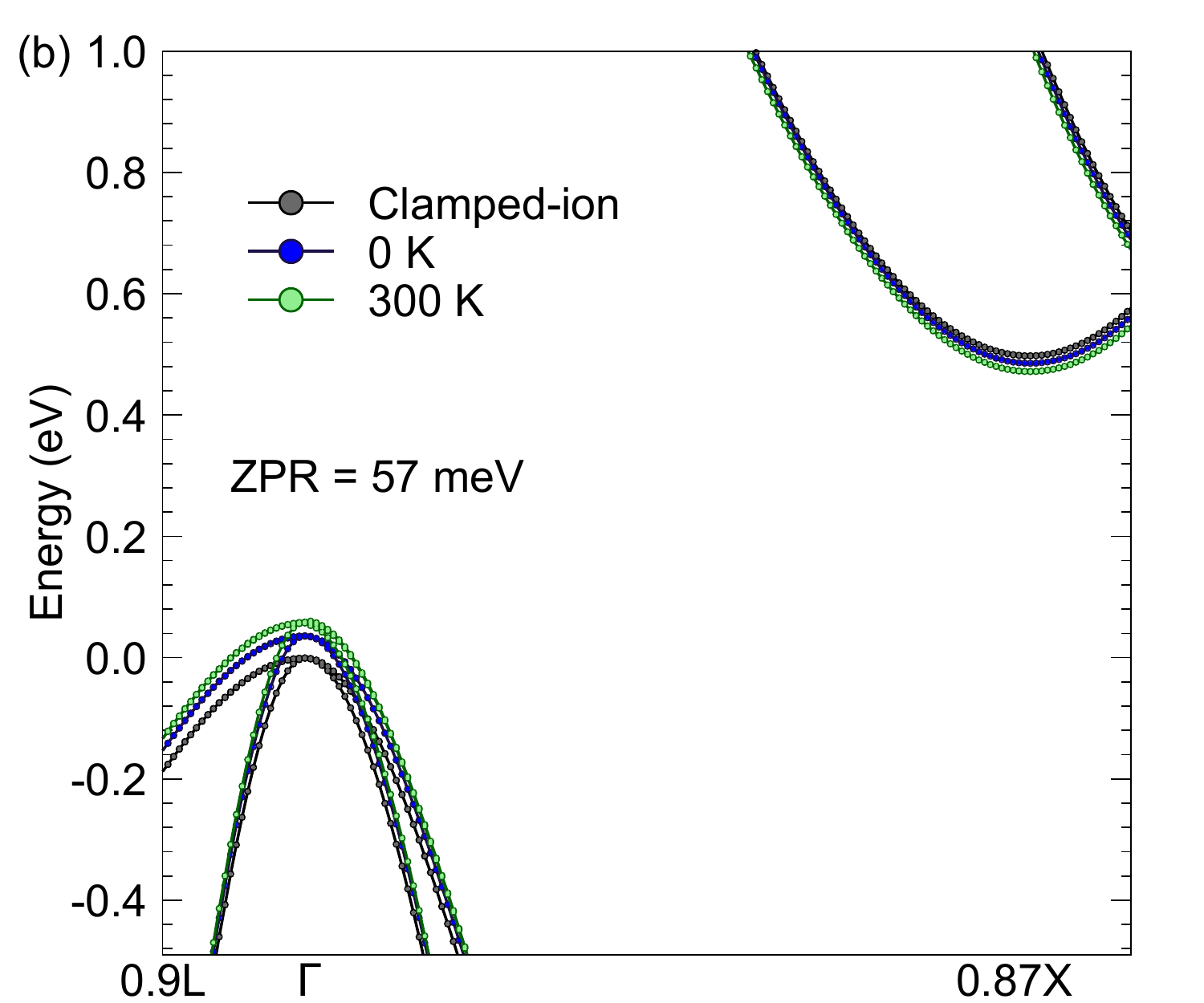}};
        \node[inner sep=0pt] (russell) at (-8,-7.2) 
{\includegraphics[width=0.455\textwidth]{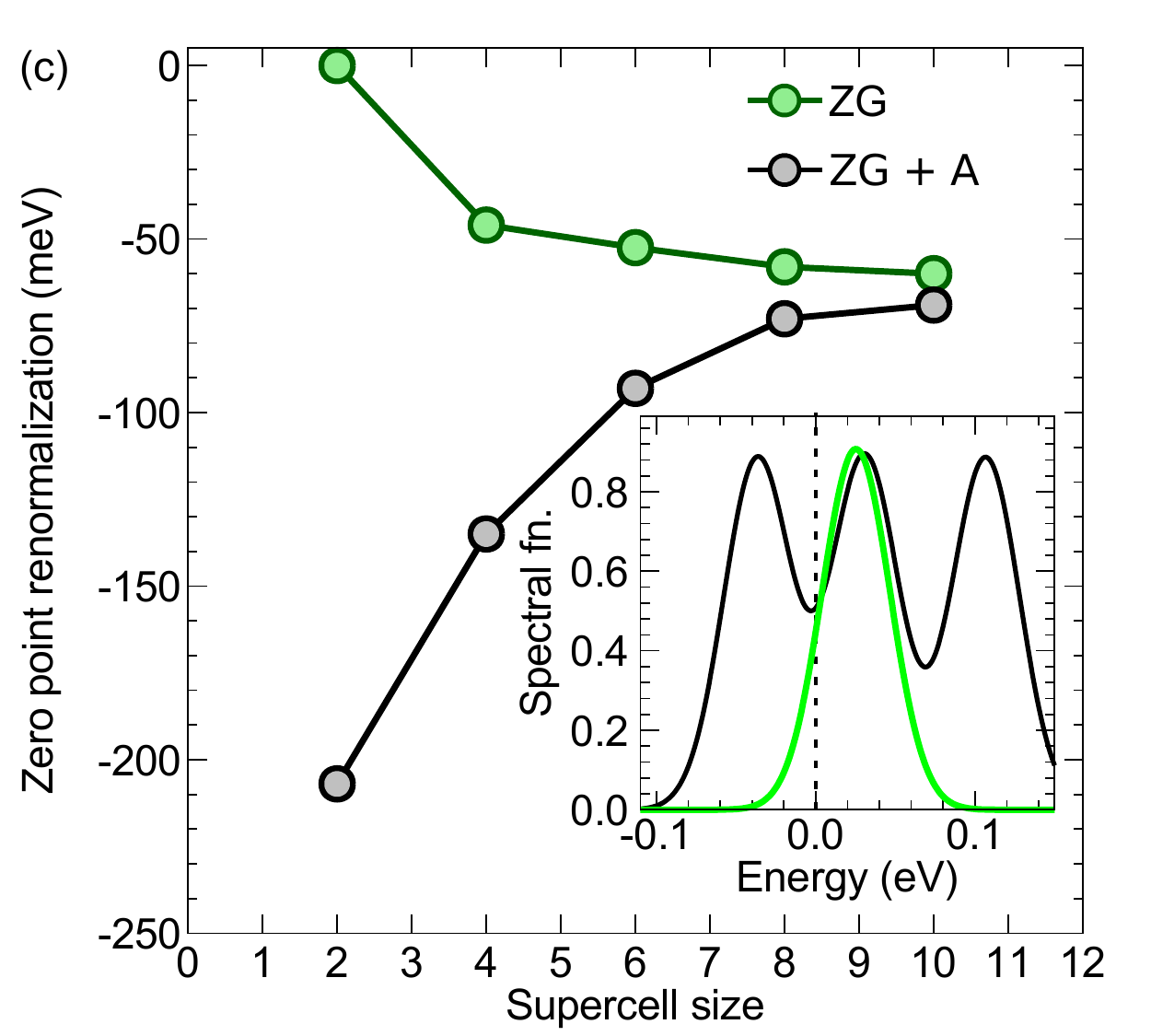}};         
        \node[inner sep=0pt] (russell) at (0.36,-7.2) 
{\includegraphics[width=0.455\textwidth]{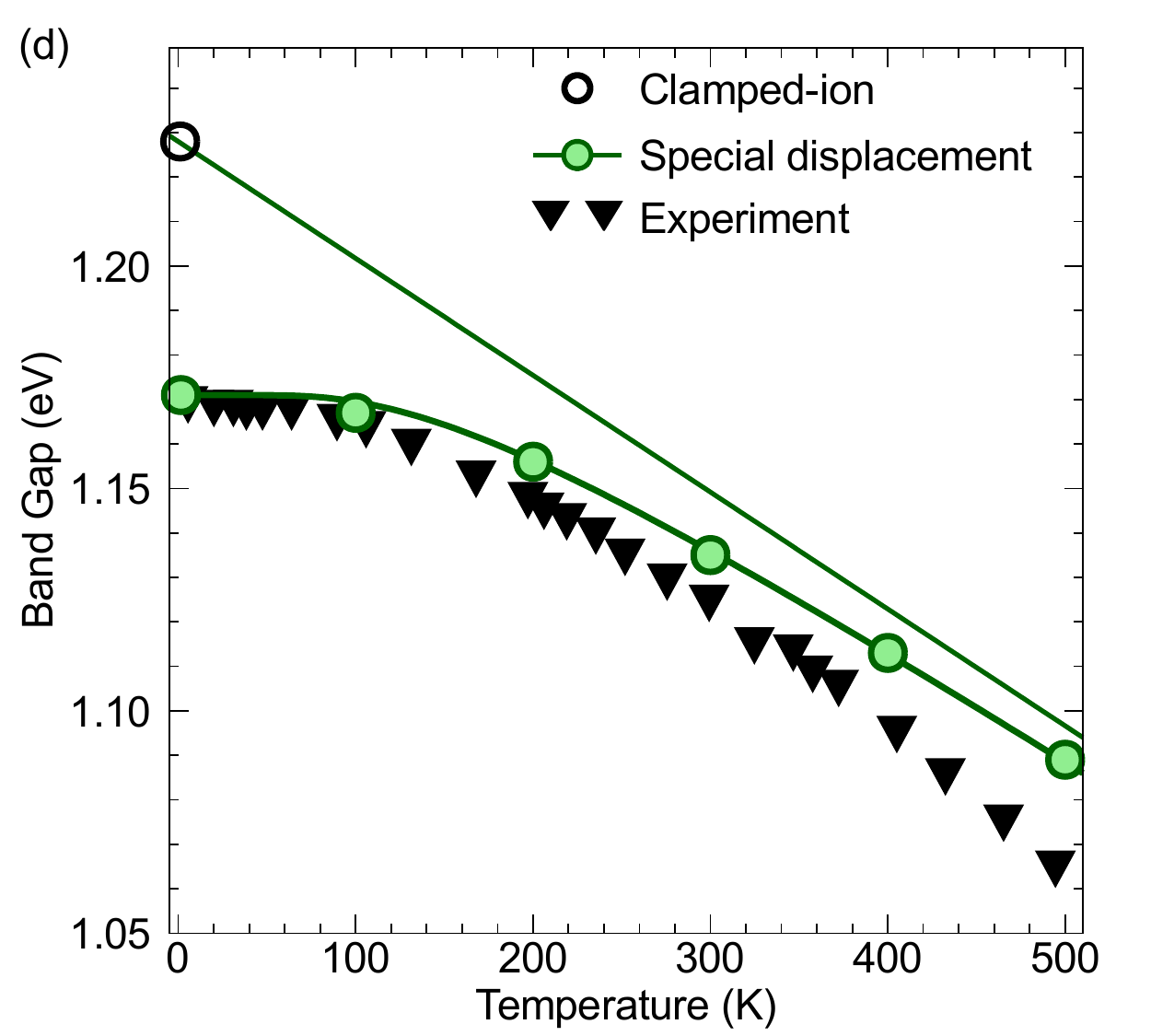}};
\end{tikzpicture}
  \caption{\label{fig_7}
   (a) Spectral function of silicon calculated using the ZG displacement at $T=0$~K.
   The calculation was performed using an 8$\times$8$\times$8 supercell and the
   unfolding procedure described in Sec.~\ref{sec.unfolding}. We sampled the
   L-$\Gamma$-X path on 280 equally-spaced $\bk$-points, and the zero of the
   energy axis is referred to the valence band top. (b) Band structures of silicon 
   at clamped ions (black), $T=0$~K (blue), and $T=300$~K (green). The bands were 
   extracted from spectral functions like the one in (a). We also report the calculated 
   zero-point renormalization (ZPR) and a ball-stick model. (c) Sensitivity of the 
   calculated ZPR of silicon to the size of the supercell. The horizontal axis indicates 
   the linear size $N$ of the $N\!\times \!N\!\times \!N$ supercell. We show both the 
   calculations performed using the ZG displacement (green), and the results obtained 
   by also including $\bq$-points in set $\mathcal{A}$ (grey). In the latter case 
   the threefold degeneracy of the valence band top is lifted (inset), and the band 
   gap is evaluated by considering the topmost valence state. (d) Temperature dependence of 
   the indirect band gap of silicon up to 500~K. We show the results of the special 
   displacement method (green circles) and experimental data from Ref.~[\onlinecite{Alex_1996}] 
   (black triangles). The calculated band gaps were scissor-shifted by 0.73~eV to 
   match the experimental value at 4~K. The straight line is the high-temperature 
   limit and intercepts the $T=0$~K axis at the clamped-ion band gap (1.23~eV, empty circle).
 }
\end{figure*}

\newpage

\begin{figure*}
\begin{tikzpicture}
        \node[inner sep=0pt] (russell) at (-8,0) 
{\includegraphics[width=0.43\textwidth]{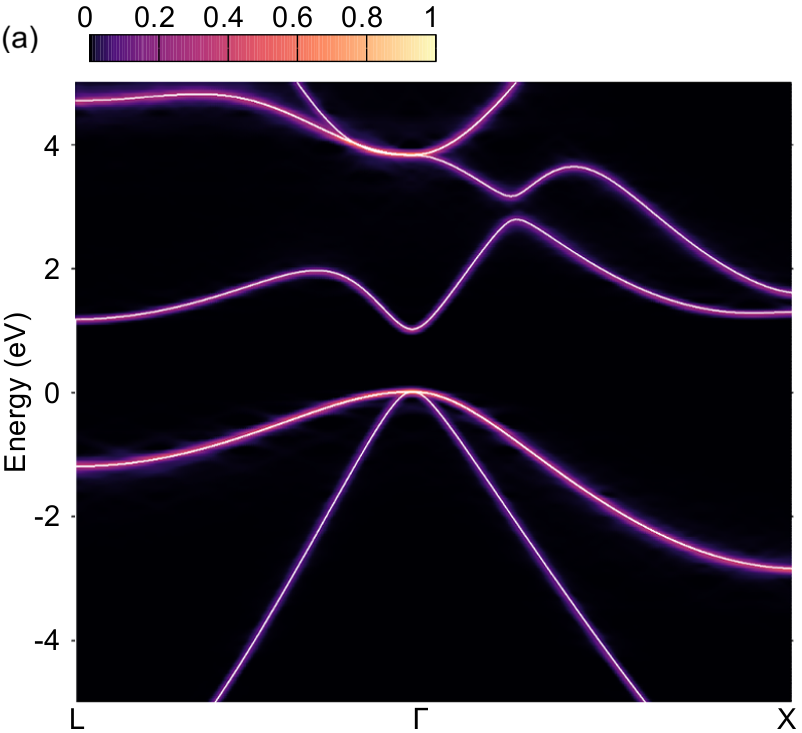}};         
        \node[inner sep=0pt] (russell) at (1.9,0) 
{\includegraphics[width=0.21\textwidth]{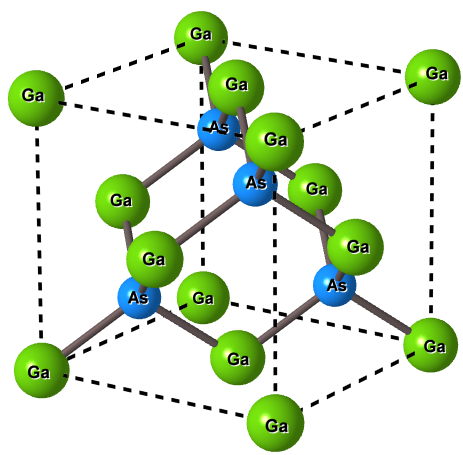}}; 
        \node[inner sep=0pt] (russell) at (0.4,0) 
{\includegraphics[width=0.45\textwidth]{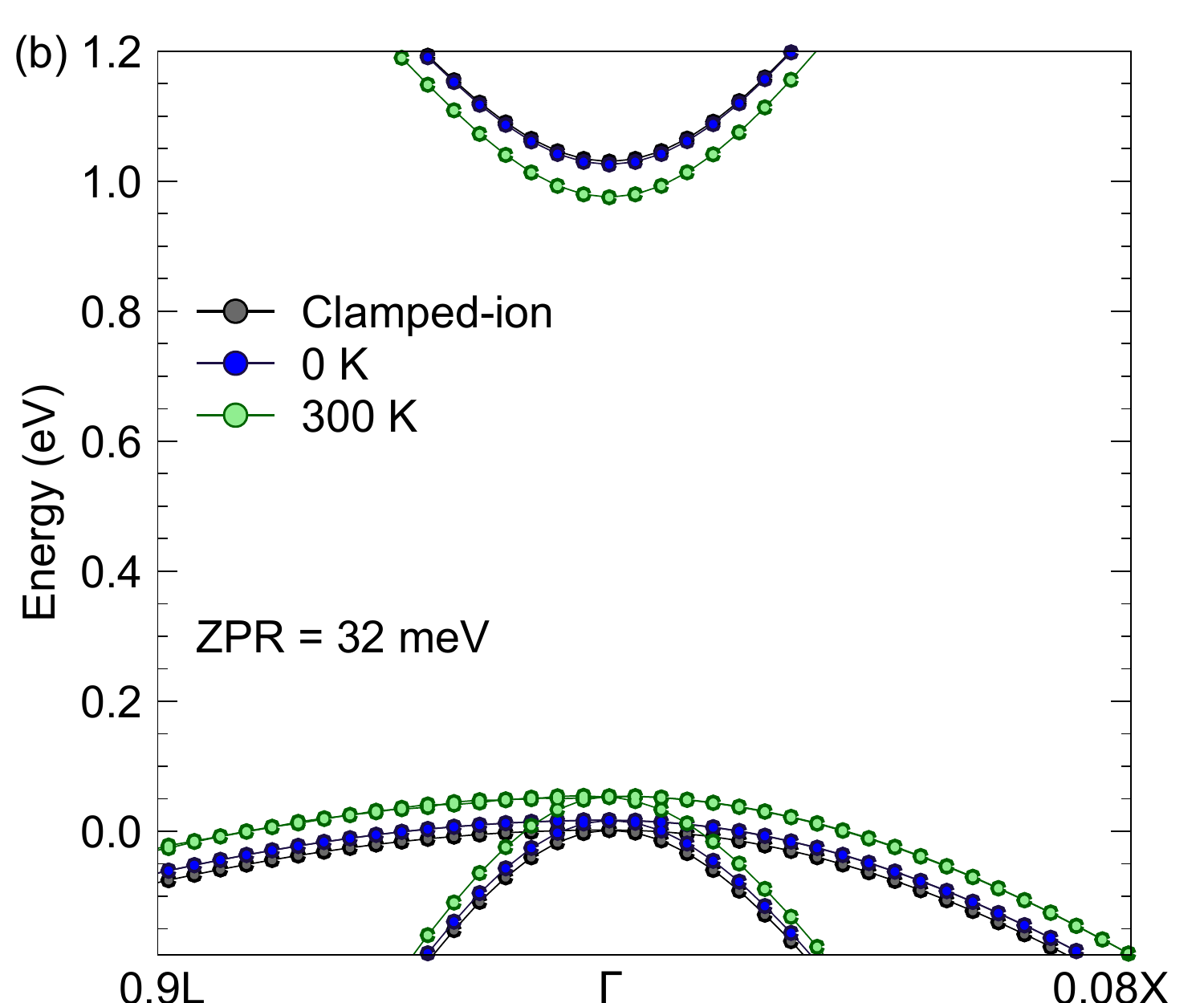}};
        \node[inner sep=0pt] (russell) at (-8,-7.22) 
{\includegraphics[width=0.455\textwidth]{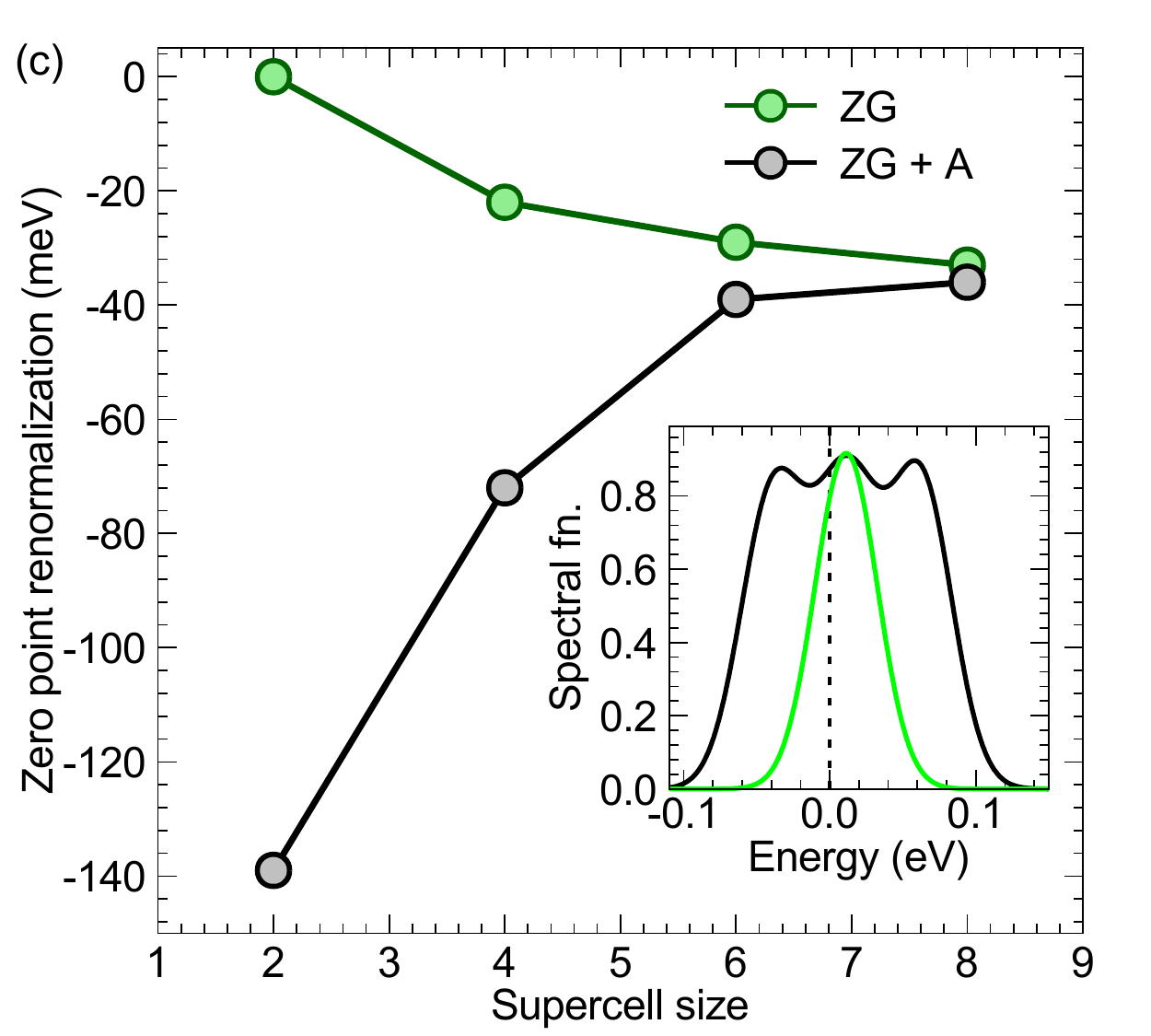}};         
        \node[inner sep=0pt] (russell) at (0.33,-7.22) 
{\includegraphics[width=0.459\textwidth]{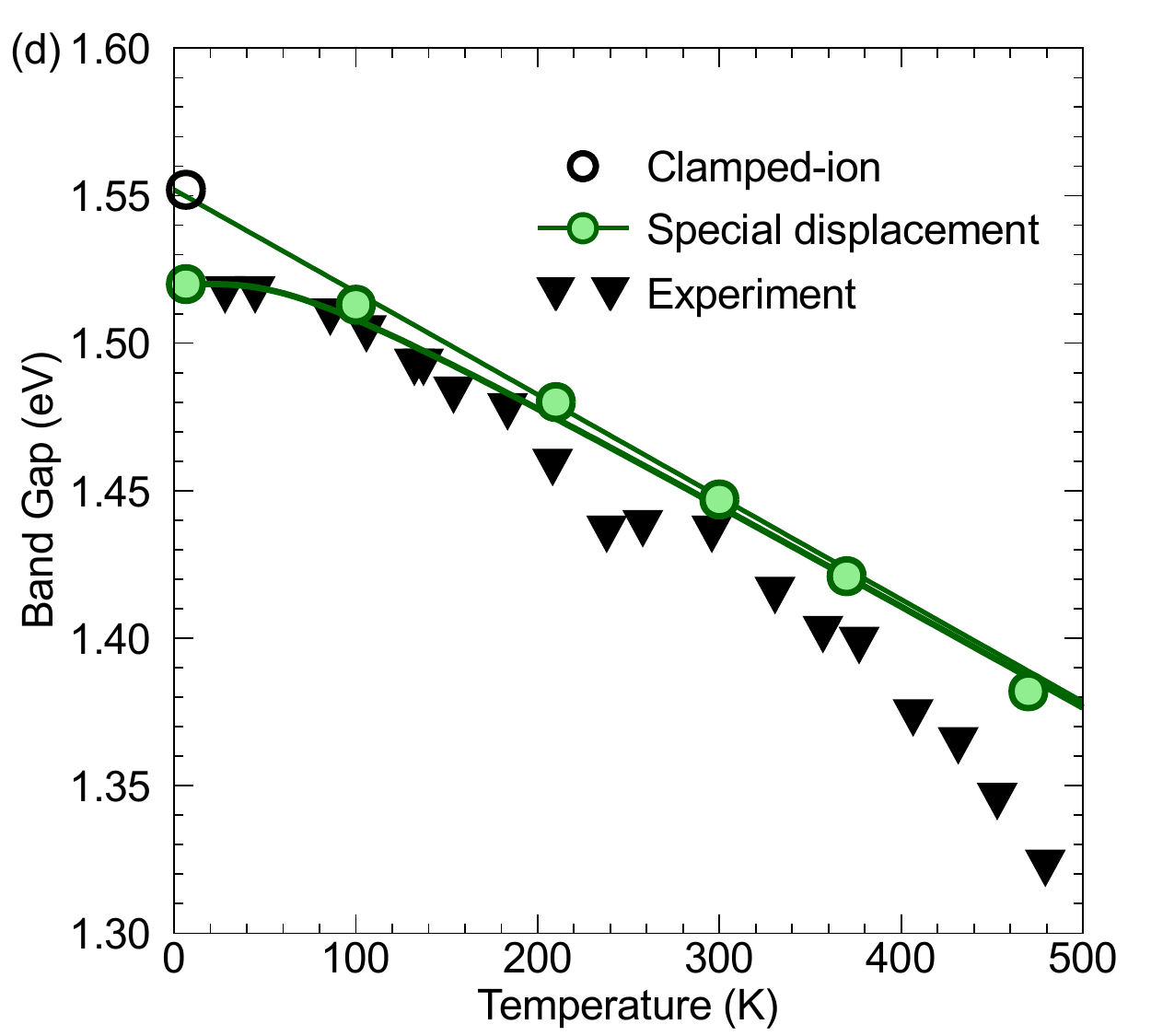}};
\end{tikzpicture}
 \caption{\label{fig_8}
   (a) Spectral function of GaAs calculated using the ZG displacement at $T=0$~K.
   The calculation was performed using an 8$\times$8$\times$8 supercell and the
   unfolding procedure described in Sec.~\ref{sec.unfolding}. We sampled the
   L-$\Gamma$-X path on 280 equally-spaced $\bk$-points, and the zero of the
   energy axis is referred to the valence band top. (b) Band structures of GaAs at 
   clamped ions (black), $T=0$~K (blue), and $T=300$~K (green). The bands were 
   extracted from spectral functions like the one in (a). We also report the calculated 
   zero-point renormalization (ZPR) and a ball-stick model. (c) Sensitivity of the 
   calculated ZPR of GaAs to the size of the supercell. The horizontal axis indicates 
   the linear size $N$ of the $N\!\times \!N\!\times \!N$ supercell. We show both 
   the calculations performed using the ZG displacement (green), and the results 
   obtained by also including $\bq$-points in set $\mathcal{A}$ (grey). In the 
   latter case the threefold degeneracy of the valence band top is lifted (inset), 
   and the band gap is evaluated by considering the topmost valence state. (d) Temperature 
   dependence of the indirect band gap of GaAs up to 500~K. We show the results of the 
   special displacement method (green circles) and experimental data from 
   Ref.~[\onlinecite{Lautenschlager_1987}] (black triangles). The calculated band gaps 
   were scissor-shifted by 0.53~eV to match the experimental value at 25~K. The 
   straight line is the high-temperature limit and intercepts the $T=0$~K axis at 
   the clamped-ion band gap (1.56~eV, empty circle).
  }
\end{figure*}

\newpage

\begin{figure*}
\begin{tikzpicture}
        \node[inner sep=0pt] (russell) at (-8,0) 
{\includegraphics[width=0.43\textwidth]{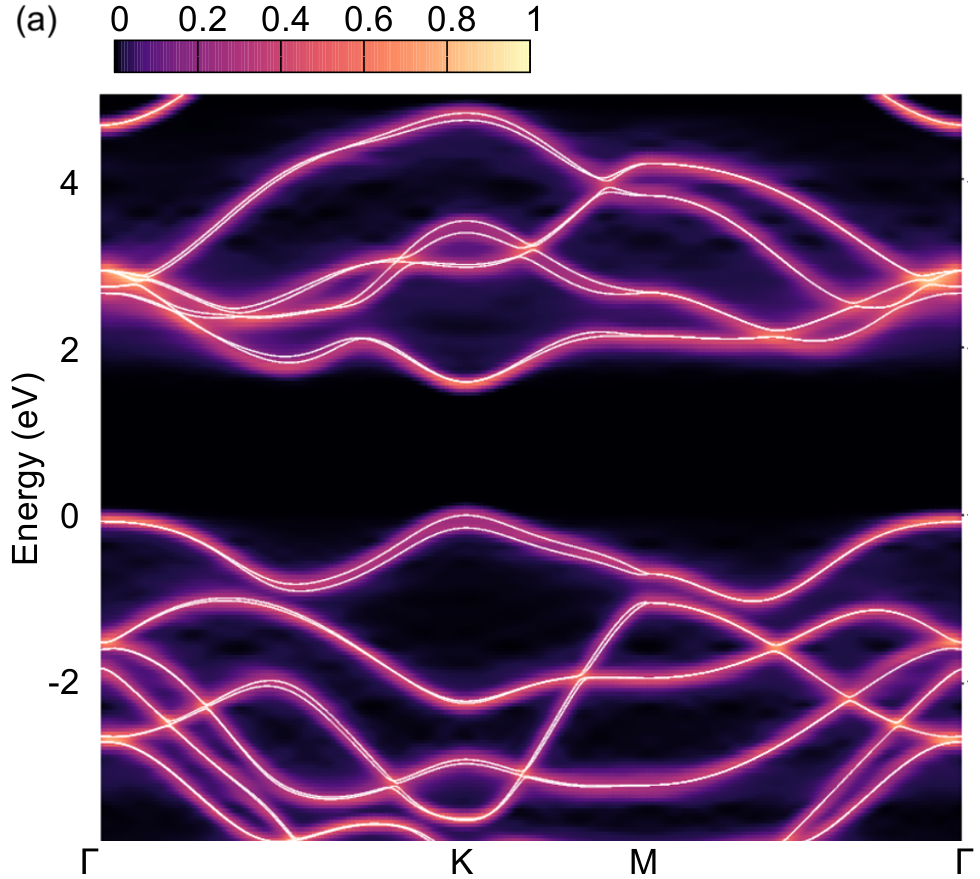}};
         \node[inner sep=0pt] (russell) at (1.2,0) 
{\includegraphics[width=0.31\textwidth]{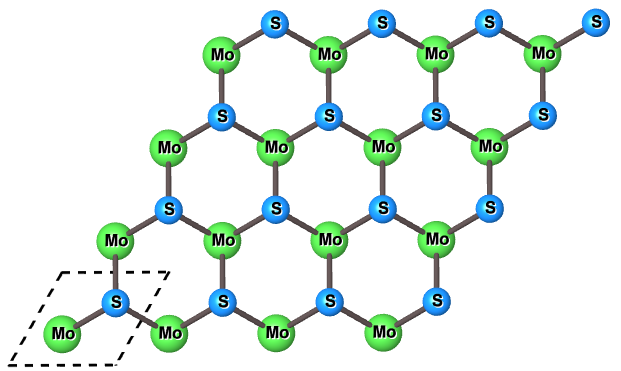}};
        \node[inner sep=0pt] (russell) at (0.4,0) 
{\includegraphics[width=0.45\textwidth]{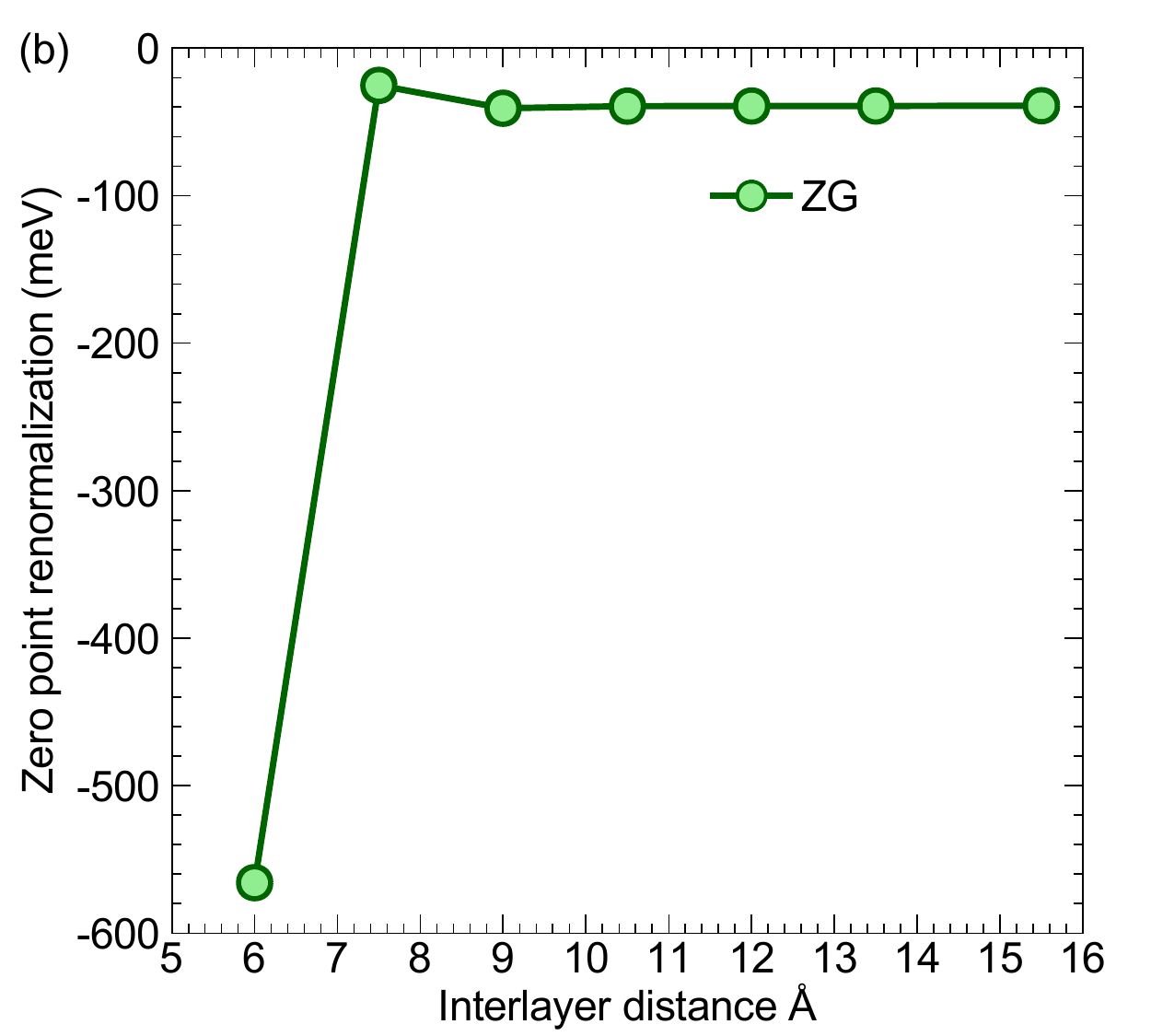}};
        \node[inner sep=0pt] (russell) at (-8,-7.22) 
{\includegraphics[width=0.45\textwidth]{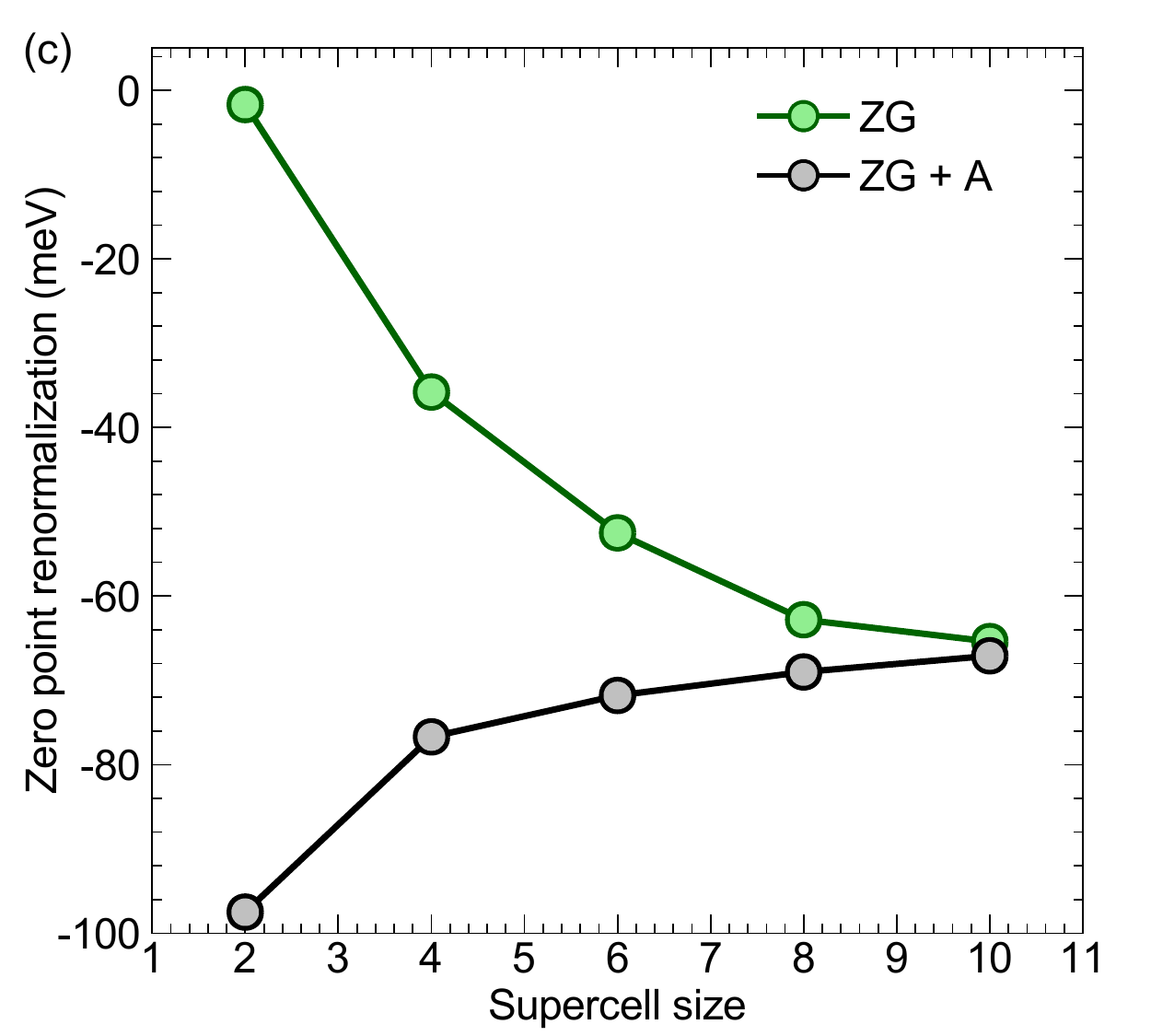}};
        \node[inner sep=0pt] (russell) at (0.4,-7.22) 
{\includegraphics[width=0.45\textwidth]{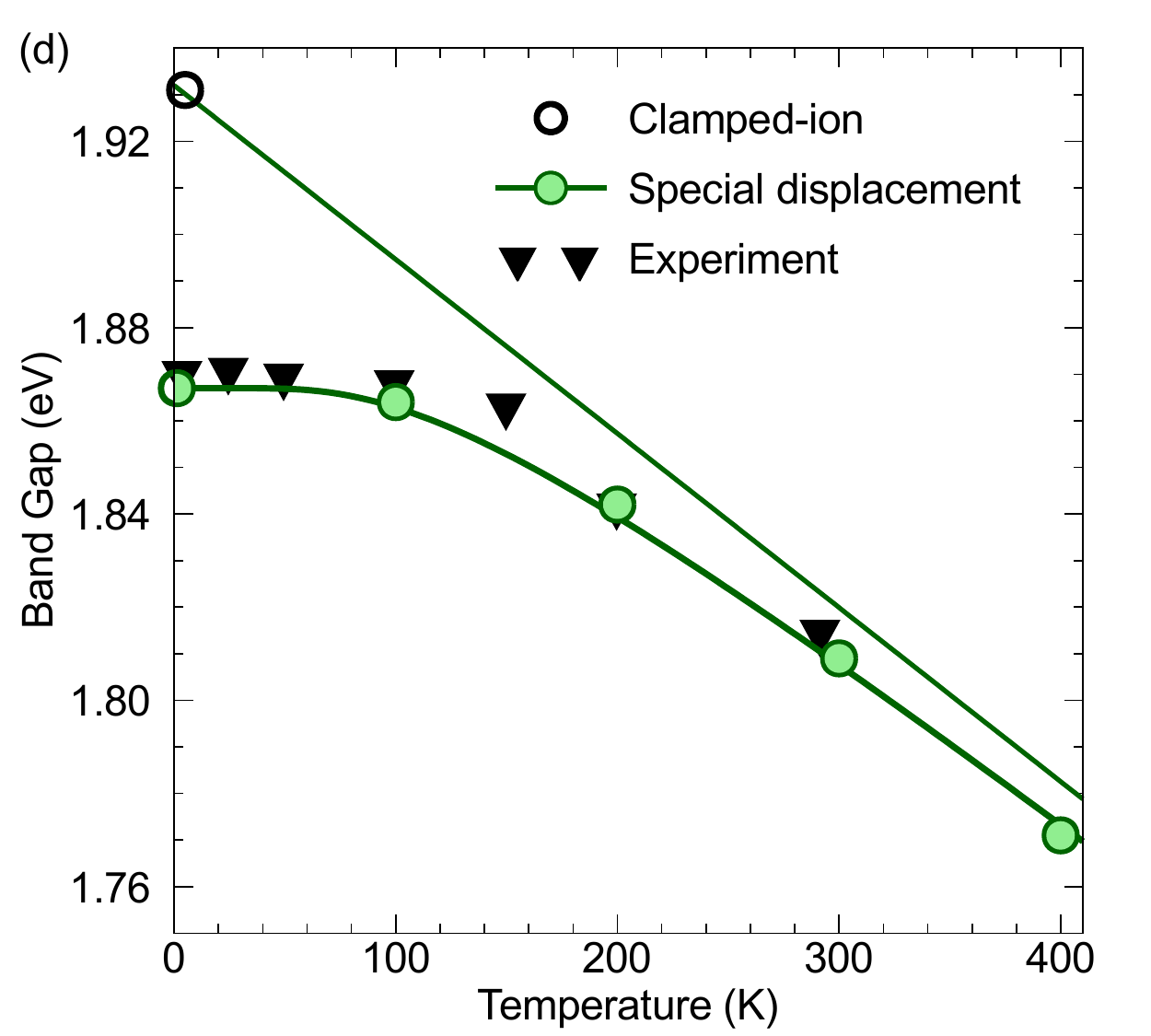}};
\end{tikzpicture}
 \caption{\label{fig_9}
   (a) Spectral function of monolayer MoS$_2$ calculated using the ZG displacement 
   at $T=0$~K. The calculation was performed using a 10$\times$10$\times$1 supercell and 
   the unfolding procedure described in Sec.~\ref{sec.unfolding}. We sampled the
   $\Gamma$-K-M-$\Gamma$ path on 237 equally-spaced $\bk$-points, 
   and the zero of the energy axis is referred to the valence band top.
   (b) ZPR of the direct band gap of monolayer MoS$_2$ vs.\ the interlayer 
   separation. The calculations were performed using a 4$\times$4$\times$1 supercell.
   Also shows is a ball-stick model of monolayer MoS$_2$. (c) Sensitivity of the calculated 
   ZPR of monolayer MoS$_2$ to the size of the supercell. The horizontal axis indicates 
   the linear size $N$ of the $N\!\times \!N\!\times \! 1$ supercell. We show both the 
   calculations performed using the ZG displacement (green), and the results obtained  
   by also including $\bq$-points in set $\mathcal{A}$ (grey). (d) Temperature dependence 
   of the indirect band gap of monolayer MoS$_2$ up to 400~K, evaluated using a 
   10$\times$10$\times$1 supercell. We show the results of the special displacement 
   method (green circles) and experimental data from Ref.~[\onlinecite{Park_2018}] 
   (black triangles). The calculated band gaps were scissor-shifted by 0.34~eV to match 
   the experimental value at 4~K. The straight line is the high-temperature limit 
   and intercepts the $T=0$~K axis at the clamped-ion band gap (1.93~eV, empty circle).
 }
\end{figure*}

\end{document}